\documentclass[prd,aps,twocolumn,showkeys,nofootinbib]{revtex4-2}

\usepackage{amsmath}
\usepackage{amsfonts}
\usepackage{amssymb}	
\usepackage{graphicx}
\usepackage{bm}
\usepackage{color}
\usepackage{ulem}
\usepackage{commath}
\usepackage[T1]{fontenc}
\usepackage{lmodern}
\usepackage{enumitem}   

\usepackage{hyperref}
\usepackage[dvipsnames]{xcolor}

\allowdisplaybreaks

\def\p{\partial}

\def\GMc2{G M_{\odot} c^{-2}}

\def\lm{{\ell m}}
\def\EOB{{\rm EOB}}

\def\lm{{\ell m}}

\def\lm{{\ell m}}

\def\ph{\varphi}

\def\F{{\cal F}}

\newcommand\be{\begin{equation}}
\newcommand\ee{\end{equation}}

\def\Qo{Q_{\omega}}
\def\PAT{1PAT1}
\def\omg_b{\omega^{\rm GSF}_{\rm break}}

\def\TEOBResumS{\texttt{TEOBResumS}}

\newcommand{\virg}[1]{``#1''}

\begin{document}

\title{Comparing second-order gravitational self-force, numerical relativity and effective one body
waveforms from inspiralling, quasi-circular and nonspinning black hole binaries}
\author{Angelica \surname{Albertini}${}^{1,2}$}
\author{Alessandro \surname{Nagar}${}^{3,4}$}
\author{Adam \surname{Pound}${}^5$}
\author{Niels \surname{Warburton}${}^6$}
\author{Barry \surname{Wardell}${}^6$}
\author{Leanne \surname{Durkan}${}^6$}
\author{Jeremy \surname{Miller}${}^7$}

\affiliation{${}^1$Astronomical Institute of the Czech Academy of Sciences,
Bo\v{c}n\'{i} II 1401/1a, CZ-141 00 Prague, Czech Republic}
\affiliation{${}^2$Faculty of Mathematics and Physics, Charles University in Prague, 18000 Prague, Czech Republic}
\affiliation{${}^3$INFN Sezione di Torino, Via P. Giuria 1, 10125 Torino, Italy} 
\affiliation{${}^4$Institut des Hautes Etudes Scientifiques, 91440 Bures-sur-Yvette, France}
\affiliation{${}^5$School of Mathematical Sciences and STAG Research Centre, University of Southampton, Southampton, United Kingdom, SO17 1BJ}
\affiliation{${}^6$School of Mathematics and Statistics, University College Dublin, Belfield, Dublin 4, Ireland, D04 V1W8}
\affiliation{${}^7$Department of Physics, Ariel University, Ariel 40700, Israel}

\begin{abstract}
We present the first systematic comparison between gravitational waveforms emitted by inspiralling, quasi-circular and nonspinning black hole binaries computed with three different approaches: 
second-order gravitational self-force (2GSF) theory, as implemented in the 1PAT1 model; 
numerical relativity (NR), as implemented by the SXS collaboration; 
and the effective one body (EOB) formalism, as implemented in the \TEOBResumS{} waveform model.
To compare the models we use both a standard, time-domain waveform alignment and a gauge-invariant analysis based on the 
dimensionless function $Q_\omega(\omega)\equiv \omega^2/\dot{\omega}$, where $\omega$ is the gravitational wave frequency. 
We analyse the domain of validity of the 1PAT1 model, deriving error estimates and showing that the effects of the final transition to 
plunge, which the model neglects, extend over a significantly larger frequency interval than one might expect. Restricting to the inspiral 
regime, we find that, while for mass ratios $q = m_1/m_2\le 10$  \TEOBResumS{} is largely indistinguishable from NR,  1PAT1 
has a significant dephasing $\gtrsim 1$~rad; conversely, for $q\gtrsim 100$, 1PAT1 is estimated to have phase errors $<0.1$~rad on a large 
frequency interval, while \TEOBResumS{} develops phase differences $\gtrsim1$~rad with it. Most crucially, on that same large frequency 
interval we find good agreement between \TEOBResumS{} and 1PAT1 in the intermediate regime $15\lesssim q\lesssim 64$, 
with $<0.5$~rad dephasing between them. A simple modification to the \TEOBResumS{} flux further improves this agreement for 
$q\gtrsim 30$, reducing the dephasing to $\approx0.27$~rad even at $q=128$. While our analysis points to the need for more highly 
accurate, long-inspiral, NR simulations for $q\gtrsim15$ to precisely quantify the accuracy of EOB/2GSF waveforms, we can clearly 
identify the primary sources of error and routes to improvement of each model. 
In particular, our results pave the way for the construction of GSF-informed EOB models for both intermediate and extreme mass 
ratio inspirals for the next generation of gravitational wave detectors.
\end{abstract}
\date{\today}

\maketitle

\section{Introduction}

The gravitational self-force (GSF) formalism deals with the two-body problem in general relativity by computing the 
deviation from geodesic motion due to the gravitational field of the smaller object. 
A recent work~\cite{Wardell:2021fyy} presented the first calculation of the waveforms obtained by 
solving Einstein's equations in second-order gravitational self-force (2GSF) theory~\cite{Pound:2015tma, Pound:2017psq, Barack:2018yvs}. 
This new result complements other recent achievements regarding the  2GSF calculation of the binding 
energy of a particle around a Schwarzschild black hole~\cite{Pound:2019lzj} and the calculation 
of the gravitational wave (GW) energy fluxes using the same approach~\cite{Warburton:2021kwk}. 
Technically, 2GSF means expanding the metric up to second order in the 
small mass ratio\footnote{GSF results are typically  
obtained via expansions in $\epsilon$, but are often re-expressed as expansions in the symmetric mass ratio 
$\nu = m_1 m_2/(m_1 + m_2)^2$.}
$\epsilon = m_2/m_1$ (with $m_2 \ll m_1$) and solving the Einstein equations order-by-order to obtain the metric perturbations while also solving for the motion of the black holes. This is often supplemented by an efficient method for handling the disparity in scales between the the slow radiation-reaction timescale on which
the orbit gradually shrinks and a fast timescale connected to orbital motion. This can be done, for instance, by employing osculating geodesics and applying 
near-identity transformations to remove the dependence on orbital phases from the equations of motion, 
a scheme recently adopted to obtain the evolution of quasi-circular
and eccentric insipirals driven by the first-order self-force~\cite{VanDeMeent:2018cgn,Lynch:2021ogr}.
A different approach (also relying on near-identity averaging transformations) is the two-timescale approximation~\cite{Hinderer:2008dm, Miller:2020bft,Pound:2021qin}, which takes 
explicit advantage of the fact that the binary evolution naturally involves two different timescales.  Although 2GSF theory is designed for extreme 
mass ratios, both Refs.~\cite{Warburton:2021kwk,Wardell:2021fyy}  showed the consistency, to some extent, 
between 2GSF results and highly accurate numerical relativity (NR) simulations for comparable-mass binaries.

Long-inspiral, highly accurate NR simulations, as those obtained 
using the SpEC code and made public via the Simulating eXtreme Spacetimes (SXS) catalog~\cite{SXS:catalog}, are currently limited to mass ratios\footnote{Actually, the $q>10$ simulations presented in Ref.~\cite{Yoo:2022erv} are not yet public,
but the $q=15$ simulation has been compared already to the EOB model we consider in this work~\cite{Nagar:2022icd},
and we also present a comparison to the GSF waveform in the following.} $q\lesssim 15$ (where $q \equiv m_1/m_2 \ge 1$).
Larger mass ratios are typically challenging
for NR methods, making NR simulations difficult to push into the natural domain of validity of the GSF approach.
However, the RIT NR group~\cite{Healy:2019jyf,Healy:2020vre,Healy:2022wdn} has recently started to explore 
larger mass ratios via NR simulations~\cite{Rosato:2021jsq}, notably achieving the successful computation 
of waveforms in the intermediate-mass-ratio (IMR) regime up to $q=128$~\cite{Lousto:2020tnb}. 

The effective one body (EOB) approach~\cite{Buonanno:1998gg,Buonanno:2000ef,Damour:2000we,Damour:2001tu,Damour:2015isa} 
to the general-relativistic two-body dynamics is a powerful analytical formalism that resums post-Newtonian (PN) results, 
obtained and strictly valid in the low-velocity, weak field regime, to make them robust and predictive also 
in the strong-field, fast velocity regime. The model is: (i) additionally informed by NR simulations to improve its
behavior through merger and ringdown and (ii) similarly benchmarked to NR data to test its accuracy all over the 
parameter space. Within the EOB approach, the two-body dynamics is a deformation of the dynamics of 
a test-particle on a Schwarzschild (or Kerr) black hole.
In particular, the spin-aligned \TEOBResumS{} is currently the waveform model that presents the highest level 
of faithfulness\footnote{Another widely used model for quasi-circular binaries, though less NR faithful by approximately 
an order of magnitude, is {\tt SEOBNRv4HM}~\cite{Bohe:2016gbl,Cotesta:2018fcv}.} 
with the largest set of NR simulations available~\cite{Nagar:2020pcj, Riemenschneider:2021ppj, Albertini:2021tbt}.

Exact results in the test-mass limit have been broadly exploited in the development
of EOB models. Historically, the first highly accurate EOB waveform templates 
were validated using Regge-Wheeler-Zerilli (RWZ) perturbation theory~\cite{Nagar:2006xv,Damour:2007xr, Damour:2008gu, Bernuzzi:2010xj}, 
and EOB dynamics in the small or extreme-mass-ratio limit were used in the numerical computation
of test-mass waveforms, numerically solving the RWZ or Teukolsky equations~\cite{Nagar:2006xv, Bernuzzi:2010ty, Bernuzzi:2011aj, 
Harms:2014dqa,  Nagar:2014kha, Harms:2015ixa, Harms:2016ctx, Lukes-Gerakopoulos:2017vkj}. 
The results from such numerical waveforms have then been especially useful in testing the resummation
choices of EOB functions and in checking some crucial elements of the EOB models~\cite{Bernuzzi:2012ku, 
Nagar:2016ayt, Messina:2018ghh, Nagar:2019wrt, Chiaramello:2020ehz}.

However, all the above-mentioned test-mass studies are limited by the fact that the underlying metric is always the Schwarzschild
or the Kerr one. In effect, the motion of the particle is driven only by the time-averaged dissipative part of the
self-force (i.e. the fluxes) ignoring conservative contributions. In this sense, results coming from GSF theory
would be extremely useful to further tune EOB quantities in the small and extreme-mass-ratio regime.
In particular, the conservative part of the self-force allows the evaluation of several quantities that may 
inform the EOB conservative sector, for instance the ISCO shift~\cite{Isoyama:2014mja} or Detweiler's 
redshift variable. The latter has already been exploited to extract higher-order PN 
information~\cite{Bini:2013rfa, Bini:2014nfa, Bini:2015bfb, Bini:2016cje}, and those
results have been already incorporated into EOB potentials~\cite{Barausse:2011dq, Antonelli:2019fmq,Akcay:2015pjz}.
The flexibility of the EOB approach is thus well-adapted, in principle, to give a faithful description of extreme-mass-ratio 
inspirals (EMRIs)~\cite{Yunes:2009ef, Yunes:2010zj, Albanesi:2021rby},
modulo increasing the speed and the accuracy of current models in order to meet the needs of future 
space-based detectors like LISA~\cite{LISA:2017pwj} and TianQin~\cite{TianQin:2015yph}.

%==============
% What we do here
%==============
In this paper we present a comprehensive analysis comparing the recently computed 2GSF waveforms\footnote{Although we can compute all waveform multipoles~\cite{Wardell:2021fyy}, we focus our analysis primarily on the dominant $\ell=m=2$ mode.} of~\cite{Wardell:2021fyy} 
and EOB waveforms obtained with the state-of-the-art model \TEOBResumS{}. 
The analysis spans from comparable-mass binaries to the IMR regime.
In particular, we present explicit comparisons for $q = (7, 10, 15, 32, 64, 128)$. To benchmark these results, 
we also revisit the 2GSF/NR phasing comparisons of Ref.~\cite{Wardell:2021fyy} when needed. To avoid possible systematics that
may arise when comparing waveforms in the time domain, we make crucial use of the gauge-invariant description 
of the phasing provided by the $\Qo\equiv \omega^2/\dot{\omega}$ function (the inverse of the adiabaticity parameter),
where $\omega$ is the GW frequency. This kind of analysis was introduced in the context of comparing EOB and
NR waveforms during the late inspiral of binary neutron stars (BNS) systems, with the goal of understanding the 
relevance of tidal effects during the last orbits~\cite{Baiotti:2010xh,Baiotti:2011am}. The precise calculation of this quantity 
for NR simulations proved to be challenging for BNS~\cite{Bernuzzi:2012ci,Bernuzzi:2014owa}, while it was relatively
straightforward for binary black hole (BBH) simulations produced by the SXS collaboration~\cite{Damour:2012ky}. The $\Qo$ diagnostics were useful
for understanding precisely the impact of spin-spin effects in BNS~\cite{Dietrich:2018uni} as well as the origin of other effects
coming from systematics in waveform models~\cite{Messina:2019uby}. In this work, the use of a well-controlled $\Qo$  is 
crucial in obtaining an improved quantitative understanding of the 2GSF/NR comparisons originally presented in Ref.~\cite{Wardell:2021fyy}.

This paper is organized as follows. Section~\ref{sec:GSF} outlines in some detail the basic elements of the 2GSF time-domain
waveform model \PAT{} introduced in Ref.~\cite{Wardell:2021fyy}, along with an internal analysis of the model's errors and domain of validity. The structure of  the EOB model \TEOBResumS{} is
briefly reviewed in Sec.~\ref{sec:EOB}. In Sec.~\ref{sec:gsf_nr} we present a novel 2GSF/NR comparison that updates 
the results of Ref.~\cite{Wardell:2021fyy}: the analysis is based on the gauge-invariant phasing description provided by $\Qo$ 
and uses EOB waveforms as a benchmark. In Sec.~\ref{sec:waveform} we provide a comprehensive 2GSF/NR/EOB waveform
comparison up to $q=128$. Finally, Sec.~\ref{sec:nu_dependence} digs deeper into the origin of the 2GSF/EOB differences,
clearly pointing to an (expected) lack of 1GSF information within the EOB model. Conclusions are collected in Sec.~\ref{sec:conclusions}.
The paper is then completed by a few Appendices. In Appendix~\ref{sec:Qomg_clean} we report technical details
about the procedure for removing low and high-frequency oscillations from the NR $\Qo$. Appendix~\ref{sec:exactQ0andQ1} derives an asymptotic expansion of $Q_\omega$ in the small-mass-ratio limit. In Appendix~\ref{sec:eobnrgsf_q}
we complement and update the 2GSF/NR analysis of Ref.~\cite{Wardell:2021fyy} for comparable-mass binaries with
mass ratios from $q=1$ to $q=6$. Finally, in Appendix~\ref{sec:flux} we perform a comprehensive EOB/GSF/NR analysis
of the energy fluxes, also complementing the findings of Ref.~\cite{Warburton:2021kwk}.

We use natural units with $c=G=1$. In terms of our conventions for the individual masses, we denote the total mass 
 and symmetric mass ratio as $M\equiv m_1+m_2$ and $\nu = m_1 m_2/M^2$.

%==============
% GSF
%==============
\section{GSF dynamics and the 1PAT1 model}
\label{sec:GSF}

We compute 2GSF waveforms following Ref.~\cite{Wardell:2021fyy}. The approach is based on the multiscale (or two-timescale) expansion of the Einstein equations in Ref.~\cite{Miller:2020bft} (specifically Appendix A of that reference) with three additional approximations described below. To help explain the additional approximations, we first review the exact 1PA formalism in Sec.~\ref{sec:exact 1PA}. The additional approximations are then described in Sec.~\ref{sec:1PA approximations}. Section~\ref{sec:1PA accuracy} discusses a 1PA model's intrinsic level of error and domain of validity, and Sec.~\ref{sec:1PAT1 errors} discusses the expected error from the additional approximations.

\subsection{Exact 1PA waveforms}\label{sec:exact 1PA}

The multiscale expansion assumes the binary's metric, in the limit $\epsilon\to0$, only depends on time through its dependence on the binary's mechanical variables: the two black holes' trajectories, masses, spins, etc. All functions, including the metric, are treated as functions of spatial coordinates $x^i$ and of the mechanical phase space coordinates, and they are all expanded in powers of $\epsilon$ at fixed values of those coordinates~\cite{Pound:2021qin}. 

Restricted to the case of quasicircular orbits, with a slowly spinning primary and nonspinning secondary, this corresponds to the following expansion (through order $\epsilon^2$)~\cite{Miller:2020bft}:
\begin{multline}
\label{eq:metric}
\textbf{g}_{\alpha\beta} = g_{\alpha\beta} + \epsilon h^{(1)}_{\alpha\beta}(\phi_p,J_A, x^i) + \epsilon^2 h^{(2)}_{\alpha\beta} (\phi_p,J_A, x^i).
\end{multline}
Here $g_{\alpha\beta}$ represents the spacetime of the primary as if it were in isolation, meaning a Schwarzschild metric with constant mass $m_1^0$ and vanishing spin $s^0_1=0$. The variables $x^i=(r,\theta,\phi)$ are the usual Schwarzschild spatial coordinates, and $(\phi_p,J_A)$ are the phase space coordinates. Concretely, $\phi_p$ is the orbital azimuthal angle of the secondary (with the subscript $p$ denoting it as the ``particle''), and $J_A=(\Omega,\delta m_1,\delta s_1)$ are the binary's slowly evolving parameters: the orbital frequency $\Omega \equiv d\phi_p/dt$, a correction $\delta m_1\equiv (m_1-m_1^0)/\epsilon$ to the primary's mass, and the primary's rescaled spin $\delta s_1\equiv s_1/\epsilon$. Because the mass and spin only change by an amount $\sim\epsilon$ over the inspiral time $\sim 1/\epsilon$, they are treated perturbatively rather than altering $g_{\alpha\beta}$, and the parameters $\delta m_1$ and $\delta s_1$ are scaled by $\epsilon$ to make them order unity. In this section only, we use $\epsilon=m_2/m_1^0$ and work in units with $m^0_1=1$.

Since $\phi_p$ is a periodic coordinate on phase space, the metric is assumed to be periodic in it, allowing us to use a discrete Fourier series
\begin{equation}\label{eq:metric Fourier}
h^{(n)}_{\alpha\beta} = \sum_{m=-\infty}^\infty h^{(n,m)}_{\alpha\beta}(J_A, x^i)e^{-im\phi_p}.
\end{equation}
This expansion divides the metric perturbation into slowly evolving amplitudes and rapidly oscillating phase factors. The amplitudes $h^{(n,m)}_{\alpha\beta}$, orbital frequency, and orbital radius evolve on the radiation-reaction time $t_{rr} \sim 1/(\epsilon\Omega)$, while $\phi_p$ evolves on the orbital timescale $t\sim 1/\Omega$. 

In Eqs.~\eqref{eq:metric} and \eqref{eq:metric Fourier}, $\phi_p$ and $J_A$ are functions of a hyperboloidal time $s$ that is equal to Schwarzschild time $t$ at the secondary's worldline, advanced time $v$ at the large black hole's horizon, and retarded time $u$ at future null infinity. The binary's evolution, through order $\epsilon^2$, is then given by expansions of the form
\begin{align}
\frac{d\phi_p}{ds} &= \Omega,\label{phidot}\\
\frac{d \Omega}{ds} &= \epsilon\left[F_0^\Omega(\Omega) + \epsilon F_1^\Omega(J_A)\right],\label{eq:Omegadot-v1}\\
\frac{d \delta m_1}{ds} &= \epsilon\mathcal{F}^{(1)}_{\cal H}(\Omega), \qquad \frac{d\delta s_1}{ds} = \epsilon\,\Omega^{-1}\,\mathcal{F}^{(1)}_{\cal H}(\Omega),\label{s1dot}
\end{align}
where $\mathcal{F}^{(1)}_{\cal H}$ is the leading-order energy flux through the black hole's horizon (i.e., the flux due to $h^{(1)}_{\mu\nu}$). The orbital radius is given in terms of $J_A$ as $r_p = r_0(\Omega) + \epsilon r_1(J_A)$, where $r_0=m_1^0(m_1^0\Omega)^{-2/3}$ is the test-mass relationship.  It follows from these equations that $dr_p/ds$ has an expansion of the form $dr_p/ds=\epsilon[F^r_0(\Omega)+\epsilon F^r_1(J_A)]$.

Within this framework, the $n^{\rm th}$ PA order includes all terms contributing up to $\epsilon^{n+1}$ to the evolution of the orbital frequency, consistently with the terminology introduced in Ref.~\cite{Hinderer:2008dm}. $F^\Omega_0$ is the adiabatic (0PA) dissipation-driven rate of change, determined by the first-order dissipative GSF or energy flux, and the 1PA term $F^\Omega_1$ is determined by the full (conservative and dissipative) first-order GSF and second-order dissipation. 

Substituting the expansions~\eqref{eq:metric}--\eqref{s1dot} into Einstein's equations, 
one finds Fourier-domain equations for the amplitudes $h^{(n,m)}_{\alpha\beta}$~\cite{Miller:2020bft}, 
which are solved in the Lorenz gauge, order by order in $\epsilon$ for fixed values of $J_A$. (Note that in this process we never set $dJ_A/ds=0$; the nonzero $dJ_A/ds$ is fully accounted for everywhere it appears.) The amplitudes are further decomposed
on a basis of tensor spherical harmonics to reduce the Einstein equations to radial ordinary differential equations for each $\ell m$ mode. At future null infinity, the $\ell m$ mode of the waveform is extracted by transforming from the Lorenz gauge (in which $h^{(2)}_{\alpha\beta}$ is singular at null infinity~\cite{Pound:2015wva}) to a Bondi-Sachs gauge (in which $h^{(2)}_{\alpha\beta}$ is smooth there). In the usual basis of $s=-2$ spin-weighted harmonics ${}_{-2}Y_{\ell m}$, the $\ell m$ mode of the resulting (dimensionless) waveform can be written as $H_{\ell m}= R_{\ell m}(J_A,\epsilon)e^{-im\phi_p}$, or
\begin{align}\label{eq:1PA waveform}
H_{\ell m}=\left[\epsilon R^{(1)}_{\ell m}(\Omega)+\epsilon^2 R^{(2)}_{\ell m}(J_A)\right]e^{-im\phi_p}.
\end{align} 

In this waveform construction, one first computes the amplitudes $h^{(1,m)}_{\alpha\beta}$ for a set of $\Omega$ values; from $h^{(1,m)}_{\alpha\beta}$, one computes $F^\Omega_0$ and $r_1$; from $F^\Omega_0$, $r_1$, and $h^{(1,m)}_{\alpha\beta}$, one computes $h^{(2,m)}_{\alpha\beta}$; and from all of the above, one computes $F^\Omega_1$. These are all computed and stored as functions of $\Omega$ prior to solving for $\phi_p(s)$ and $\Omega(s)$. Using the stored amplitudes $R^{(n)}_{\ell m}$ and driving forces $F^\Omega_n$, one can then rapidly generate the waveform modes \eqref{eq:1PA waveform} by solving the evolution equations~\eqref{phidot}--\eqref{s1dot}.

\subsection{The approximate 1PAT1 model}\label{sec:1PA approximations}

The 1PAT1 model in Ref.~\cite{Wardell:2021fyy} closely follows the exact 1PA waveform construction but with three simplifying approximations.

We start by expressing $F^\Omega_0$ and $F^\Omega_1$ in terms of energy fluxes rather than the local GSF. We define the binding energy as a function of $s$, 
\begin{equation}\label{Ebind def}
E_{\rm bind}(s)\equiv M_{\rm Bondi}(s)-m_1(s)-m_2. 
\end{equation}
The Bondi mass $M_{\rm Bondi}$ and primary mass $m_1$ can be directly calculated as functions of $J_A$ from the amplitudes $h^{(n,m)}_{\mu\nu}$ as described in Ref.~\cite{Pound:2019lzj} (see also \cite{Bonetto:2021exn}). Differentiating Eq.~\eqref{Ebind def} with respect to $s$, using the Bondi-Sachs mass-loss formula $dM_{\rm Bondi}/ds = -{\cal F}_\infty$ and the flux-balance law $dm_1/ds = {\cal F}_{\cal H}$~\cite{Ashtekar:2004cn}, and applying the chain rule $dE_{\rm bind}/ds = (\partial E_{\rm bind}/\partial J_A)dJ_A/ds$, we can rearrange for $d\Omega/ds$ to find
\begin{equation}
\frac{d\Omega}{ds} = -\frac{{\cal F}_\infty+{\cal F}_{\cal H}+\frac{\partial E_{\rm bind}}{\partial \delta m_1}\frac{d\delta m_1}{ds}+\frac{\partial E_{\rm bind}}{\partial \delta s_1}\frac{d\delta s_1}{ds}}{\partial E_{\rm bind}/\partial\Omega}.\label{Omegadot-exact}
\end{equation}
Here ${\cal F}_\infty$ is the energy flux to infinity, which is given in terms of the asymptotic amplitudes as
\begin{align}
\mathcal{F}_{\infty} &= \frac{1}{16\pi}\sum_{\ell m}\left|\frac{d}{ds}(R_{\ell m}e^{-im\phi_p})\right|^2 \nonumber\\
&= \frac{1}{16\pi}\sum_{\ell m}\left\{\epsilon^2 |m \Omega R^{(1)}_{\ell m}|^2+2\epsilon^3 {\rm Re}\left[m^2\Omega^2 R^{(2)}_{\ell m}R^{(1)*}_{\ell m}\right.\right.\nonumber\\
&\qquad\qquad\quad\left.\left. +im\Omega F^\Omega_0 R^{(1)*}_{\ell m}\partial_\Omega R^{(1)}_{\ell m}\right] + O(\epsilon^4) \right\}\label{flux}
\end{align}
(still in units $m_1^0=1$). This computation of $\mathcal{F}_{\infty}$ was carried out in Ref.~\cite{Warburton:2021kwk}. The other terms on the right-hand side of Eq.~\eqref{Omegadot-exact} can also be straightforwardly expanded, leading to an expression of the form~\eqref{eq:Omegadot-v1}. 

So far we have made no approximations. The flux-based evolution equation for $\Omega$ follows directly from exact laws of GR together with our multiscale expansion; the formulas for $F^\Omega_0$ and $F^\Omega_1$ in terms of fluxes must necessarily agree with the formulas in terms of the local GSF (though numerically verifying the equality of the two formulas for $F^\Omega_1$ will be a crucial check in the future). 

We now apply our three approximations:
\begin{enumerate}[label=(\roman*)]
\item We neglect 1PA terms involving $dm_1/ds$ and $d\delta s_1/ds$ in Eq.~\eqref{Omegadot-exact}. Specifically, we use only the leading-order horizon flux, ${\cal F}_{\cal H}=\epsilon^2{\cal F}^0_{\cal H}(\Omega)$, and we discard $\frac{\partial E_{\rm bind}}{\partial \delta m_1}\frac{d\delta m_1}{ds}$ and $\frac{\partial E_{\rm bind}}{\partial \delta s_1}\frac{d\delta s_1}{ds}$. This is motivated by the facts that (i) the subleading horizon flux has not yet been computed, and (ii) the horizon flux is numerically small compared to the flux to infinity.
\item We neglect the evolution of the black hole, setting $\delta m_1=\delta s_1=0$ and ignoring the evolution equations~\eqref{s1dot}, such that $m^0_1=m_1$ and $s_1=0$. This is motivated by the change in the black hole parameters having negligible effect on the asymptotic fluxes in Ref.~\cite{Warburton:2021kwk}.
\item Rather than using Ref.~\cite{Pound:2019lzj}'s direct measurement of $E_{\rm bind}$ from the Bondi mass and black hole mass, we use the binding energy obtained from the first law of compact binary mechanics~\cite{LeTiec:2011dp, LeTiec:2011ab}. This is motivated by the facts that (i) the $E_{\rm bind}$ computed in Ref.~\cite{Pound:2019lzj} was calculated for a different choice of time function $s$ than the fluxes computed in Ref.~\cite{Warburton:2021kwk}, and (ii) the first-law binding energy was found to be numerically very close to Ref.~\cite{Pound:2019lzj}'s directly measured binding energy.
\end{enumerate}

In addition to applying these approximations, we also rewrite $m_1$ and $m_2$ as functions of the total mass $M$ and symmetric mass ratio $\nu$, and then re-expand all quantities in powers of $\nu$ at fixed dimensionless frequency $\hat\Omega\equiv M\Omega$, truncating the re-expansion at 1PA order. This enforces the system's symmetry under interchange of the two masses, and it substantially improves the accuracy of the small-mass-ratio expansion for non-extreme mass ratios. It is unrelated to the three approximations above; the re-expansions could equally well be done in the exact 1PA formulas. To facilitate the re-expansion, we restore factors of $m_1$ and make all dependence on the masses explicit [such that $R^{(n)}_{\ell m}(\Omega)$ becomes $R^{(n)}_{\ell m}(m_1\Omega)$ before re-expansion, for example]. 

After these steps, the full set of evolution equations~\eqref{phidot}--\eqref{s1dot} are replaced by the simplified set
\begin{align}
\frac{d \phi_p}{ds} &= \Omega ,\\
\frac{d \Omega}{ds} &= \frac{\nu}{M^2} \left[ F_0(x) + \nu F_1(x) \right] ,\label{eq:Omegadot}
\end{align}
where $x \equiv (M\Omega)^{2/3} = \hat{\Omega}^{2/3}$
and
\begin{align}
F_0(x) &= a(x) \mathcal{F}^{(1)}(x) = a(x) \left[ \mathcal{F}^{(1)}_{\infty}(x) + \mathcal{F}^{(1)}_{\cal{H}}(x)  \right] , \\
F_1(x) &= a(x)  \mathcal{F}^{(2)}_{\infty} (x) - a^2(x)  \mathcal{F}^{(1)}(x) \partial_{\hat{\Omega}} \hat{E}_{\rm SF}.
\end{align}
Here all functions of $x$ are dimensionless functions of the dimensionless variable $x$. We have expanded the flux~\eqref{flux} as ${\cal F}_\infty = \nu^2\mathcal{F}^{(1)}_{\infty}(x)+\nu^3\mathcal{F}^{(2)}_{\infty}(x)+O(\nu^4)$, where the superscripts indicate if the quantities are computed from the first-order amplitudes $h^{(1, m)}_{\mu\nu}$ or from the second-order ones, $h^{(2, m)}_{\mu\nu}$. We have similarly expanded the binding energy as $E_{\rm bind}=\nu M[\hat E_0(x)+\nu\hat E_{\rm SF}(x)+O(\nu^2)]$ and defined
\begin{align}
a(x) &\equiv - \left( \frac{\partial \hat{E}_0}{\partial \hat{\Omega}} \right)^{-1}
        = \frac{3 x^{1/2} (1 - 3x)^{3/2}}{1-6x}.\label{a(x)}
\end{align}
The leading-order specific binding energy $\hat{E}_0(x)=\frac{1-2x}{\sqrt{1-3x}}-1$ is identical to the circular geodesic orbital energy of a test particle on a Schwarzschild background of mass $M$ (dependence on the nonzero $\dot r$ enters into the binding energy at subleading orders in $\nu$). We have changed notation from $F^\Omega_n$ to $F_n$ to distinguish between coefficients of $\epsilon$ and coefficients of $\nu$, but we note $F_0(x)=F^\Omega_0(\hat\Omega(x))$.

In Ref. ~\cite{Wardell:2021fyy}, two additional 1PA waveform models were presented: a second time-domain model, 1PAT2; and a frequency-domain model, 1PAF1. They used the same three approximations but alternative expansions. Here we restrict our attention to 1PAT1 as the more accurate of the two time-domain models.

Before moving to the next section, to fix conventions we write the final strain waveform as
\be
h_+ - i h_\times = \dfrac{M}{D_L}\sum_\ell \sum_{m=-\ell}^{\ell}h_\lm{}_{-2}Y_\lm(\theta,\phi),
\ee
where $D_L$ indicates the luminosity distance.
As an expansion in powers of $\nu$ at fixed $(\phi_p,x)$, the modes read
\begin{equation}\label{1PAT1 hlm - nu}
h_{\ell m} = \left[\nu R^{(1)}_{\lm}+\nu^2\left(R^{(2)}_{\ell m}+R^{(1)}_{\ell m}-\hat\Omega \partial_{\hat\Omega}R^{(1)}_{\ell m}\right)\right]e^{-im\phi_p},
\end{equation}
where $R^{(n)}_{\ell m}=R^{(n)}_{\ell m}(\hat\Omega)$ are the amplitudes in Eq.~\eqref{eq:1PA waveform}.\footnote{To derive this, note that $H_{\ell m}$ in Eq.~\eqref{eq:1PA waveform} is defined with the luminosity distance in units of $m_1$, as the $\ell m$ mode of a component of $\lim_{r\to\infty}\frac{r}{m_1} (\epsilon h^{(1)}_{\mu\nu} +\epsilon^2 h^{(2)}_{\mu\nu})$, while $h_{\ell m}$ is instead defined from the limit $\lim_{r\to\infty}\frac{r}{M}$. This implies $h_{\ell m} = \frac{m_1}{M}H_{\ell m}$.} In practice, we work with
the Regge-Wheeler-Zerilli normalization convention and address the waveform as $\Psi_\lm\equiv h_\lm/\sqrt{(\ell+2)(\ell+1)\ell(\ell-1)}$. This waveform is separated into amplitude and phase with the convention
\be
\label{eq:RWZnorm}
\Psi_{\ell m} = A_{\ell m} e^{-i \phi_{\ell m}},
\ee
and we define the frequency as $\omega_{\ell m} \equiv \dot{\phi}_{\ell m}$. In practice we will always state values of $\omega_{\ell m}$ in units of $1/M$, but for clarity we will sometimes distinguish between the dimensionful quantity $\omega_{\ell m}$ and the dimensionless quantity $M\omega_{\ell m}$.

Although we can compute all waveform multipoles~\cite{Wardell:2021fyy}, in this paper we restrict our analysis to $\ell=m=2$. In our computation of the fluxes, we include modes up to $\ell=30$ in ${\cal F}^{(1)}$ and up to $\ell=5$ in ${\cal F}^{(2)}_\infty$. We use a large-$\ell$ fit to approximate the ($\lesssim 1\%$) contribution of higher-$\ell$ modes to ${\cal F}^{(2)}_\infty$. Specifically, since the flux modes fall off exponentially with $\ell$ \cite{Barack:2007tm} we consider a model of the form $\alpha e^{\beta \ell}$ and determine the constants $\alpha$ and $\beta$ by fitting this model to the $\ell=\{3,4,5\}$ modes of ${\cal F}^{(2)}_\infty$. We then verify the robustness of our fit by computing modes $\ell=\{6, \ldots, 10\}$ in a couple of representative test cases and comparing against the model. The net result of using this fit is a ${\cal F}^{(2)}_\infty$ that is at least an order of magnitude more accurate than without the fit.

\subsection{Intrinsic error and domain of validity}\label{sec:1PA accuracy}

We first assess the domain of validity of an exact 1PA model before discussing the uncertainty that arises from our three additional approximations.

By construction, a complete 1PA model, expressed in terms of $(\hat\Omega,\nu)$, has errors $O(\nu^3)$ in the waveform amplitude and $O(\nu)$ in the waveform phase. This error estimate follows immediately from the structure of the expansions~\eqref{eq:metric}--\eqref{s1dot} (re-expanded in powers of $\nu$). It applies both pointwise at each fixed frequency and uniformly on any fixed interval $[\hat\Omega_i,\hat\Omega_f]$ with $0<\hat\Omega_i<\hat\Omega_f<\hat\Omega_{\rm LSO}$. Here  $\hat\Omega_{\rm LSO}= 6^{-3/2}$ is the Schwarzschild geodesic frequency of the last stable orbit.

However, a 1PA model is {\it not} uniformly accurate over the whole interval $(0,\hat\Omega_{\rm LSO})$. Near $\hat\Omega_{\rm LSO}$, $d\Omega/dt$ grows large due to the divergent factor in Eq.~\eqref{a(x)}, and  the particle transitions into a plunge trajectory; near $\hat\Omega=0$, missing small-$\hat\Omega$ terms will cause large cumulative phase shifts. This lack of uniformity can have significant  impact at finite $\nu$, particularly due to the transition to plunge.

We determine the domain of validity of a complete 1PA model by excluding the boundary regions where missing PN or transition-to-plunge effects dominate over 1PA effects. More precisely, we define the domain of validity $(\hat\Omega^*_{i},\hat\Omega^*_{f})$ as the interval in which (i) the error is small compared to 1PA terms, and (ii) all omitted terms in the phase vanish in the limit $\nu\to0$. Note that these two conditions are distinct because condition (i) on its own could allow a large error in the phase as long as that error remained small compared to the 1PA contribution. Also note that, importantly, $\hat\Omega^*_i$ and $\hat\Omega^*_f$ will depend on $\nu$.

To find $\hat\Omega^*_{i}$ and $\hat\Omega^*_{f}$, we write $\phi_p$ as a function of $\hat\Omega$, $\phi_p = \int\frac{\hat\Omega}{d\hat\Omega/d\hat s}d\hat\Omega$, with $\hat s\equiv s/M$. The integrand can be expanded to 2PA order as 
\begin{align}
\phi'&\equiv\frac{\hat\Omega}{d\hat\Omega/d\hat s}\nonumber\\
&= \frac{\hat\Omega}{F_0 \nu}-\frac{\hat\Omega F_1}{(F_0)^2}+\frac{\left[(F_1)^2-F_0 F_2\right]\hat\Omega \nu}{(F_0)^3}+O(\nu^2)\nonumber\\
&\equiv \frac{1}{\nu}\phi'_0(\hat\Omega)+\phi'_1(\hat\Omega)+\nu\phi'_2(\hat\Omega)+O(\nu^2),\label{phiprime}
\end{align}
and the associated phase as
\begin{equation}\label{phi expansion}
\phi_p = \frac{1}{\nu}\phi^0_p(\hat\Omega) + \phi_p^1(\hat\Omega) + \nu\phi_p^2(\hat\Omega)+O(\nu^2),
\end{equation}
where $\phi^n_p=\int \phi'_n d\hat\Omega$ and we have started from an expansion of the form $d\hat\Omega/d\hat s = \nu\sum_{n\geq0}\nu^n F_n(x)$ (but assumed none of the three additional approximations). In our 1PA approximation, the error should be dominated by the 2PA term $\nu\phi'_2$ in $\phi'$ and associated $\nu\phi_p^2$ term in $\phi_p$. We are therefore interested in the size of $\phi'_2$ near $\hat\Omega=0$ and $\hat\Omega_{\rm LSO}$. 

Near $\hat\Omega=0$, we use a PN expansion to estimate the behaviour of the 1PA and 2PA terms.
 It is straightforward to derive $\phi'$, via the balance law, from Eqs.~(232) and (313) of Ref.~\cite{Blanchet:2013haa}. %The 2PA term begins at 2PN order, entering through both the binding energy and energy flux. 
Explicitly,
\begin{multline}\label{dphidOmega - small omega}
\phi' = \frac{5}{96 \nu  x^4}\left\{\nu^0\left[1+O(x)\right] +\nu\left[\dfrac{11}{4}x+O(x^2)\right]\right.\\
\left.+\nu^2\left[\dfrac{617}{144} x^2+O(x^{5/2})\right]+O(\nu^3)\right\}.
\end{multline}
The 2PA term begins at 2PN order, behaving as $\nu\phi'_2 \sim \nu x^{-2}=\nu \hat\Omega^{-4/3}$, while the 1PA term begins at 1PN order, behaving as $\phi'_1\sim x^{-3}=\hat\Omega^{-2}$. We can see that the conditions $\hat\Omega\ll1$ and $\nu\ll1$ automatically enforce $\nu\phi'_2\ll \phi'_1$. However, we {\it cannot} take $\hat\Omega^*_i$ to be arbitrarily small: the phase error behaves as $\int^{\hat\Omega}_{\hat\Omega^*_i}\nu\phi'_2d\hat\Omega\sim \nu\int^{\hat\Omega}_{\hat\Omega^*_i}  \hat\Omega^{-4/3} d\hat\Omega$, implying that it diverges in the limit $\hat\Omega^*_i\to 0$. Our requirement that the phase error vanishes when $\nu\to0$ implies $\nu(\hat\Omega^*_i)^{-1/3}\xrightarrow{\nu\to0}0$, or 
\begin{align}\label{Omega_i}
\hat\Omega^*_i\sim \nu^{\delta_i} \quad \text{with } 0<\delta_i<3.
\end{align}

%========================================
% Table of near-ISCO behaviour
%========================================
\begin{table*}[t]
\begin{center}
\begin{ruledtabular}
\begin{tabular}{c c c c c c}
 &   $F^{\Delta\Omega}_{0}$ & $F^{\Delta\Omega}_{1}$ & $F^{\Delta\Omega}_{2}$ & $F^{\Delta\Omega}_{3}$ & $F^{\Delta\Omega}_{4}$ \\
\hline
\hline
$F_0$ & $(\Omega-\Omega_{\rm LSO})^{-1}$ & 0 & $(\Omega-\Omega_{\rm LSO})^{0}$ & 0 & $(\Omega-\Omega_{\rm LSO})$\\
$F_1$ & 0 & $0\times (\Omega-\Omega_{\rm LSO})^{-3}$ & 0 &  $(\Omega-\Omega_{\rm LSO})^{-2}$ & 0\\
$F_2$ & $(\Omega-\Omega_{\rm LSO})^{-6}$ & 0 & $(\Omega-\Omega_{\rm LSO})^{-5}$  & 0 & $(\Omega-\Omega_{\rm LSO})^{-4}$\\
$F_3$ & 0 & $0\times (\Omega-\Omega_{\rm LSO})^{-8}$ & 0 &  $(\Omega-\Omega_{\rm LSO})^{-7}$ & 0
\end{tabular}
\end{ruledtabular}
\end{center} 
\caption{\label{tab:near-LSO}Near-LSO behavior of the first few $n$PA driving forces $F_{n}$. In the near-LSO limit, each $F_n$ is a sum of the terms in its row. In the $\nu\ll(\hat\Omega-\hat\Omega_{\rm LSO})$ limit, each $F^{\Delta\Omega}_n$ is a sum of the terms in its column. $F^{\Delta\Omega}_1$ identically vanishes, but we keep the terms in its column to illustrate the generic structure of the expansions.}
\end{table*}
%=====================================

Near $\hat\Omega_{\rm LSO}$, we carry out a similar analysis. The dynamics during the transition to plunge is well known and has recently been developed with a systematic asymptotic expansion~\cite{Compere:2021zfj}. We consider a variant of that expansion that allows us to directly examine the dependence on $\hat\Omega$. In a region of width $|\hat\Omega-\hat\Omega_{\rm LSO}|\sim\nu^{2/5}$, the evolution timescale changes from the long radiation-reaction time $\sim 1/(\nu\Omega)$ of the inspiral to the much shorter transition time $\sim 1/(\nu^{1/5}\Omega)$~\cite{Buonanno:2000ef,Ori:2000zn,Compere:2021zfj}. We can therefore change the frequency variable to $\Delta\hat\Omega\equiv \nu^{-2/5}(\hat\Omega-\hat\Omega_{\rm LSO})$, which is $O(\nu^0)$ in the transition region, and adopt an expansion
\begin{equation}\label{transition expansion}
\frac{d\Delta\hat\Omega}{d\hat s} = \nu^{1/5}\sum_{n\geq0}\nu^{n/5} F_n^{\Delta\Omega}(\Delta\hat\Omega) ,
\end{equation}
along with, e.g., 
\begin{equation}
r_p=6M+\nu^{2/5}\sum_{n\geq0}\nu^{n/5}R_n(\Delta\hat\Omega). 
\end{equation}
We will only require a small amount of information from this expansion, leaving a complete development to a separate paper.  Specifically, we will appeal to the equation governing $F_0^{\Delta\Omega}$:
\begin{equation}\label{FDOmega0}
\left[9 \sqrt{6}\frac{d}{d\Delta\hat\Omega}\left(F^{\Delta\Omega}_0 \frac{dF^{\Delta\Omega}_0}{d\Delta\hat\Omega}\right)-\Delta\hat\Omega\right]\! F^{\Delta\Omega}_0 = - \frac{f^t_{(1)}}{48},
\end{equation}
where $f^\mu_{(1)}$ is the self-force due to $h^{(1)}_{\mu\nu}$ evaluated at the LSO. This equation is straightforwardly found by substituting the above expansions into the self-forced equation of motion $\frac{D^2x^\mu_p}{d\tau^2}=\nu f^\mu_{(1)}$, with $x^\mu_p=\{t,r_p(\Delta\Omega,\nu^{1/5}),\pi/2,\phi_p\}$.

To extract the relevant information from the expansion~\eqref{transition expansion}, we note that it must agree with the inspiral expansion~\eqref{eq:Omegadot-v1} in the following sense: If we re-express Eq.~\eqref{transition expansion} in terms of $\nu$ and $\hat\Omega$ and re-expand it for small $\nu$ at fixed $\hat\Omega$, and if we expand Eq.~\eqref{eq:Omegadot-v1} near the LSO, then in both cases we arrive at a double expansion for small $\nu$ and small $(\hat\Omega-\hat\Omega_{\rm LSO})$. Since they are both expansions of the same function, these two double expansions must agree term by term. 

The  re-expansion of~\eqref{transition expansion} for small $\nu$ at fixed $\hat\Omega$, written as an expansion of $d\hat\Omega/d\hat s=\nu^{2/5}d\Delta\hat\Omega/d\hat s$, has the structure
\begin{align}
\frac{d\hat\Omega}{d\hat s} &=\nu^{3/5}\sum_{n\geq0}\nu^{n/5}F_n^{\Delta\Omega}(\Delta\Omega)\nonumber\\
&=\nu^{3/5}\sum_{n\geq0}\nu^{n/5}\sum_{k}\frac{\nu^{2k/5}}{(\hat\Omega-\hat\Omega_i)^k}F^{\Delta\Omega}_{n,k},
\end{align}
where the coefficients $F^{\Delta\Omega}_{n,k}$ are constants. On the other hand, the re-expansion of $d\hat\Omega/d\hat s = \nu\sum_{n_{\rm PA}\geq0}\nu^{n_{\rm PA}} F_{n_{\rm PA}}(x)$ for small $(\hat\Omega-\hat\Omega_{\rm LSO})$ has the structure 
\begin{equation}
d\hat\Omega/d\hat s = \nu\sum_{n_{\rm PA}\geq0}\nu^{n_{\rm PA}}\sum_{k}\frac{F_{n_{\rm PA},k}}{(\hat \Omega-\hat\Omega_{\rm LSO})^k}. 
\end{equation}
Comparing the powers of $\nu$ in the two double expansions, we read off the relationship $3+n+2k = 5(n_{\rm PA}+1)$, or $n+2k = 5n_{\rm PA}+2$. This implies that for an even $n$PA order, all terms near the LSO must match terms in $F^{\Delta\Omega}_n$ with $n$ odd; and for an odd $n$PA order, they must match terms with $n$ even. We can also rearrange the relationship to obtain $k = 1+\frac{1}{2}(5n_{\rm PA}-n)$, which tells us the power of $(\hat\Omega-\hat\Omega_{\rm LSO})$ that can be identified with a particular $n$PA order and a given order in the transition expansion~\eqref{transition expansion}. This structure is summarized in Table~\ref{tab:near-LSO}. In the table, we have highlighted that $F^{\Delta\Omega}_1=0$. We can establish that the leading term in $F^{\Delta\Omega}_1$, $\sim (\Omega-\Omega_{\rm LSO})^{-3}$, vanishes by directly comparing to our numerical results for $F_1$; the complete analysis to be presented elsewhere shows $F^{\Delta\Omega}_1$ identically vanishes.
 
It is clear from the expansion~\eqref{phiprime} and the scalings in Table~\ref{tab:near-LSO} that near the LSO, the condition $\nu\phi'_2\ll \phi'_1$ is equivalent to $\nu^3 F_2\ll \nu^2 F_1$. Substituting the near-LSO behavior, this becomes $\nu (\hat\Omega - \hat\Omega_{\rm LSO})^{-6}F_{2,6}\ll (\hat\Omega - \hat\Omega_{\rm LSO})^{-2}F_{1,2}$, or 
\begin{equation}\label{near-LSO Omega}
|\hat\Omega - \hat\Omega_{\rm LSO}|\gg \left(\frac{F_{2,6}\nu}{F_{1,2}}\right)^{1/4}.
\end{equation}
Unlike in the PN limit, this constraint is stronger than the condition that the error in $\phi_p$ vanishes for $\nu\to0$: for $\int^{\hat\Omega^*_f}\nu\phi'_2 d\hat\Omega\sim \nu(\hat\Omega^*_f-\hat\Omega_{\rm LSO})^{-3}$ to vanish in the limit, we require $|\hat\Omega^*_f- \hat\Omega_{\rm LSO}|\gg \nu^{1/3}$, which is automatically satisfied if Eq.~\eqref{near-LSO Omega} is satisfied. Therefore, the upper limit on the frequency should satisfy
\begin{align}\label{Omega_f}
|\hat\Omega^*_f - \hat\Omega_{\rm LSO}|\sim \nu^{\delta_f} \quad \text{with } 0<\delta_f<1/4.
\end{align}

Combining the above results, we conclude that a 1PA approximation is uniformly accurate, with $o(\nu^0)$ phase errors, on a domain $(c_i\nu^{\delta_i},\hat\Omega_{\rm LSO}-c_f\nu^{\delta_f})$, where $0<\delta_i<3$ and $0<\delta_f<1/4$ and $c_i$ and $c_f$ are constants. Strictly speaking, this is a statement about scaling rather than a statement about absolute error. It says that if we can determine  that the phase error is acceptably small (through comparison with NR, for example) for one mass ratio on a specific frequency interval, then for smaller mass ratios the error will remain acceptable (and tend toward zero) not just on that interval, but on a larger interval that tends toward $(0,\hat\Omega_{\rm LSO})$ at the rates $\nu^{\delta_i}$ and $\nu^{\delta_f}$. However, the main qualitative takeaway is that because the exponent $\delta_f$ is at most $1/4$, the upper limit tends toward $\hat\Omega_{\rm LSO}$ extremely slowly: for an equal-mass system, the factor $\nu^{\delta_f}$ is 0.7 or larger; decreasing the mass ratio to $\nu=1/100$ only reduces this factor to 0.3. In other words, the effects of the transition to plunge appear to be significant in a far larger frequency interval than one might expect.
 
For our EOB-GSF-NR comparisons, we will need to choose a more definite frequency cutoff prior to the LSO. We will consider two options: (i) the critical frequency $\Omega_{\rm critical}$ at which the evolution stops (i.e., $d\Omega/ds=0$) in the 1PAT1 model, and (ii) a frequency $\Omega_{\rm break}$ at which the two-timescale approximation has broken down. The critical frequency exists because $F_0$ and $F_1$ have opposite sign, such that they cancel when $|F_1|=|\nu^{-1}F_0|$. This always occurs prior to $\Omega_{\rm LSO}$ but after $\Omega_{\rm break}$. For the latter, we will say the two-timescale expansion has broken down when the dominant phase error $\nu\phi'_2$ becomes equal in magnitude to the 1PA term $\phi'_1$. We can estimate this frequency following the analysis that led to Eq.~\eqref{near-LSO Omega}, which implies 
\begin{equation}\label{Omega break unevaluated}
\hat\Omega_{\rm break} = \hat\Omega_{\rm LSO} - \left(\frac{F_{2,6}\nu}{F_{1,2}}\right)^{1/4}. 
\end{equation}
Since $F_{2,6}=F^{\Delta\Omega}_{0,6}$, we can find this coefficient from the leading-order transition dynamics. Substituting $F^{\Delta\Omega}_0=\frac{F^{\Delta\Omega}_{0,1}}{\Delta\hat\Omega}+\frac{F^{\Delta\Omega}_{0,6}}{\Delta\hat\Omega^6}+O(\Delta\hat\Omega^{-11})$ into Eq.~\eqref{FDOmega0} and solving for the coefficients, one quickly finds $F^{\Delta\Omega}_{0,6}=\sqrt{\frac{3}{2}}\frac{M^3(f^t_{(1)})^3}{2048}$, which evaluates to $F^{\Delta\Omega}_{0,6}\approx -1.7\times10^{-12}$. We find $F_{1,2}$ by observing that in Eq.~\eqref{eq:Omegadot}, the contribution to $F_1$ from the ${\cal F}^{(1)}$ term is more than an order of magnitude larger than the contribution from the ${\cal F}^{(2)}_\infty$ term near the LSO. Since ${\cal F}^{(1)}$ is finite at the LSO, this allows us to easily read off the coefficient of $\frac{1}{(\hat\Omega-\hat\Omega_{\rm LSO})^2}$, whch we find to be $F_{1,2}\approx -3.5\times 10^{-6}$. Combining these results in Eq.~\eqref{Omega break unevaluated} gives the breakdown frequency
\begin{equation}\label{Omega break}
\hat\Omega_{\rm break} \approx \hat\Omega_{\rm LSO} - 0.026\nu^{1/4}.
\end{equation}

The corresponding waveform frequencies $\omega_{\rm critical}\approx 2\hat\Omega_{\rm critical}$ and $\omega_{\rm break}\approx 2\hat\Omega_{\rm break}$
are displayed in the second and the third column of Table~\ref{tab:Dphi}.
For the time-domain phasing (and the corresponding $\Qo$ analysis
developed in the following) we will consider the GSF evolution only up to the breakdown frequency. We stress that these frequencies represent agressive choices of cutoff: both $\hat\Omega_{\rm break}$ and (especially) $\hat\Omega_{\rm critical}$ fall outside the interval allowed by the condition~\eqref{near-LSO Omega}. More conservative choices of cutoff might be preferable, but we opt for what appears to us to be the cleanest option. Similarly, while we caution that the breakdown frequency~\eqref{Omega break} is an asymptotic approximation in the small-$\nu$ limit, not a statement based on absolute error at specific mass ratios, we find it to be a convenient choice that also correctly predicts the frequency at which divergent terms in ${\cal F}^{(2)}$ begin to qualitatively change the total flux's behaviour; see the plots in Appendix~\ref{sec:flux}, where this qualitative change is clearly visible.

Our focus here has been on the near-LSO behaviour. No analogous breakdown frequency is available in the low-frequency limit because, as explained above, the error terms in the 1PA approximation of $\phi'$ remain small compared to the 1PA terms for arbitrarily low frequencies. However, one must be cautious because the phase error diverges in the limit $\Omega^*_i\to0$ (at fixed $\nu$). In general we must first ensure the 1PA approximation is sufficiently accurate on some finite interval for some values of $\nu$ before expanding the interval for smaller $\nu$, following Eq.~\eqref{Omega_f}.

\subsection{Uncontrolled and numerical errors}\label{sec:1PAT1 errors}

The errors discussed above are intrinsic to a 1PA model. They are controlled in the usual sense of perturbation theory: we understand their scalings with the small parameter and can, in principle, reduce the small-$\nu$ error by proceeding to 2PA order. (Errors due to the transition to plunge can likewise be eliminated by developing a complete inspiral-merger-ringdown model.)

We now consider the three additional approximations described above, which are sources of {\it un}controlled errors in our 1PAT1 model. These are errors in the 1PA terms themselves, and we do not have a precise estimate of their magnitude (though of course, like all 1PA terms, they make an order-$\nu^0$ contribution to the phase). Comparisons with NR suggest that these errors are numerically small, but their precise impact cannot be assessed without reference to a complete 1PA model. 

We first consider the approximations related to ignoring 1PA terms that arise from the black hole's evolution. These terms enter into the frequency evolution~\eqref{Omegadot-exact} in three ways: through the terms $\frac{\partial E_{\rm bind}}{\partial \delta m_1}\frac{d\delta m_1}{ds}$ and $\frac{\partial E_{\rm bind}}{\partial \delta s_1}\frac{d\delta s_1}{ds}$ in Eq.~\eqref{Omegadot-exact}, which behave as $\sim \nu^3 {\cal F}_{\cal H}^{(1)}$;  through corrections $\propto\delta m_1$ and $\propto\delta s_1$ to the leading-order binding energy and leading-order flux (at both the horizon and infinity);  and through the contribution of the subleading horizon-flux $\nu^3{\cal F}^{(2)}_{\cal H}$. The first two of these could be almost immediately included in our evolution, but they have negligible impact. The terms proportional to ${\cal F}^{(1)}_{\cal H}$ are suppressed by a factor $\sim{\cal F}^{(1)}_{\cal H}/{\cal F}^{(1)}_\infty$ relative to the other 1PA terms; this factor is very small, reaching $\approx3.3\times10^{-4}$ at the LSO and decaying rapidly away from the LSO, with a PN scaling $\sim x^4$. Similarly, the terms directly proportional to $\delta m_1$ and $\delta s_1$ are highly suppressed. Over the entire inspiral up to the LSO, the change in the black hole mass is $\nu\delta m_1\approx\nu^2\int_0^{\Omega_{\rm LSO}} ({\cal F}^{(1)}_{\cal H}/\dot\Omega)d\Omega\approx 1.2\times10^{-5} \nu M$, and if the black hole starts with zero spin then it accumulates a spin $s_1\approx 2.7\times 10^{-4} \nu M^2$. Hence, the resulting 1PA terms are suppressed by a factor $\lesssim 10^{-4}$ relative to other 1PA terms. 

The last neglected black-hole-evolution term in Eq.~\eqref{Omegadot-exact}, stemming from $\nu^3{\cal F}^{(2)}_{\cal H}$, is harder to estimate and would require a significant new calculation to incorporate into our model. However, we can obtain a rough estimate from a PN analysis. At the first two orders in the mass ratio, the known PN terms in the horizon flux are~\cite{Taylor:2008xy} 
\begin{multline}
{\cal F}_{\cal H} = \frac{32}{5}\nu^2 x^9\left\{\nu^0[1-5x+O(x^2)]\right.\\
\left.-2\nu[2-9x+O(x^2)]+O(\nu^2)\right\}.
\end{multline}
The fluxes to infinity, restricted to the same number of PN terms, are~\cite{Blanchet:2013haa}
\begin{multline}
{\cal F}_\infty=\frac{32}{5} \nu^2  x^5\left\{\nu^0\left[1-\frac{1247}{336}x+O(x^2)\right]\right. \\
\left.- \nu\left[\frac{35}{12} x - \frac{9271}{504}x^2+O(x^3)\right]+O(\nu^2)\right\}.
\end{multline} 
We can gain some confidence in the accuracy of these expressions by noting that they correctly predict the ratio of the leading-order-in-$\nu$ fluxes to one digit at the LSO, ${\cal F}^{(1)}_{\cal H}/{\cal F}^{(1)}_{\infty}\approx 3\times 10^{-4}$. At the first subleading order in $\nu$, they predict ${\cal F}^{(2)}_{\cal H}/{\cal F}^{(2)}_{\infty}\approx 3\times 10^{-2}$ at the LSO, decaying rapidly to $\approx 2\times 10^{-4}$ at $x=1/20$. Moreover, as we pointed out in the previous section, the contribution of ${\cal F}^{(2)}_{\infty}$ is already significantly smaller than other 1PA contributions to the phase evolution in the strong field. % we know that it will necessarily be constructed from terms $\propto \sum_m|h^{(2,m)}_{\mu\nu}h^{(1,m)}_{\mu\nu}|$ and $\propto \sum_m|h^{(1,m)}_{\mu\nu}|^3$ evaluated at the horizon. Both types of terms will be small compared to other 1PA contributions because the metric amplitudes at the horizon are very small compared to those at infinity. We find that $\ell m$ modes of $h^{(2)}_{\mu\nu}$ are typically a factor \AP{Niels to insert number} smaller than the amplitudes $R^{(2)}_{\ell m}$ at infinity, and the cubic terms $\propto |h^{(1,m)}_{\mu\nu}|^3\sim ({\cal F}^{(1)}_{\cal H})^{3/2} \lesssim 10^{-4}({\cal F}^{(1)}_{\infty})^{3/2}$ can be expected to be roughly four orders of magnitude smaller than other 1PA contributions. 
We therefore conclude that all 1PA effects of the black hole evolution would not materially impact any of our comparisons in this paper.

%========================================================
% Q1 comparisons with different choices of binding energy
%========================================================
\begin{figure}[t]
\center
\includegraphics[width=\columnwidth]{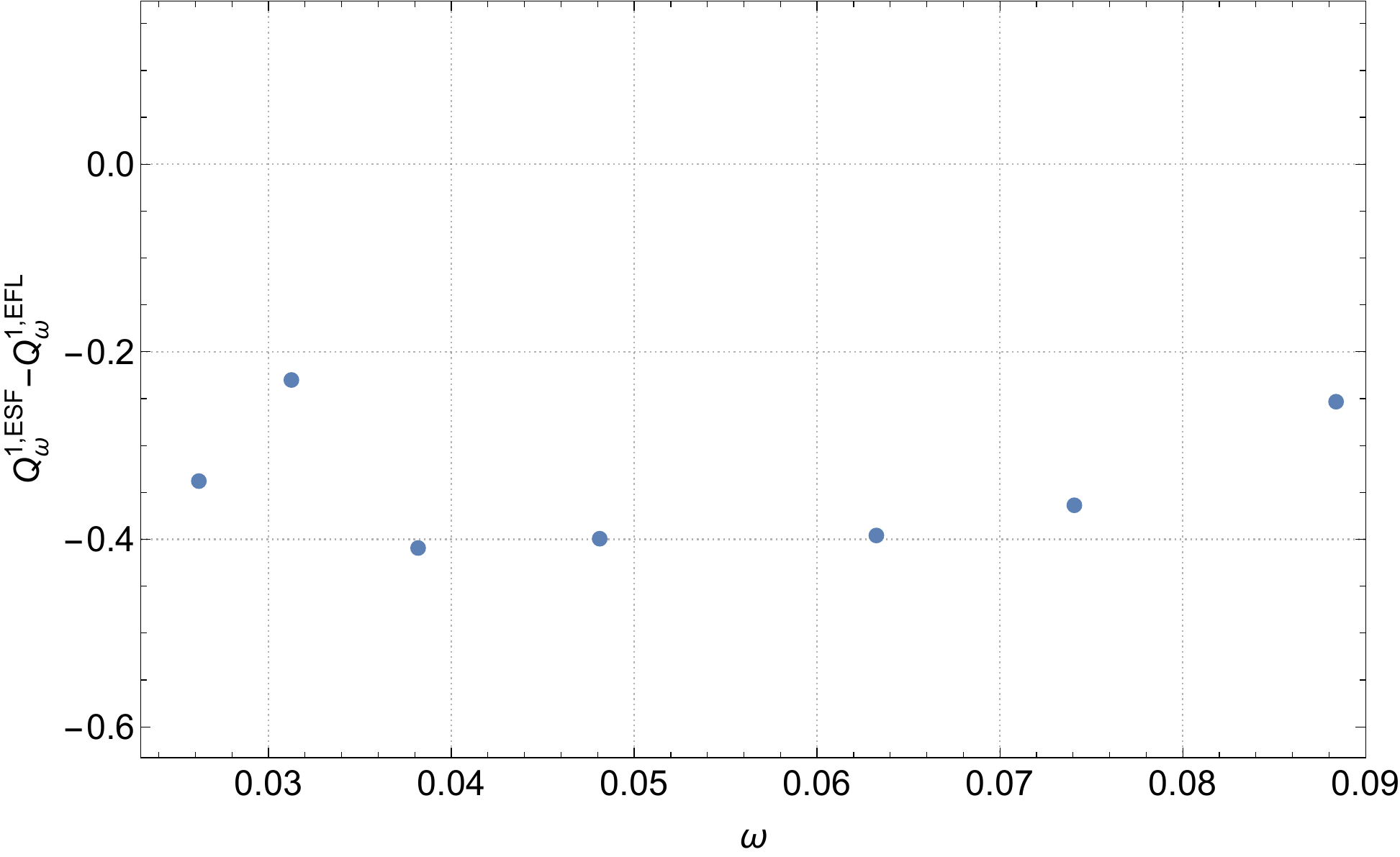}
\caption{\label{fig:Q1-EFL-ESF} Difference between $\Qo^1$ computed using the first-law binding energy and the second-order self-force binding energy from Ref.~\cite{Pound:2019lzj}. The frequency range is chosen for easy comparison with Fig.~\ref{fig:Q0andQ1}.}
\end{figure}

Our use of the first-law binding energy may have a substantially larger effect. This is the final of the three approximations outlined in Sec.~\ref{sec:1PA approximations}. Assessing its  impact is difficult. As a  rough guide, we compare the phase evolution using two versions of the binding energy (while noting it is unclear which of them lies closer to the true result): (i) the first-law binding energy, and (ii) the direct calculation from the Bondi mass in Ref.~\cite{Pound:2019lzj}, specifically the ``conservative'' approximant calculated there. Rather than using $\phi'$ to estimate errors in the phase evolution, as we did in the previous section, to assist the comparisons in later sections we use the dimensionless quantity $Q_\omega=\omega^2/\dot\omega$, which is $\approx \omega \phi'$. This has an asymptotic expansion
\be\label{Q expansion}
\Qo(\omega,\,\nu) = \frac{\Qo^{0}(\omega)}{\nu} + \Qo^1(\omega) + \Qo^2(\omega) \nu + O(\nu^2),
\ee
with the first two terms derived explicitly in Appendix~\ref{sec:exactQ0andQ1}. The different binding energies first enter in the 1PA term, $Q^1_\omega$. Figure~\ref{fig:Q1-EFL-ESF} displays the absolute difference between the two results for $Q^1_\omega$  (the relative difference, for comparison, is $<2\%$ across all frequencies considered). As we discuss later, this difference is appreciable, and it is comparable to the EOB-GSF difference shown later in Fig.~\ref{fig:Q0andQ1}. Only in the late inspiral does the EOB-GSF difference grow significantly larger.

In addition to the above approximations, our 1PAT1 model also contains numerical error. This enters primarily in the calculation of ${\cal F}^{(2)}_\infty$; all other sources of numerical error are negligible. We estimate our error in ${\cal F}^{(2)}_\infty$ to vary from $\sim0.01\%$ (near the LSO) to $\lesssim 1\%$ (for $x\approx 0.02$). This may be comparable to the 1PA effects of the black hole's evolution but is  subdominant compared to the uncertainty due to the binding energy.

All of the errors described in this section have an impact comparable to or smaller than other differences considered in Ref.~\cite{Wardell:2021fyy}. In particular, for mass ratios $q\lesssim10$ there are larger differences between the various formulations of the 1PA phase evolution: 1PAT1, 1PAT2, 1PAF1, or leaving the fraction in Eq.~\eqref{Omegadot-exact} unexpanded. However, those  differences all vanish in the limit $\nu\to0$, while the 1PA sources of error described here leave a $\nu$-independent impact on the phase.

%==============
% EOB
%==============
\section{EOB dynamics and waveform}
\label{sec:EOB}
We work here with the most advanced version of the \TEOBResumS{}~\cite{Nagar:2020pcj,Riemenschneider:2021ppj} 
EOB waveform model for nonprecessing quasi-circular binaries (see Ref.~\cite{Gamba:2021ydi} for 
the spin-precessing version). 

All technical details of the model are discussed extensively in Refs.~\cite{Nagar:2018zoe,Nagar:2020pcj,Riemenschneider:2021ppj},
so that we report here only the main conceptual elements to orient the reader. The conservative dynamics 
is described by a Hamiltonian $H_{\rm EOB}$~\cite{Nagar:2018zoe},
depending on the EOB potentials $A(R)$ and $B(R)$ (that include spin-spin interactions~\cite{Damour:2014sva}), 
given as a function of the EOB mass-reduced phase-space variables $(r,\varphi,p_\varphi,p_{r_*})$,  
related to the physical ones by $r=R/M$ (relative separation), $p_{r_*}=P_{R_*}/\mu$ 
(radial momentum), $\varphi$ (orbital phase), 
$p_\varphi=P_\varphi/(\mu M)$ (angular momentum) and $t=T/M$ (time),
and we replace the conjugate momentum $p_r$ with the
 \virg{tortoise} rescaled variable
$p_{r_*}\equiv (A/B)^{1/2}p_r$.
The Hamiltonian equations for the relative dynamics read
\begin{subequations}
\begin{align}
\dot{\ph} &= \Omega = \p_{p_\ph} \hat{H}_\EOB, \\
\dot{r} &= \left( \frac{A}{B} \right)^{1/2} \p_{p_{r_*}} \hat{H}_\EOB, \\
\dot{p}_\ph &= \hat{\F}_\ph , \\
\dot{p}_{r_*} &= - \left( \frac{A}{B} \right)^{1/2} \p_{r} \hat{H}_\EOB,
\end{align}
\end{subequations}
where $\Omega$ is the orbital frequency 
and $\hat{\F}_\ph$ is the radiation reaction force accounting for mechanical 
angular momentum losses due to GW emission\footnote{Note that within this context 
we are assuming that the radial force $\hat{\F}_r=0$, that is equivalent to a 
gauge choice for circular orbits~\cite{Buonanno:2000ef}.}, notably including 
both the asymptotic and the horizon contribution~\cite{Nagar:2011aa, Damour:2012ky}. 
The flux at infinity includes all multipoles up to $\ell=8$ in a special factorized 
and resummed form~\cite{Damour:2008gu,Nagar:2019wds, Nagar:2020pcj} so to improve
the behavior of the original PN series in the strong-field, fast velocity regime.
The complete quadrupole EOB waveform is written as
\be
\label{eq:h_eob}
h_{22} = h_{22}^{\rm Newt} \hat{h}_{22} \hat{h}_{22}^{\rm NQC}
\ee
where $h_{22}^{\rm Newt}$ is the Newtonian contribution, $\hat{h}_{22}$ the higher-order
PN correction in factorized and resummed form~\cite{Damour:2008gu} 
and $\hat{h}_{22}^{\rm NQC}$ the next-to-quasi-circular factor informed by NR simulations. 
We do not give additional details on $(\hat{h}_{22},\hat{h}_{22}^{\rm NQC})$ but rather 
direct the reader to Refs.~\cite{Nagar:2021gss,Riemenschneider:2021ppj}.
Here it is sufficient to recall that the purpose of the NR-informed NQC factor is to correct
the purely analytical waveform so that it is consistent with the NR one around merger,
an approach originally introduced in the extreme-mass-ratio limit~\cite{Damour:2007xr}.
Although our focus here will be on the inspiral, and not on the ringdown, let us remember
that the model provides a complete analytical description of the ringdown waveform that is
informed by NR simulations~\cite{Damour:2014yha,DelPozzo:2016kmd,Nagar:2021gss}.
For the purposes of this paper, we use a private {\tt MATLAB} implementation of \TEOBResumS{},
instead of the publicly available $C$ one~\cite{teobresums}, in which NQC corrections
are usually determined by iterating the evolution 3 times~\cite{Damour:2009kr}.

%================
% GSF/NR using Qomg
%================

\section{Numerical relativity, gravitational self-force  and the $Q_\omega$ diagnostic}
\label{sec:gsf_nr}
Before comparing EOB and GSF results it is useful to discuss direct GSF/NR phasing 
comparisons, complementing the discussion of Ref.~\cite{Wardell:2021fyy}.
To do so, we focus here on two specific NR datasets from the SXS catalog: $q=7$, SXS:BBH:0298, and
a 20-orbit long $q=10$ binary, SXS:BBH:0303, that has a {\it rather small} initial eccentricity ($\sim 10^{-5}$).
Note that Ref.~\cite{Wardell:2021fyy} selected the SXS:BBH:1107 dataset for $q=10$, that 
is 30-orbits long, but it is also marred by a larger eccentricity.

The main purpose of this section is to use an intrinsic measure of the NR phase evolution to 
obtain careful GSF/NR comparisons. To do so, we use the $Q_\omega$ function, defined as
\begin{equation}
\label{eq:Qomg}
Q_{\omega} = \frac{\omega^2}{\dot{\omega}},
\end{equation}
where $\omega\equiv \omega_{22}$ is the waveform frequency. 
From this, the accumulated phase difference 
in the time-domain in the frequency interval $(\omega_1,\omega_2)$ is given by the integral
\be
\label{eq:Dphi_from_Q}
\Delta\phi_{(\omega_1,\omega_2)}=\int_{\omega_1}^{\omega_2}Q_\omega d\log\omega \ .
\ee
The use of this
diagnostic was essential to produce reliable EOB/NR phasing comparisons for coalescing binary
neutron star systems~\cite{Baiotti:2010xh, Baiotti:2011am, Bernuzzi:2014owa}. In that particular
case, the $Q_\omega$ analysis was an important check on the reliability of standard time-domain 
phasing comparisons that depend on two shift ambiguities: an arbitrary phase shift and an arbitrary
time shift. 

For BBH systems, a systematic phasing analysis involving the $Q_\omega$ function dates
back to Ref.~\cite{Damour:2012ky}, which focused on EOB/NR phasing comparisons for nonspinning
binaries. For the highly accurate SXS data, Ref.~\cite{Damour:2012ky} demonstrated the equivalence
of the time-domain and frequency-domain analyses. In particular it showed that one can rely on
a time-domain phasing analysis to inform the EOB model using SXS data because of the excellent
EOB/NR agreement found during the inspiral.

At the technical level, Ref.~\cite{Damour:2012ky} pointed out that the extraction of a quantitatively
useful $Q_\omega$ function from NR data is a challenging process. In particular, one has to remove
the {\it many} numerical oscillations of spurious origin (either high-frequency or low-frequency) that
prevent any quantitatively reliable comparison with any other alternative representation of the binary
phasing (for example, EOB or GSF).
%===============
% Qomg comparisons
%===============
\begin{figure*}[t]
\center
\includegraphics[width=0.42\textwidth]{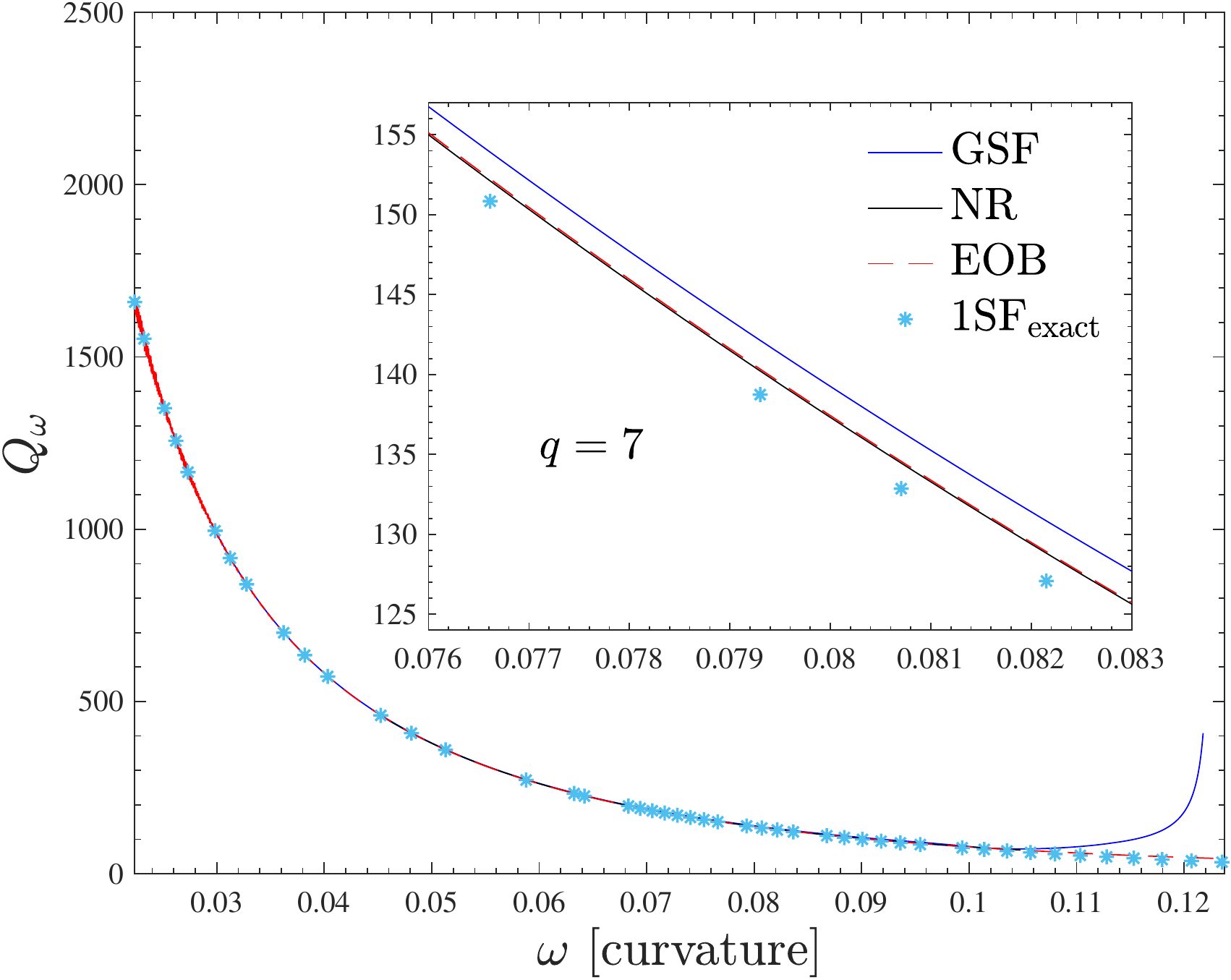}
\hspace{5 mm}
\includegraphics[width=0.42\textwidth]{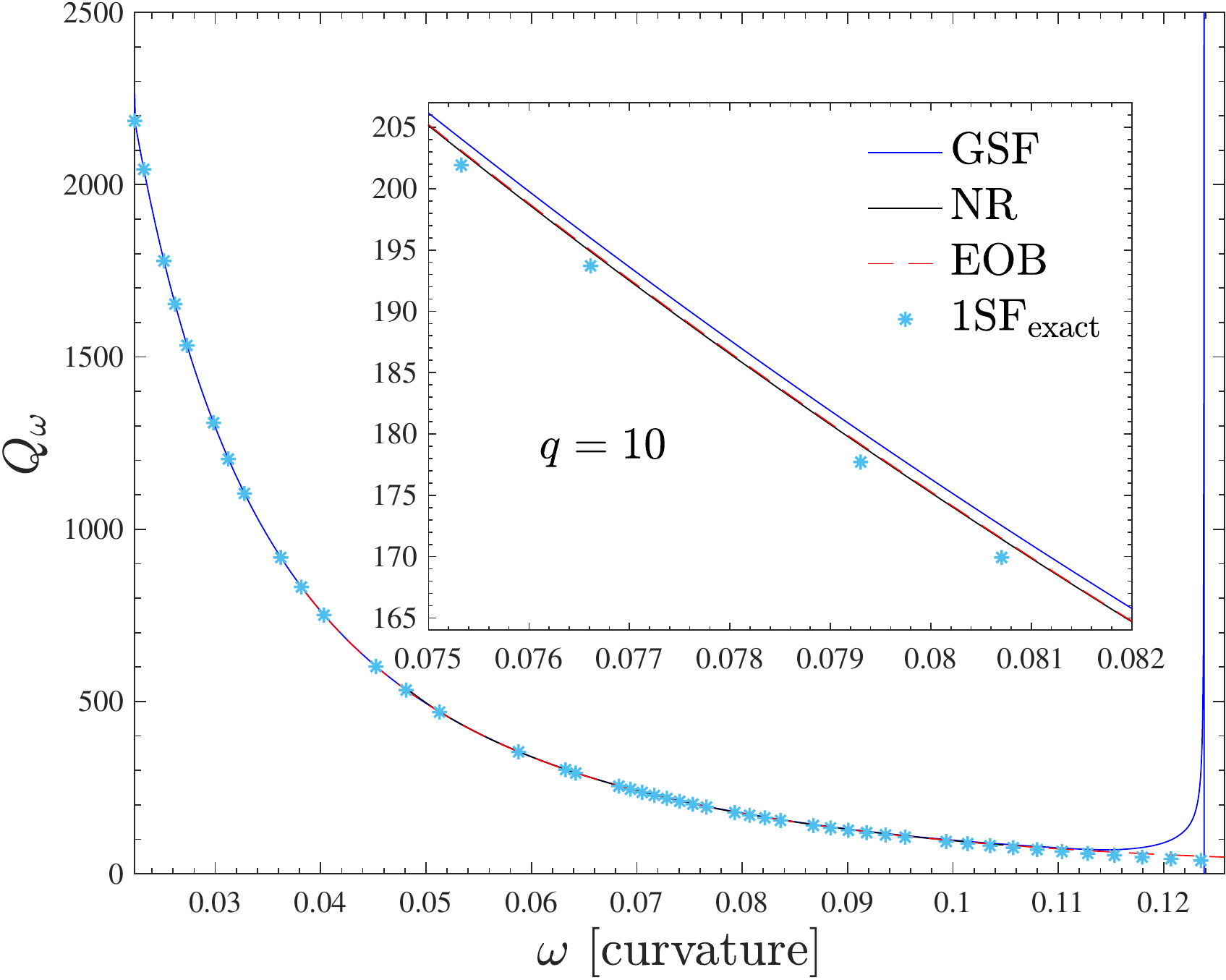} 
\caption{\label{fig:Qomg_all}NR, GSF and EOB comparison of $Q_\omega$ functions obtained from the
phase of $\psi_4^{22}$. Note that the GSF curve is consistently {\it above} both the NR and EOB ones. For comparison we also include the small-$\nu$ expansion truncated at 1PA order, $Q_\omega=\nu^{-1}Q_\omega^0(\omega)+Q_\omega^1(\omega)$, with the exact coefficients calculated from 1GSF and 2GSF data.}
%
%=========
% Differences
%=========
\end{figure*}
\begin{figure*}[t]
\center
\includegraphics[width=0.42\textwidth]{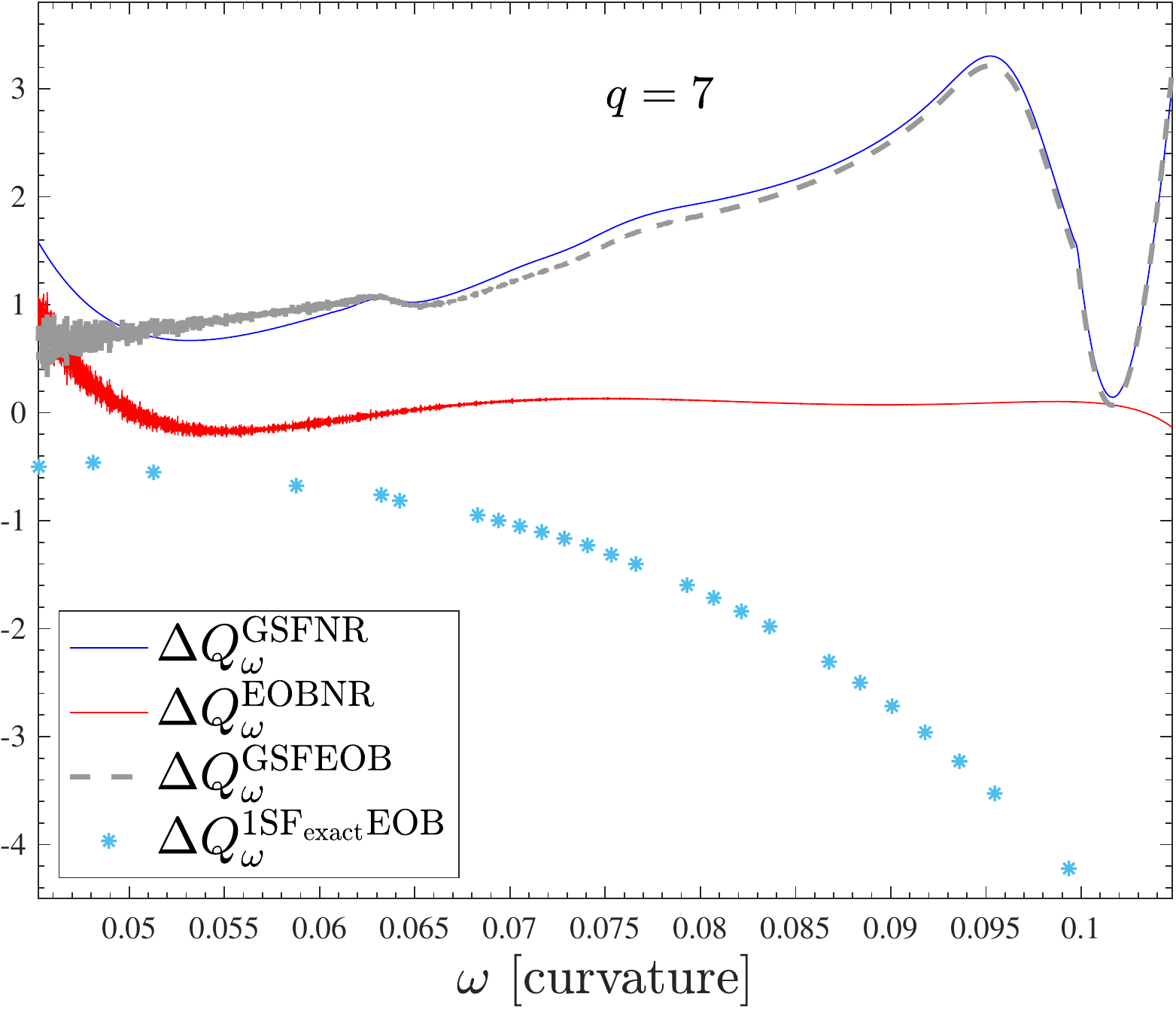}
\hspace{5 mm}
\includegraphics[width=0.42\textwidth]{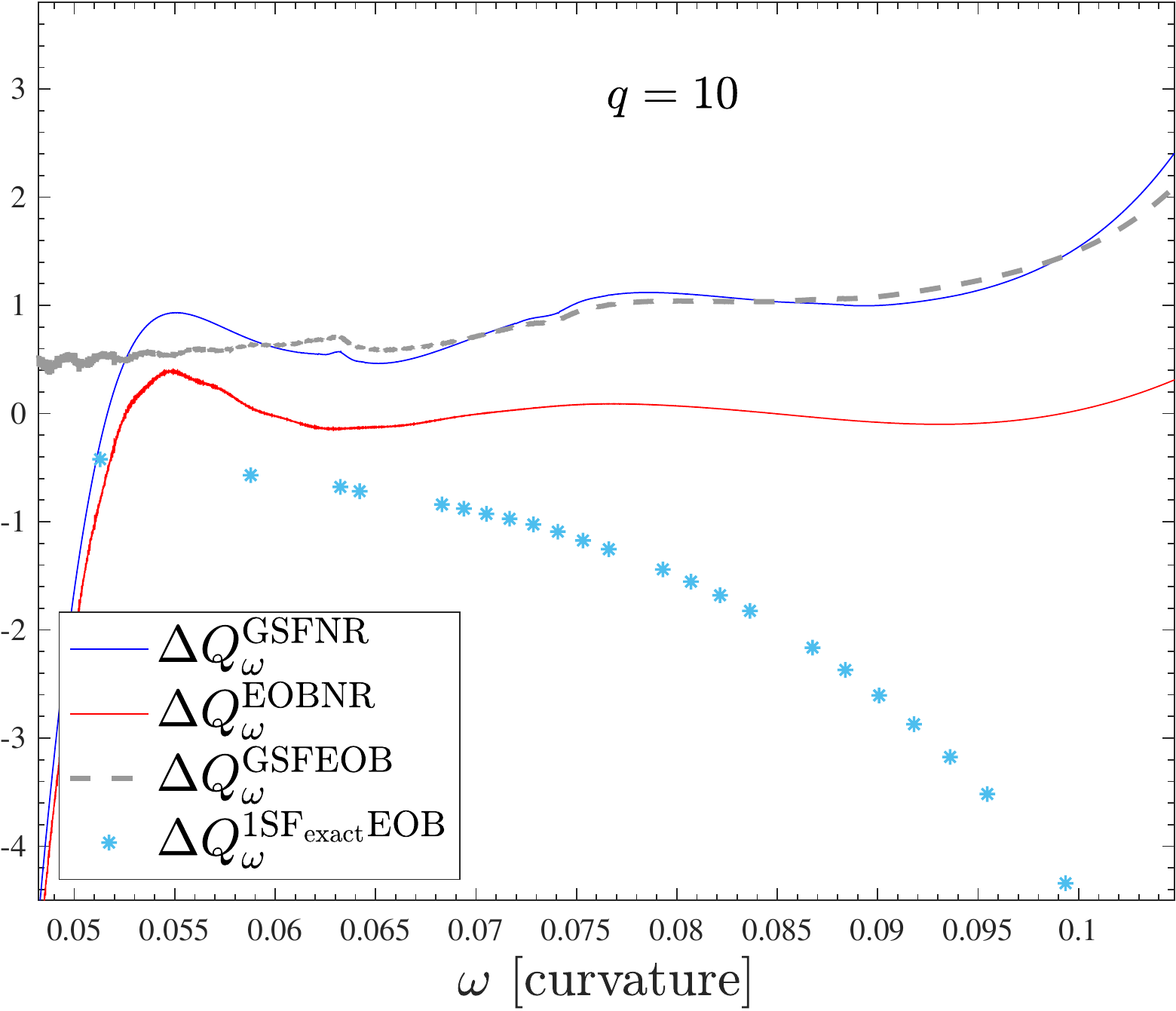} 
\caption{\label{fig:DQomg}Differences between the $Q_\omega$ curves from Fig.~\ref{fig:Qomg_all}.
The logarithmic integral of these yields the phase differences given in Table~\ref{tab:Dphi_Q}.
Note that below $\omega\approx0.055$ the computation of $\Qo^{\rm NR}$
is not reliable.} 
\end{figure*}
%========
The successful computation of $\Qo$ in Ref.~\cite{Damour:2012ky} was based not on the strain quadrupole 
waveform, but rather on the {\it curvature} waveform, i.e. the Weyl scalar\footnote{Note that we simplify here 
the notation and define $\psi_4^{22}\equiv R\psi_4^{22}$, where $R$ is the extraction radius.} $\psi_4^{22}$.
We use $\psi_4^{22}$ instead
of $h$ because the former is less affected by various kinds of high-frequency and low-frequency noise and
it is simpler to obtain a $\Qo$ that is qualitatively and quantitatively reliable.
For our $Q_\omega$ analysis\footnote{Later, we will revert to using $\omega$ to denote the frequency of the $\ell = m = 2$ strain multipole.} we thus adopt the phase convention
\be\label{psi4 phase}
\psi_4^{22}=|\psi_4^{22}|e^{-i\phi_{22}} , 
\ee
and define the corresponding frequency $\omega\equiv \dot{\phi}_{22}$.

\begin{figure*}[t]
\center
\includegraphics[width=0.32\textwidth]{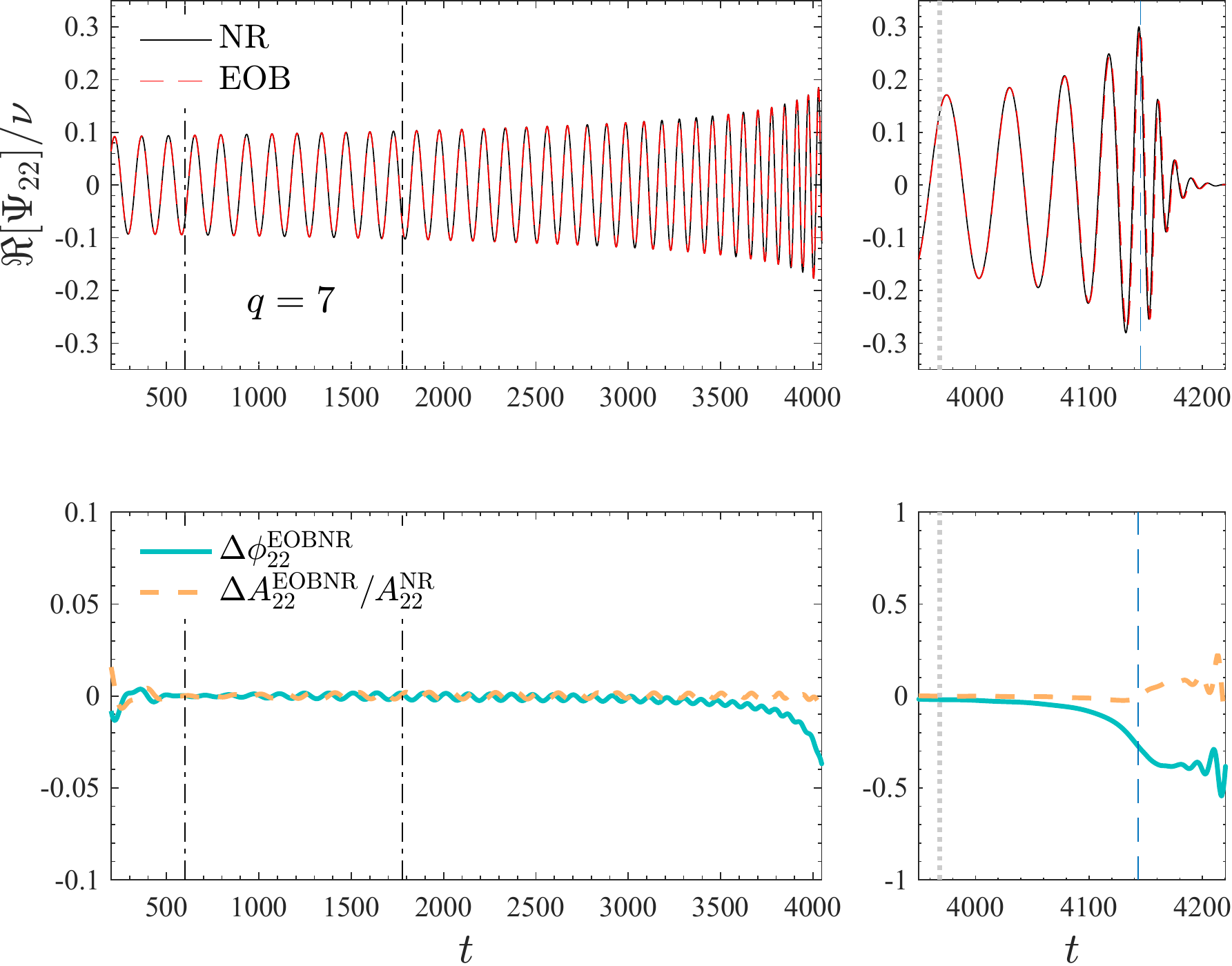}
\includegraphics[width=0.32\textwidth]{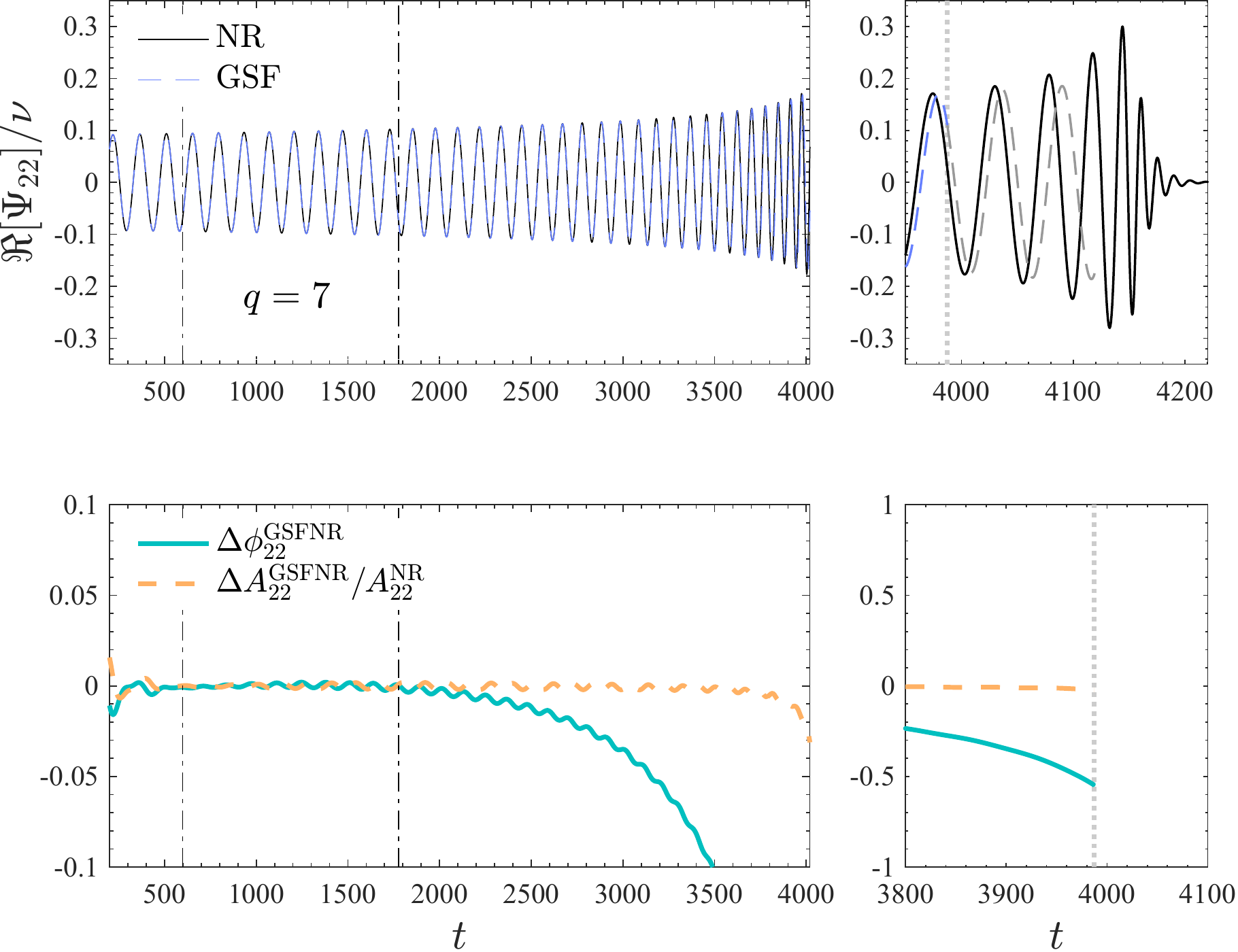} 
\includegraphics[width=0.32\textwidth]{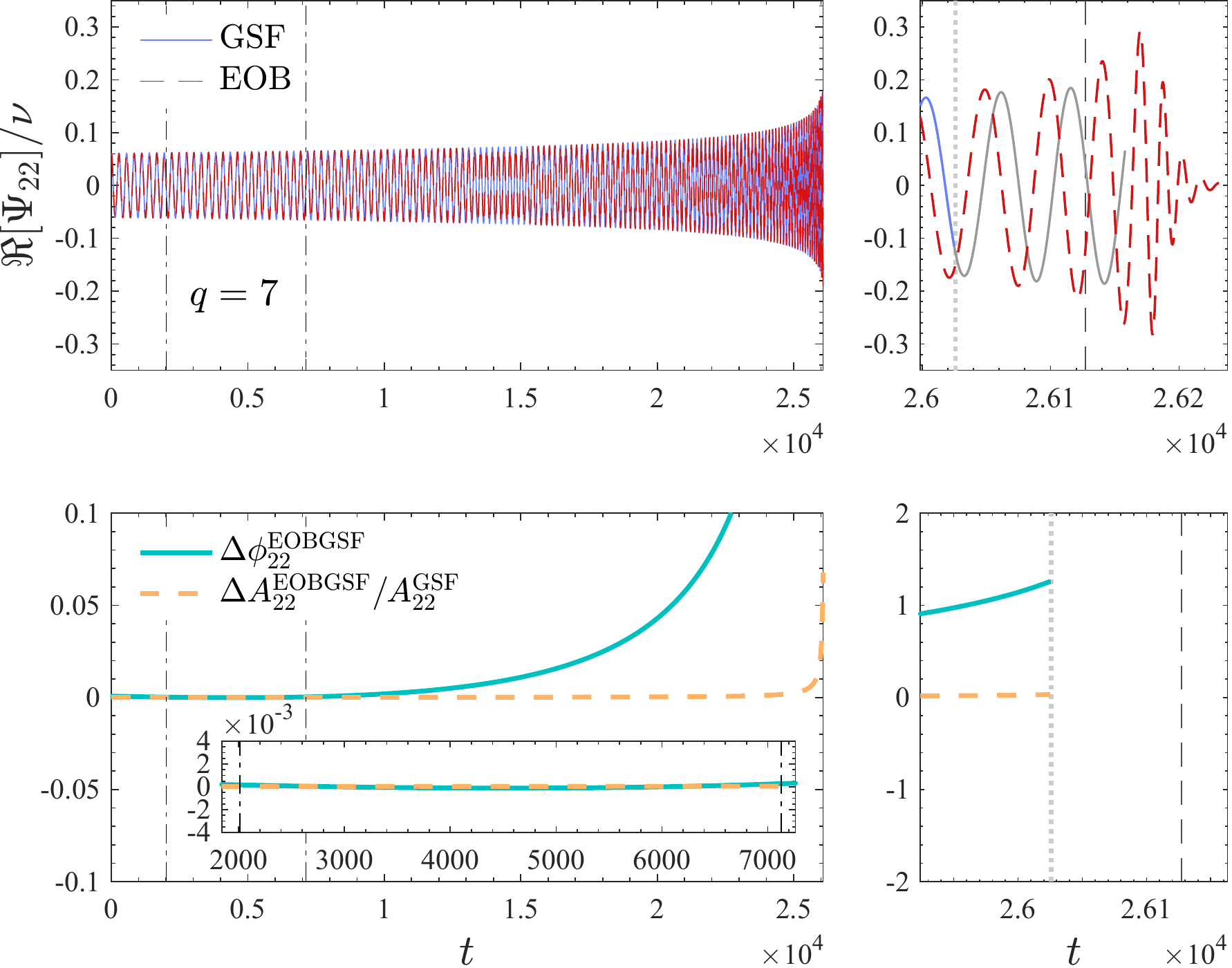} 
\caption{\label{fig:phasings_q7} 
Triple EOB/NR/GSF comparison for $q=7$. Left: EOB/NR phasing using SXS:BBH:0298.
The alignment frequency window is $[\omega_{\rm L}, \omega_{\rm R}] = [0.044, 0.05]$ (indicated by vertical dash-dotted lines), 
and the phase difference accumulated at the NR merger (dashed blue line) is $\Delta\phi^{\rm EOBNR}_{22}=-0.27$. 
The dotted vertical grey line indicates the time at which $\omega_{22}^{\rm NR} = \omega_{22}^{\rm GSF_{break}}$, 
and the EOB/NR phase difference at that point is $-0.02$.
Middle: GSF/NR phasing comparison using the {\it same} alignment window. One gets $\Delta\phi^{\rm GSFNR}_{22}\simeq -0.55$
at $\omega_{22}^{\rm GSF_{break}}$, and here the dotted vertical grey line indicates indeed the time at which 
$\omega_{22}^{\rm GSF} = \omega_{22}^{\rm GSF_{break}}$. Note that we show in grey the last part of the GSF waveform up to the critical frequency, 
but evaluate the phase difference at a time corresponding to the breakdown frequency (see Table~\ref{tab:Dphi}).
Right: EOB/GSF phasing with alignment window $[\omega_{\rm L}, \omega_{\rm R}] = [0.023, 0.025]$,
that yields $\Delta\phi^{\rm EOBGSF}_{22}=1.26$ at $\omega_{22}^{\rm GSF_{break}}$. 
Here the dashed grey line indicates the EOB last stable orbit (LSO).}
\end{figure*}

\begin{figure*}[t]
\center
\includegraphics[width=0.32\textwidth]{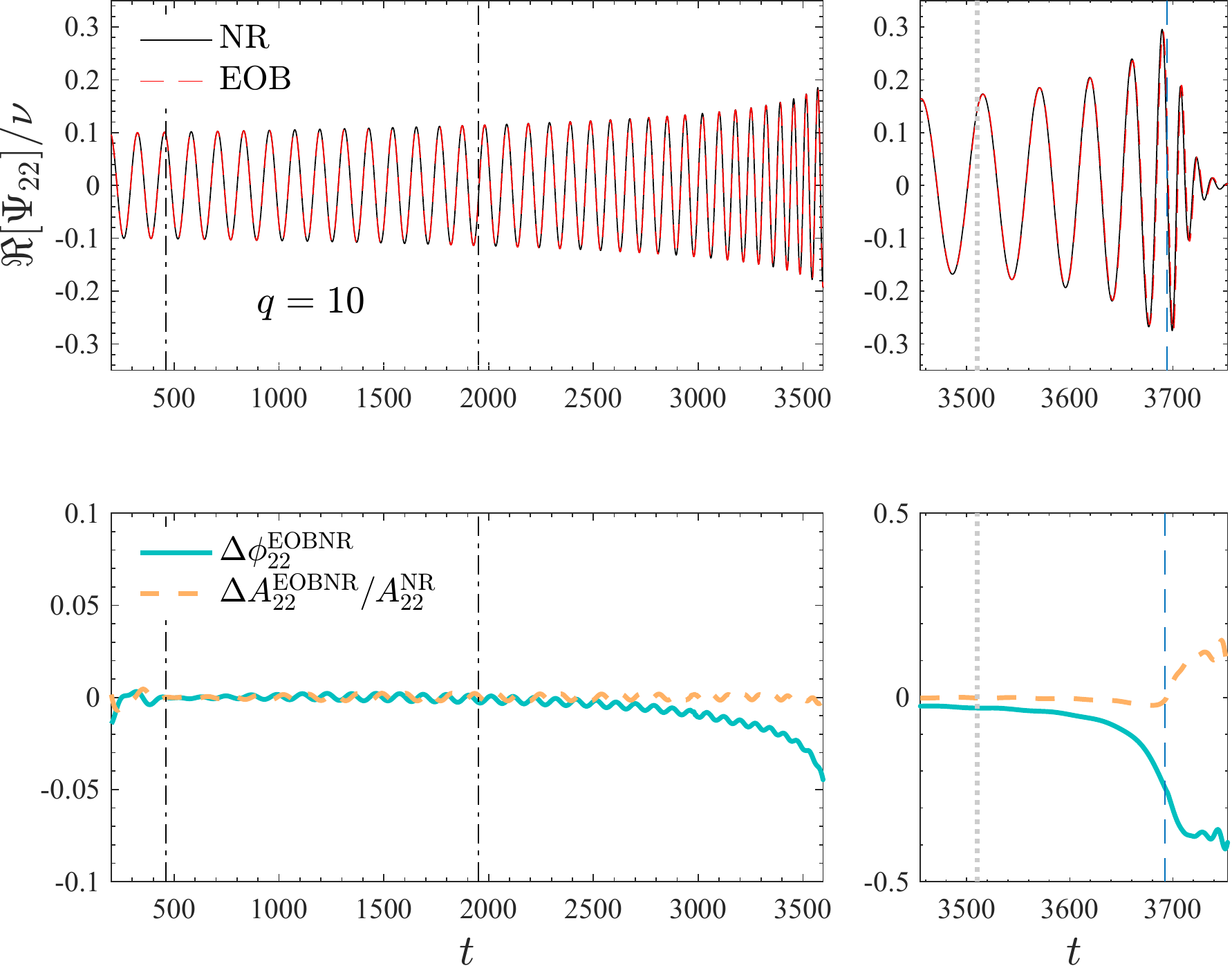}
\includegraphics[width=0.32\textwidth]{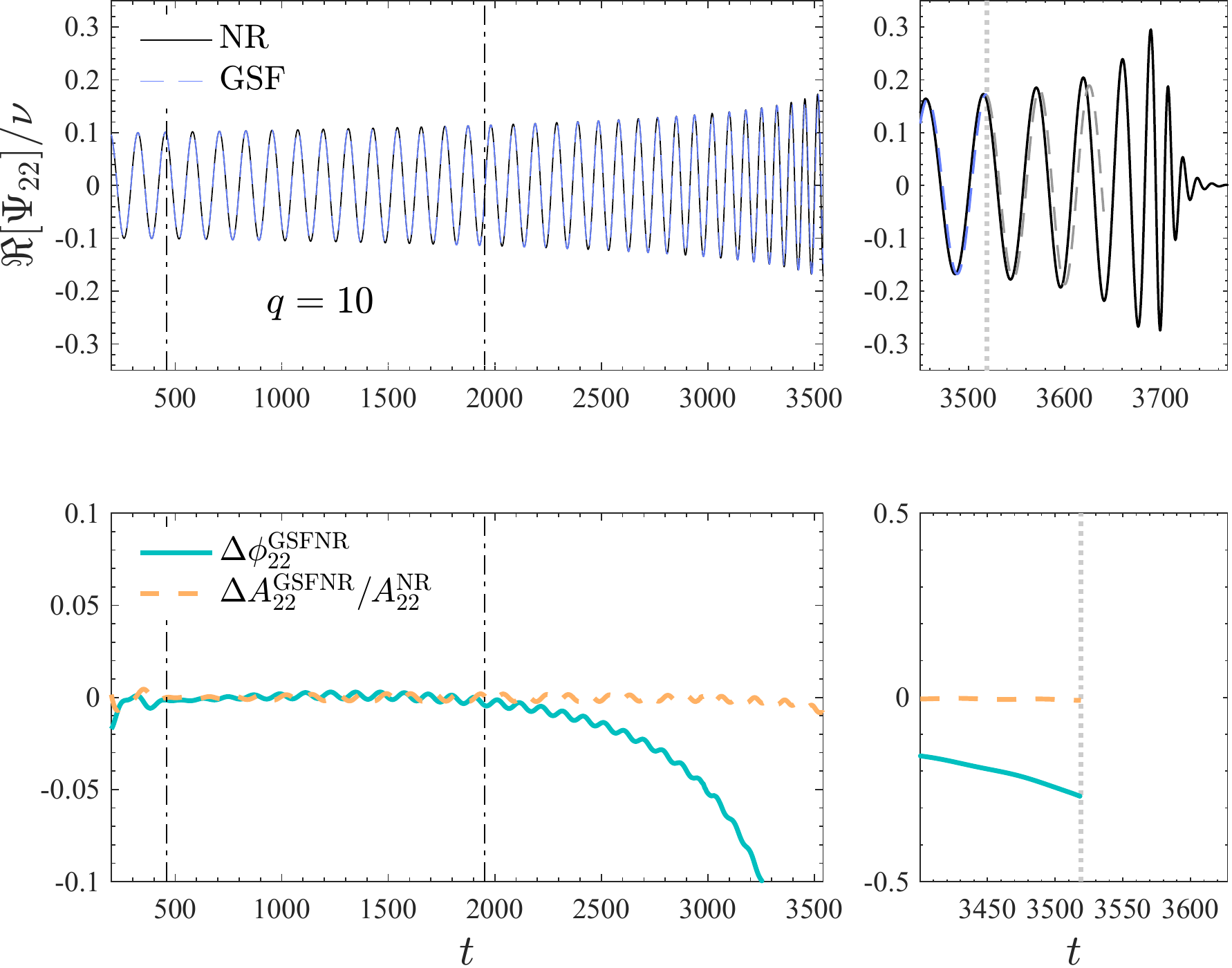} 
\includegraphics[width=0.32\textwidth]{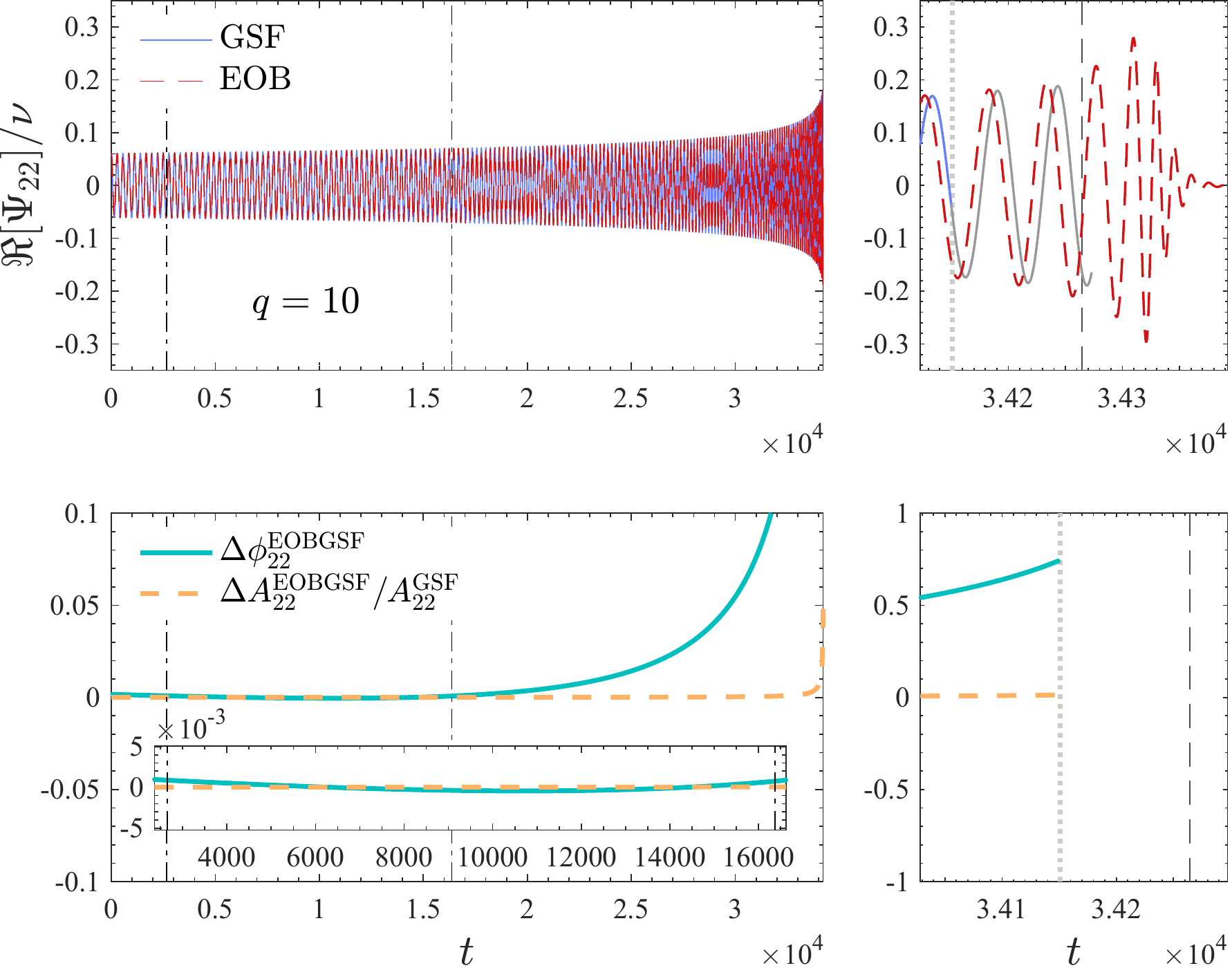} 
\caption{\label{fig:phasings_q10} 
Triple EOB/NR/GSF comparison for $q=10$. Left: EOB/NR phasing using SXS:BBH:0303.
The alignment frequency window is $[\omega_{\rm L}, \omega_{\rm R}] = [0.049, 0.059]$ (indicated by vertical dash-dotted lines), 
and the phase difference accumulated at the NR merger (dashed blue line) is $\Delta\phi^{\rm EOBNR}_{22}=-0.24$. 
The dotted vertical grey line indicates the time at which $\omega_{22}^{\rm NR} = \omega_{22}^{\rm GSF_{break}}$, 
and the EOB/NR phase difference at that point is $-0.03$.
Middle: GSF/NR phasing comparison using the {\it same} alignment window. One gets $\Delta\phi^{\rm GSFNR}_{22}\simeq -0.27$
at $\omega_{22}^{\rm GSF_{break}}$, and here the dotted vertical grey line indicates indeed the time at which 
$\omega_{22}^{\rm GSF} = \omega_{22}^{\rm GSF_{break}}$. Note that we show in grey the last part of the GSF waveform up to the critical frequency, 
but evaluate the phase difference at a time corresponding to the breakdown frequency (see Table~\ref{tab:Dphi}).
Right: EOB/GSF phasing with alignment window $[\omega_{\rm L}, \omega_{\rm R}] = [0.023, 0.028]$,
that yields $\Delta\phi^{\rm EOBGSF}_{22}=0.75$ at $\omega_{22}^{\rm GSF_{break}}$.
Here the dashed grey line indicates the EOB last stable orbit (LSO).}
\end{figure*}

For each SXS dataset, we take $\psi_4^{22}$ data from the SXS catalog, corrected for the spurious motion 
of the center of mass and extrapolated to infinity with extrapolation order $N=3$.
Although $N=4$ extrapolation order would be the ideal choice for the inspiral, we work with $N=3$
to be consistent with the time-domain phasings shown in Figs.~\ref{fig:phasings_q7} and \ref{fig:phasings_q10},
for which the choice is always $N=3$ as a compromise between the early evolution and the merger. 
The NR $\psi_4^{22}$ $\Qo$ is computed using the technique described in Sec.~IIIB of Ref.~\cite{Damour:2012ky},
that aims at removing various kind of spurious oscillations that emerge when taking finite-difference time derivatives of $\phi_{22}$.
More precisely, after the successive application of Savitzky-Golay filters on $\omega$ and $\dot{\omega}$ to
remove the high-frequency noise, the final result is obtained by fitting the Newton-normalized $Q_\omega$
with a suitably chosen rational function. 
Following Ref.~\cite{Damour:2012ky}, we define the Newtonian part of $Q_\omega$ as
\be
Q_\omega^N(\omega)=\dfrac{5}{3\nu}2^{-7/3}\omega^{-5/3} \ ,
\ee
and the Newton-normalized function reads
\be
\label{eq:hatQ}
\hat{Q}_\omega(\omega) = Q_\omega/Q_\omega^N \ .
\ee
The function $\hat{Q}_\omega$ is finally fitted on 
a given frequency interval with a rational function of the form
\be
\label{eq:Qom_ratio}
\hat{Q}_\omega^{\rm fit}=\dfrac{1 + n_1 x + n_2 x^{3/2} + n_3 x^2 + n_4 x^{5/2}+n_5 x^3}{1 + d_1 x + d_2 x^2 + d_3 x^3}
\ee
where $x\equiv (\omega/2)^{2/3}$. Although we are just following step-by-step the
technique applied in Ref.~\cite{Damour:2012ky}, for completeness we collect
all useful technical details in Appendix~\ref{sec:Qomg_clean}.

Figure~\ref{fig:Qomg_all} compares the results of computing three different $Q_\omega$'s for $q=7$ (left) and $q=10$ (right): (i) the NR one computed
using the technique described thus far (black solid line); (ii) the GSF one, simply obtained by taking
the time-derivatives of the strain waveform and applying a low-pass filter to remove high-frequency
noise (blue solid line) and  (iii) the EOB one (red dashed line). We also display the small-$\nu$ expansion truncated at 1PA order, $Q_\omega=\nu^{-1}Q_\omega^0(\omega)+Q_\omega^1(\omega)$, calculated from GSF data using Eqs.~\eqref{Q0 exact} and \eqref{Q1 exact}. The main panel of the figure shows the full $Q_\omega$ 
functions, while the inset focuses on a smaller frequency interval in order to highlight the difference between
the three curves. The figure is quantitatively complemented by Fig.~\ref{fig:DQomg}, which shows
various differences between $Q_\omega$'s, that is: $\Delta Q_\omega^{XY}\equiv Q_\omega^X-Q_\omega^Y$
where $(X,Y)$ can be EOB, GSF or NR.
Fig.~\ref{fig:DQomg} illustrates that the estimate of the NR $Q_\omega$ is not reliable before
$\omega\sim 0.055$, due to boundary effects related to fitting procedure.
If we focus on the part of the plot for $\omega>0.055$, the GSF description 
yields a $Q_\omega$ that is noticeably different from the other two, with a somewhat smaller difference for $q=10$ than for $q=7$. Even for the $q=10$ case, $\Delta Q^{\rm GSFNR}_\omega$ remains of order unity on a large frequency interval. By contrast,
$\Delta Q^{\rm EOBNR}_\omega$ remains consistently close to zero across all frequencies. Finally, Fig.~\ref{fig:DQomg} also indicates that, although agreement between $Q_\omega^{\rm GSF}$ and $Q_\omega^{\rm EOB}$ improves at lower frequencies, there is still a noticeable difference, even outside the frequency interval where it was possible to reliably compute $Q_\omega^{\rm NR}$.

The effect that all of this has on the waveform phasing is made quantitative in Table~\ref{tab:Dphi_Q}, which lists the phase differences accumulated 
on the $(\omega_1,\omega_2)$ frequency interval evaluated using Eq.~\eqref{eq:Dphi_from_Q}. These dephasings can be compared to (and are compatible with) Fig.~4 in Ref.~\cite{Wardell:2021fyy}; the frequency interval $(0.055,0.095)$ we use here roughly corresponds to the interval between the square and the circle in the third panel of that figure or between the downward and upward triangle in the fourth panel, for example.

The fact that $Q_\omega^{\rm GSF}$ is always above $Q_\omega^{\rm NR}$ (or $Q_\omega^{\rm EOB}$) physically
means that the system is inspiralling {\it more slowly} than it should according to the NR prediction, and is reflected in the fact that the phase differences are positive. One should be careful not to read too much into this as, for example, a similar analysis with the 1PAF1 model yields the opposite result. In that case $Q_\omega^{\rm GSF}$ {\it underestimates} the true value, making the system inspiral {\it more quickly} than it should. In the next section we will rephrase this finding 
also in terms of more intuitive waveform comparisons in the time-domain.

The data for the truncated expansion $\nu^{-1}Q_\omega^0+Q_\omega^1$ in these plots also reveals valuable information. Because $Q_\omega$ is a nonlinear function of $\dot\Omega$,  $Q_\omega$ as calculated from the 1PAT1 model contains contributions at all orders in $\nu$. The difference between $Q^{\rm GSF}_\omega\equiv Q^{\rm 1PAT1}_\omega$ and $\nu^{-1}Q_\omega^0+Q_\omega^1$ in these figures suggests that these higher-order effects in $Q^{\rm GSF}_\omega$ are significant at these mass ratios; and in particular, the behaviour of $\nu^{-1}Q_\omega^0(\omega)+Q_\omega^1(\omega)$ near the LSO tells us that the higher-order effects are entirely responsible for the divergence of $Q^{\rm GSF}_\omega$ at the LSO. On the other hand, these higher-order contributions in $Q^{\rm GSF}_\omega$ will differ from the true values of $Q^n_\omega$ for $n>1$, as the true values will receive contributions from $n$PA terms in $\dot\Omega$. The difference between $Q^{\rm NR}_\omega$ and $\nu^{-1}Q_\omega^0+Q_\omega^1$ tells us that these true higher-order terms are also significant (and significantly different than those in 1PAT1) at these mass ratios. We return to these points in Sec.~\ref{sec:nu_dependence}.

%========================================
% Table of GSF frequencies and phasing accumulated
%========================================
\begin{table}[t]
\begin{center}
\begin{ruledtabular}
\begin{tabular}{c  c c c  c}
$q$ & $(\omega_1,\omega_2)$ &$\Delta\phi^{\rm GSFNR}$ & $\Delta\phi^{\rm GSFEOB}$ & $\Delta\phi^{\rm EOBNR}$ \\
\hline
\hline
7 & $(0.055,0.095)$ & 0.8703  & 0.8486 &  0.0217  \\
10 & $(0.055,0.095)$ & 0.4627 & 0.4581  &  0.00463  \\
\end{tabular}
\end{ruledtabular}
\end{center} 
\caption{\label{tab:Dphi_Q}Accumulated phase differences (in radians) obtained by integrating the $\psi_4^{22}$ $\Qo$ 
curves of Fig.~\ref{fig:DQomg} between frequencies $(\omega_1,\omega_2)$.
Note that for these mass ratios and over this frequency range the EOB/NR phase differences are smaller than GSF/NR by
more than an order of magnitude.}
\end{table}
%=========================================

%===========
% Figure for q=7
%===========

%============
% Figure for q=10
%============
%=================
% q=10: SXS:BBH:1107
%=================
\begin{figure}[t]
\center
\includegraphics[width=0.42\textwidth]{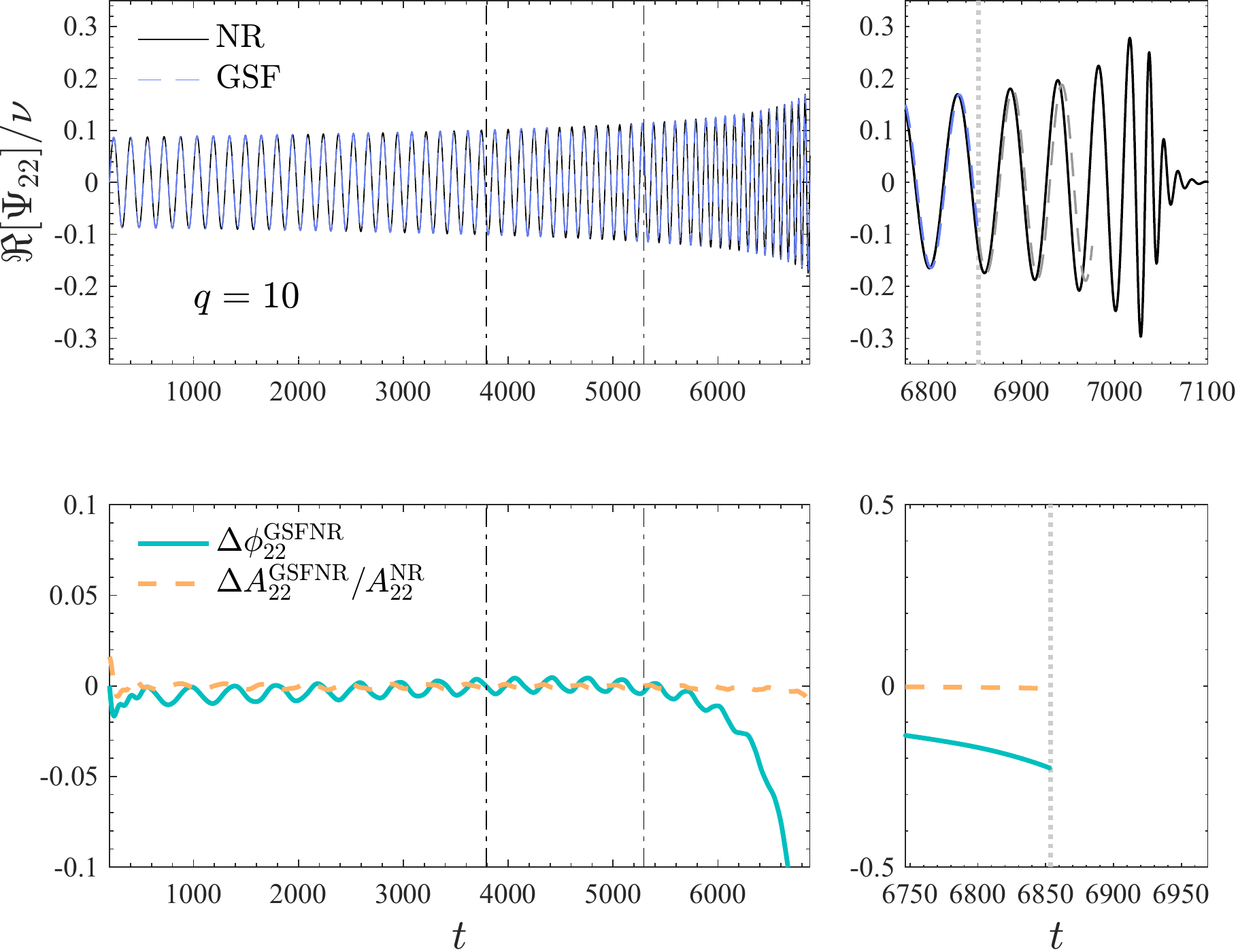}
\caption{\label{fig:phasing_q10_1107} Comparison agaist the $q=10$ simulation SXS:BBH:1107,
using the same alignment interval as was used for SXS:BBH:0303, namely $[\omega_{\rm L}, \omega_{\rm R}] = [0.049, 0.059]$.
The final accumulated phase difference is $-0.23$. If the alignment interval is moved to lower frequencies,
$[\omega_{\rm L}, \omega_{\rm R}] = [0.040, 0.047]$,  we get $\Delta\phi^{\rm GSFNR}_{22}= -0.26$.
As previously, we show in grey the last part of the GSF waveform up to the critical frequency, 
but evaluate the phase difference at a time corresponding to the breakdown frequency.}
\end{figure}

%==================
% Comparing waveforms
%==================
\section{Comparing waveforms in the time domain}
\label{sec:waveform}

\subsection{EOB/NR/GSF: comparable-mass case}

Let us now complement the $Q_\omega$-based analysis with additional information
obtained using more standard phasing comparisons in the time domain. Unlike 
the gauge-invariant $Q_\omega$ phase analysis, to align the two waveforms in 
the time-domain we need to specify an arbitrary phase shift and an arbitrary time shift. 

We follow here a well-tested procedure analogous to the one described in Sec.~V A of Ref.~\cite{Baiotti:2011am},
which in turn stems from Sec.~VI A of Ref.~\cite{Boyle:2008ge}. In the latter it was pointed out 
that by simply matching the GW phase and frequency at a fiducial time in an NR simulation, one does not obtain a robust 
estimate of the phase difference, especially when the chosen time corresponds to a low frequency where the NR waveform
is contaminated by noise and residual eccentricity.
One needs instead to consider an interval and to minimize the phase difference over this interval. 

Given two $\ell =m =2$ waveform strain multipoles in the form~\eqref{eq:RWZnorm},
and considering the frequency $\omega = \omega_{22}$, we choose a frequency interval $[\omega_L, \omega_R]$ which we use to define
a common time interval $[t_L, t_R]$ for the two waveforms. Since a given frequency interval will not necessarily correspond to the same time interval in two difference waveforms, we here set the time interval using the NR waveform when comparing EOB or GSF to NR, 
and using the GSF waveform when comparing EOB to GSF. We then interpolate the other waveforms onto a common grid of time steps within this interval. Given that the time interval is made up of $N$ numerical points, we have two timeseries
of the phase $\phi_1(t_i)$ and $\phi_2(t_i)$, where $i = 1, ..., N$, that allow us to define the quantity
\be
\Delta \phi (t_i , \tau, \alpha) = \left[ \phi_2 (t_i - \tau) - \alpha \right] - \phi_1(t_i)  \, .
\ee
We then determine $\tau$ and $\alpha$, respectively the time and the phase shift, so that they minimize 
the root-mean-square deviation of $\Delta \phi$ over $[t_L, t_R]$,
\be
\sigma = \sqrt{\frac{1}{N} \sum_{i = 1}^{N}\left[\Delta \phi (t_i , \tau, \alpha)\right]^2} \, .
\ee
For a given value of $\tau$, the minimization of $\sigma$ is faster if one optimizes $\alpha$ by defining it analytically as 
$\alpha = \frac{1}{N} \sum_{i=1}^{N} \phi_2 (t_i - \tau) - \phi_1(t_i)$.
We note in passing that $\sigma$ also gives a useful estimate of phase errors, and in the
waveform alignment considered in the following it is always of order $10^{-4}$.
Finally, the two obtained waveforms are
\begin{subequations}
\begin{align}
\Psi_{22}^1 &= A_{22}^1(t_1) e^{- i \phi_1(t_1)} \, , \\
\Psi_{22}^2 &= A_{22}^2(t_2 - \tau) e^{- i [\phi_2(t_2 - \tau) - \alpha ]} \, ,
\end{align}
\end{subequations}
and the second one is again interpolated onto the time grid of the first.

Evidently, any computation of the phase difference between two waveforms will depend on the frequency interval over which the comparison is made. For the purposes of GW data analysis, it is the phase error over a fixed frequency interval that is most relevant.  One may also wish to compute a {\it total} accumulated phase difference by aligning the waveforms in the infinite past and computing the phase difference at some time near the end of the waveform. However, as described in Sec.~\ref{sec:1PA accuracy}, this is not sensible when using a 1PA GSF model: the phase error in the model will be larger for larger frequency intervals, and it will ultimately become infinite if the frequency interval starts in the infinite past, at $\omega=0$. Restating the discussion in Sec.~\ref{sec:1PA accuracy} in terms of $Q_\omega$, we can say that the phase error $\int_0^\omega \Delta Q_\omega d \log \omega$ will diverge unless $\Delta Q_\omega$ tends to zero as $\omega \to 0$. From the analysis around Eq.~\eqref{dphidOmega - small omega}, we find $\Delta Q_\omega\sim \omega \nu\phi'_2\sim \nu \omega^{-1/3}$, blowing up in the $\omega\to0$ limit. %This is not guaranteed to give a finite result unless the rate of accumulated difference falls off sufficiently fast at early times. One way to assess this is by considering the 
%$Q_\omega$ analysis described previously: if the difference in $Q_\omega$ (for example, as shown in Fig.~\ref{fig:DQomg}) tends to zero as $\omega \to 0$ then the integral $\int_0^\omega \Delta Q_\omega d \log \omega$ will converge. Unfortunately, this is not the case for the comparisons that we are interested in here. This can be seen by considering the PN expansion of $Q_\omega$ \cite{Damour:2012ky}
%\begin{align}
%  Q^{\rm PN}_\omega &= \dfrac{5}{3\nu}2^{-7/3}\omega^{-5/3}  \left[ 1 + \left(\frac{743}{336} + \frac{11}{4} \nu\right)x - 4 \pi  x^{3/2} \right. \nonumber \\
%  & \quad + \left. \left(\frac{3058673}{1016064} + \frac{5429}{1008}\nu + \frac{617}{144} \nu^2\right) x^2 + \mathcal{O}(x^{5/2})\right].
%\end{align}
%By definition, the 2GSF approximation neglects the $O(\nu^2)$ term in the $O(x^2)$ coefficient so when one takes into account the fact that $x = (M \Omega)^{2/3} \approx (M \omega/2)^{2/3}$ it is clear that the difference $\Delta Q_\omega^{\rm GSFEOB}$ will be non-zero in the limit $\omega \to 0$. 
We therefore focus here on computing phase differences over a finite portion of the inspiral and consider how those differences depend on how much of the inspiral is included.

Figure~\ref{fig:phasings_q7} focuses on the $q=7$ binary and shows the time-domain 
phasing comparison between EOB, NR and GSF, where the alignment frequency interval 
is $[\omega_L,\omega_R]=[0.044,0.05]$ for the EOB/NR and GSF/NR comparisons (left and middle panels), 
while $[\omega_L,\omega_R]=[0.023,0.025]$ for the EOB/GSF one (rightmost panel). The 
dotted line in the part of the figure including the EOB and NR mergers indicates the point 
corresponding to the  breakdown of the two-timescale approximation that GSF calculations are based on (see Table~\ref{tab:Dphi}).
The left panel of Fig.~\ref{fig:phasings_q7} illustrates the EOB/NR phase agreement. We see that $\Delta\phi^{\rm EOBNR}_{22}\equiv \phi^{\rm EOB}_{22}-\phi^{\rm NR}_{22}$ remains {\it flat} (oscillating around zero) for most of the inspiral, then it is 
$-0.02~\text{rad}$ when $\omega_{22}^{\rm NR} = \omega_{22}^{\rm GSF_{break}}$
%less than $0.05~\text{rad}$ at $\omg_b$, 
and it remains less than $0.5~\text{rad}$ through plunge.  Note that the estimated NR phase uncertainty at merger\footnote{This uncertainty is estimated by comparing the simulation with the highest available resolution 
to the one with next-to-highest resolution.} for this dataset is rather small, $\delta\phi_{\rm mrg}^{\rm NR}=-0.0775$~rad. The middle panel of Fig.~\ref{fig:phasings_q7} displays 
the corresponding 2GSF/NR phase comparison, obtained using the same alignment window. 
We see that $\Delta \phi_{22}^{\rm GSFNR}$ is oscillating around zero initially, but then 
decreases to reach $\simeq -0.55$~rad at $\omg_b$, a value significantly larger the EOB/NR dephasing.

Since we do not have longer NR simulations at hand, we use a longer EOB waveform to gain
some more insights on the dephasing over a larger portion of the inspiral. The fact that the top-left panel of Fig.~\ref{fig:phasings_q7} 
indicates that the \TEOBResumS{} model offers an excellent description of the phasing over the full
inspiral of SXS:BBH:0298 suggests that it will give a similarly good representation
of the true waveform also at lower frequencies.
In the right panel of Fig.~\ref{fig:phasings_q7} we show an EOB/GSF comparison with
the alignment interval chosen in the very early inspiral, $[\omega_L,\omega_R]=[0.023,0.025]$.
In this case the phase difference accumulated up to $\omg_b=0.109$
is $\sim 1.2646$.

%========================================
% Table of GSF frequencies and phasing accumulated
%========================================
\begin{table*}[t]
\begin{center}
\begin{ruledtabular}
\begin{tabular}{c  c c c | c c c | c}
$q$ &  $\omega_{22}^{\rm GSF_{break}}$ &  $\omega_{22}^{\rm GSF_{critical}}$ & $\omega_{22}^{\rm EOB_{LSO}}$ 
       & $[\omega_L, \omega_R]$ 
      & $\Delta\phi_{22,t}^{\rm EOBGSF}$ & $\Delta\phi^{\rm EOBGSF}_{22,Q_\omega}$
      & $\Delta\phi_{22,t}^{\rm EOB_{6PN}GSF}$  \\
\hline
\hline
%7     & 0.109341 & 0.12032 & 0.15707 & [0.023, 0.025] &  1.3958 & 1.4068 \\
%10 & 0.114402  &0.123602 &0.15127 & [0.023, 0.028] &  0.86978 & 0.86804 \\ 
%15 & 0.119589 & 0.12678 & 0.14644 &   [0.023, 0.028] & 0.433  &  0.46925 \\
%32 & 0.126963 & 0.127465  & 0.14104 & [0.023, 0.033] &  $-0.25052$ & $-0.22515$ \\ 
%64 & 0.12738 & 0.127429 &0.13858 &  [0.023, 0.033] &  $-0.67611$ & $-0.63741$  \\
%128 &  0.12769 & 0.12778 & 0.13733 &  [0.023, 0.027] &  $-1.3134$ & $-1.3126$ 
% improved script that integrates Qomg
%7     & 0.109341 & 0.12032 & 0.15707 & [0.023, 0.025] &  1.3958 & 1.3951 \\
%10 & 0.114402  &0.123602 &0.15127 & [0.023, 0.028] &  0.86978 & 0.86806 \\ 
%15 & 0.119589 & 0.12678 & 0.14644 &   [0.023, 0.028] & 0.433  & 0.43222 \\
%32 & 0.126963 & 0.127465  & 0.14104 & [0.023, 0.033] &  $-0.25052$ & $-0.25058$ \\ 
%64 & 0.12738 & 0.127429 &0.13858 &  [0.023, 0.033] &  $-0.67611$ & $-0.67551$  \\
%128 &  0.12769 & 0.12778 & 0.13733 &  [0.023, 0.027] &  $-1.3134$ & $-1.3125$ 
% latest with updated breakdown frequencies
7     & 0.10618 & 0.12032 & 0.15707 & [0.023, 0.025] & 1.2646 & 1.2639  & \dots \\
10 & 0.10820  &0.12360 &0.15127 & [0.023, 0.028] & 0.7455 & 0.7438 & \dots  \\ 
15 & 0.11050 & 0.12678 & 0.14644 &   [0.023, 0.028] & 0.3782 & 0.3775 & 0.4772 \\
32 & 0.11455 & 0.12747  & 0.14104 & [0.023, 0.033] & $-0.1267$ & $-0.1266$ & 0.0656 \\ 
64 & 0.11784 & 0.12743 &0.13858 &  [0.023, 0.033] & $-0.5091$ & $-0.5085$ & $-0.1213$ \\
128 & 0.12068 & 0.12778 & 0.13733 &  [0.023, 0.027] & $-1.1287$  & $-1.1278$ & $-0.2677$
\end{tabular}
\end{ruledtabular}
\end{center} 
\caption{\label{tab:Dphi}From left to right: the mass ratio $q$; 
the frequency related to the breakdown of the two-timescale approximation, $\omega_{22}^{\rm GSF_{break}}$; 
the frequency at which the first-order and second-order forcing terms in the GSF evolution cancel each other, $\omega_{22}^{\rm GSF_{critical}}$;
the adiabatic LSO GW frequency $\omega_{22}^{\rm EOB_{LSO}}$; the phase difference, computed up to $\omega_{22}^{\rm GSF_{break}}$, either
using the time-domain alignment or the $Q_\omega$ analysis. The consistency between the two values confirms the robustness of the EOB/GSF phasings.
The last column shows the time-domain phase difference  obtained by improving the $\ell=m=2$  \TEOBResumS{} resummed radiation reaction 
with a 6PN test-mass term~\cite{Nagar:2022icd}.}
\end{table*}
%=====================================

To check the possible presence of systematics related to the alignment ambiguities, we also computed the
corresponding dephasing using the EOB and GSF $\Qo$'s.
Our interest, as per the right panel of Fig.~\ref{fig:phasings_q7}, is on the EOB-GSF phase difference
between initial time, $t_1$, and final time, $t_2$, corresponding to $\omg_b$.
With both $\Qo^{\rm EOB}$ and $\Qo^{\rm GSF}$ at hand, the equivalent of the time-domain
phasing up to $t_2$ is obtained as
\begin{align}\label{eq:Qint - fixed t interval}
\Delta\phi_{Q_{\omega}}^{\rm EOBGSF} &= \int_{\omega^{\rm EOB}(t_1)}^{\omega^{\rm EOB}(t_2)} Q_{\omega}^{\rm EOB} d\log \omega^{\rm EOB} \nonumber\\
& \qquad - \int_{\omega^{\rm GSF}(t_1)}^{\omega^{\rm GSF}(t_2)}  Q_{\omega}^{\rm GSF} d\log \omega^{\rm GSF}.
\end{align}
The result of this calculation is shown in Table~\ref{tab:Dphi} and is in excellent agreement with the time-domain dephasing
also given in the same table. This confirms, a posteriori, the reliability of our dephasing estimates.

The same procedure and conclusions we drew for $q =7$ also hold for the  $q=10$ case: the various time-domain phasings are 
shown in Fig.~\ref{fig:phasings_q10}.
Here the accumulated GSF phase difference, compared to either EOB or NR  (see middle panel of Fig.~\ref{fig:phasings_q10}) 
is a factor of $\sim 2 $ smaller than the $q=7$ case.

To benchmark our analysis, we can also check the robustness of our conclusions using a different $q=10$ dataset
available in the SXS catalog, SXS:BBH:1107. This simulation was also considered in Ref.~\cite{Wardell:2021fyy}; 
it has a larger initial eccentricity but also starts from a larger initial separation than SXS:BBH:0303. Figure~\ref{fig:phasing_q10_1107} 
shows the GSF/NR phasing comparison using the same alignment interval as for SXS:BBH:0303.
If the alignment interval is lowered to $[\omega_L,\omega_R]=[0.040,0.047]$, the phase difference up
to $\omg_b$ increases by $\sim 10\%$, from $-0.23$ to $-0.26$. This supports our previous understanding
that the accumulated phase difference increases as a larger portion of the inspiral is considered.

A reader might note that the GSF-NR dephasings reported in this section are substantially smaller than those in the previous section. This difference is not due to our use of $Q_\omega$ in one analysis and direct measurements of $\phi(t)$ in the other. Instead the distinction is between dephasings on a fixed \textit{time} interval or on a fixed \textit{frequency} interval. If we integrate $Q_\omega$ over a fixed time interval, as in Eq.~\eqref{eq:Qint - fixed t interval}, then the resulting dephasing will agree with a direct measurement of $\phi(t)$ on that interval. % An important point to note here is that the integration of the difference in $\Qo$ yields different results when it is
%done on a fixed \textit{time} interval or on a fixed \textit{frequency} interval. 
This equivalence is shown by the results
in Table~\ref{tab:Dphi}, where the time interval corresponds to the one used for the waveforms (after the alignment).
But the EOB and GSF frequency intervals are not the same on this time interval, namely $\omega^{\rm EOB}(t_1) \ne \omega^{\rm GSF}(t_1)$
and $\omega^{\rm EOB}(t_2) \ne \omega^{\rm GSF}(t_2)$. The phase difference obtained by this integration
can be compared to the one yielded by the waveform aligned in the time domain, and correspondently
brings informations about the waveform dephasing. By contrast, integrating the difference in $\Qo$ on a fixed
\textit{frequency interval}, as done in Table~\ref{tab:Dphi_Q}, yields an accumulated phase that gives information
about the adiabaticity of the models on that frequency interval. Since for each model a fixed frequency
interval corresponds to a different time interval, namely $t^{\rm EOB}(\omega_1) \ne t^{\rm GSF}(\omega_1)$
and $t^{\rm EOB}(\omega_2) \ne t^{\rm GSF}(\omega_2)$, the phase differences evaluated in this way \textit{cannot}
be compared to those of the time-domain alignment.

In conclusion, our comprehensive analysis here complements Ref.~\cite{Wardell:2021fyy}, and it (i) demonstrates the limitations of the 1PAT1 model for comparable mass binaries, and (ii) reaffirms the high fidelity of \TEOBResumS{} for these mass ratios. %suggests that the 1PAT1 model is
%less accurate than \TEOBResumS{} for comparable mass binaries.

%==============
% Fig. 1 - phasings
%==============
\begin{figure*}[t]
\center
\includegraphics[width=0.43\textwidth]{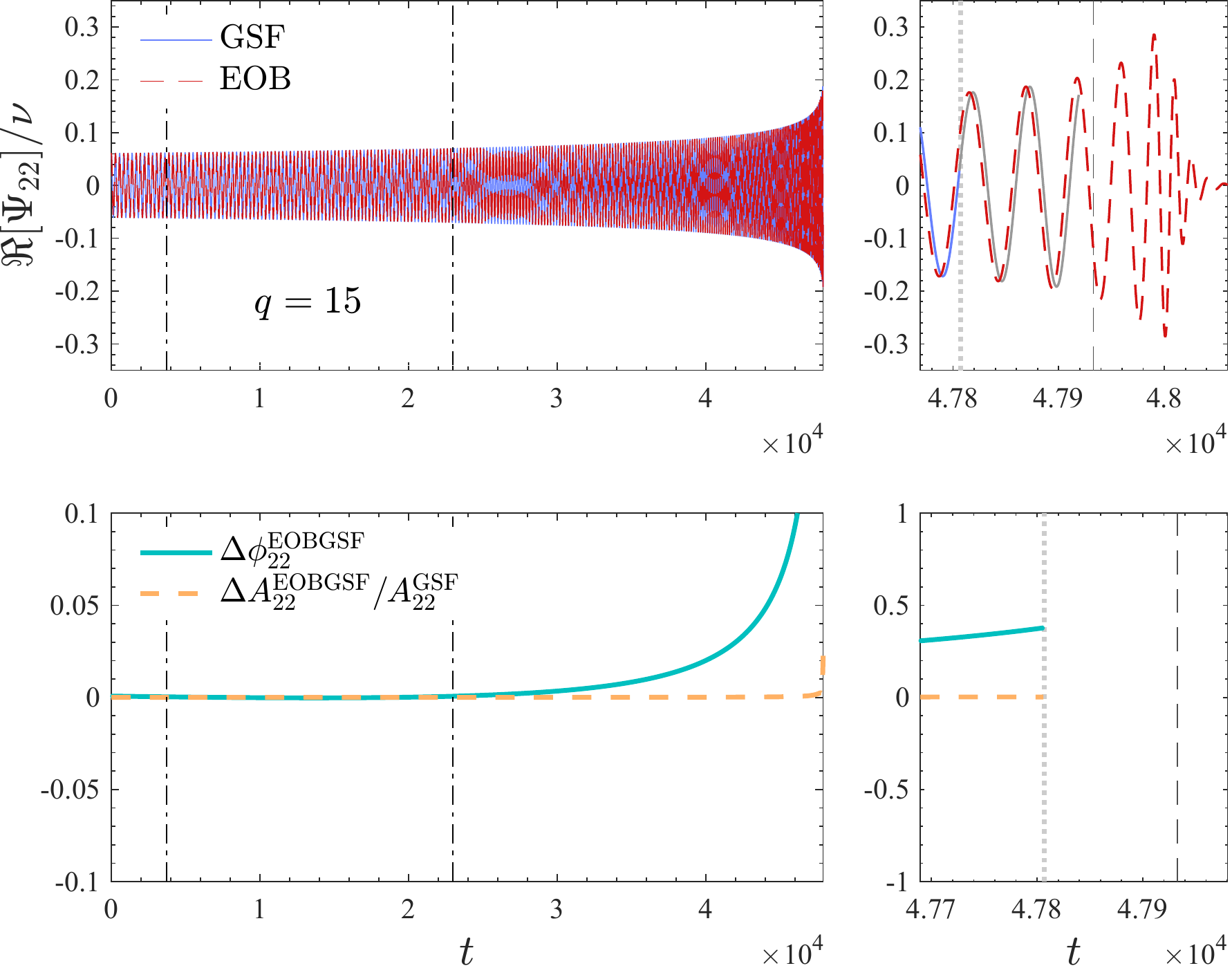} 
\hspace{8mm}
\includegraphics[width=0.43\textwidth]{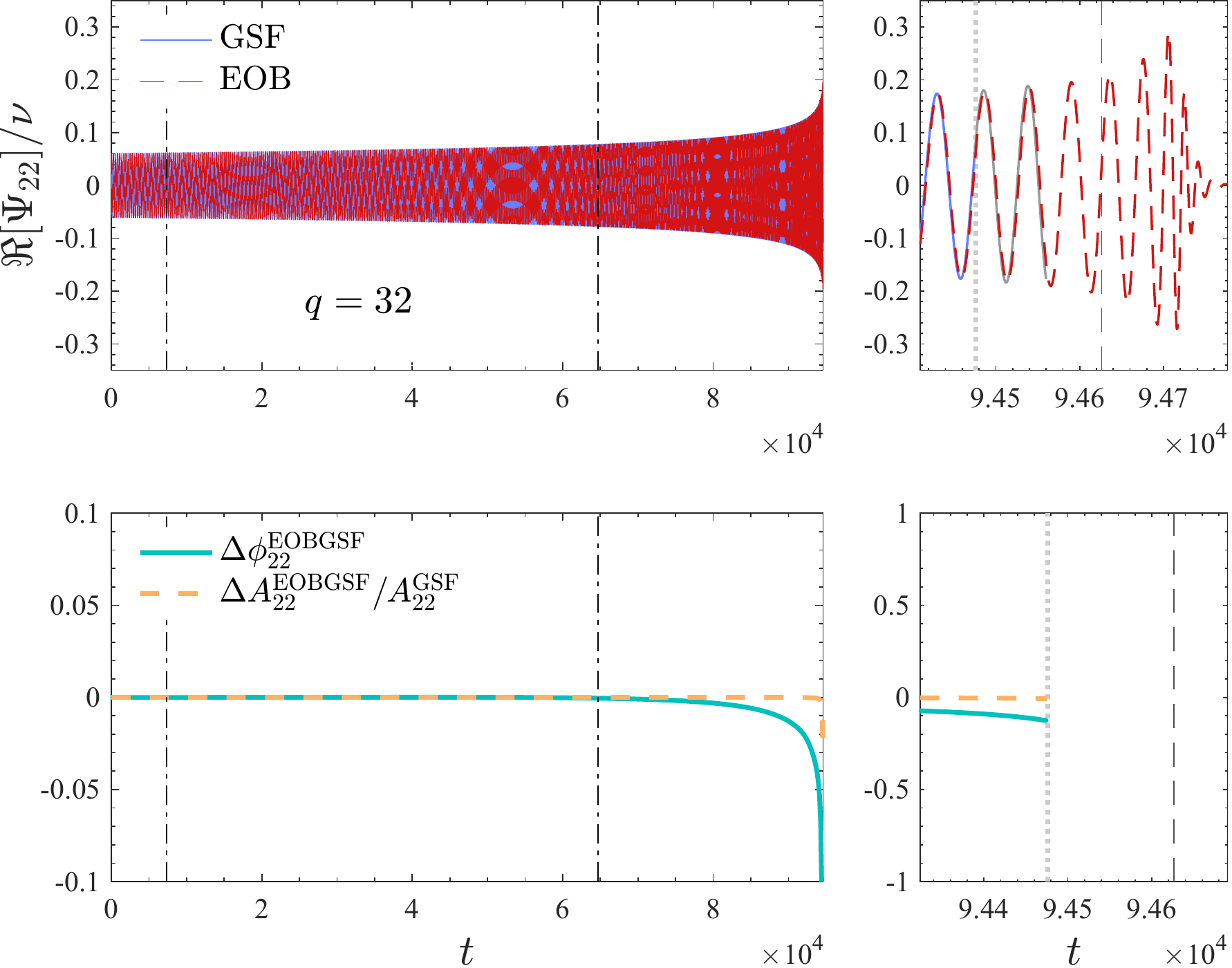}\\ 
\vspace{8mm}
\includegraphics[width=0.43\textwidth]{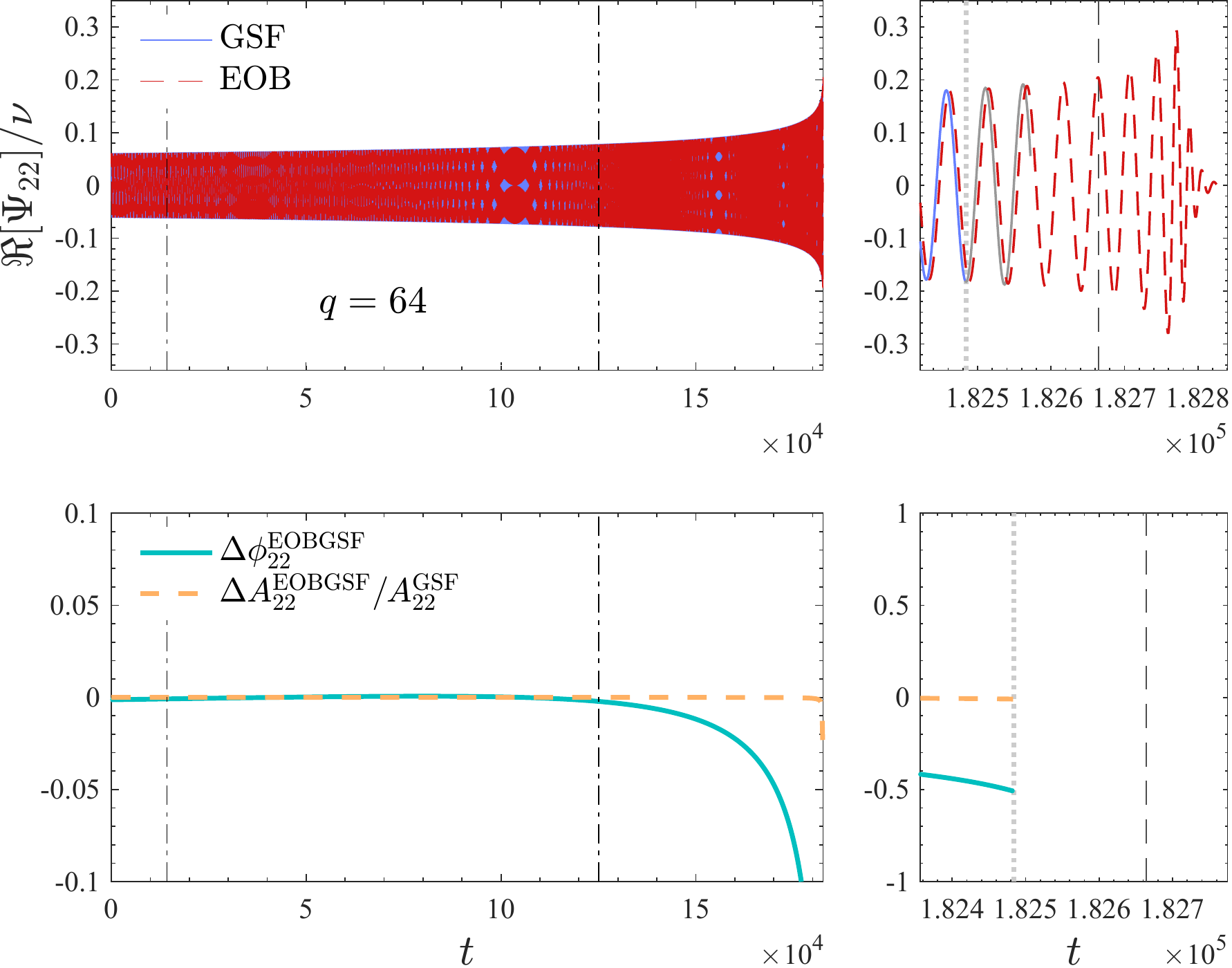} 
\hspace{8mm}
\includegraphics[width=0.43\textwidth]{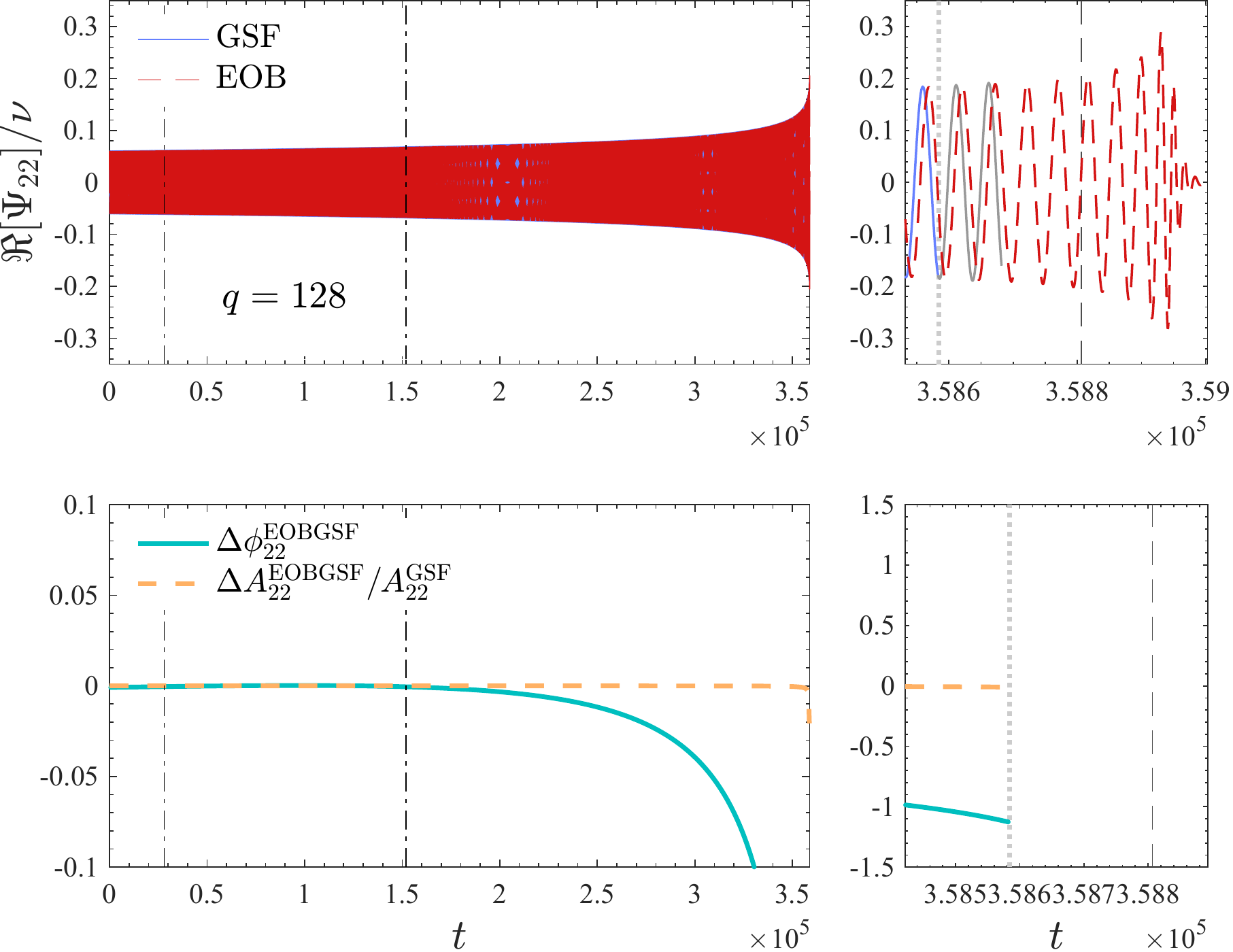} 
\caption{\label{fig:phasingsEOBGSF} EOB/GSF phasings for $q = \{15, 32, 64, 128\}$ binaries. 
The vertical dash-dotted lines in the left panels indicate the times corresponding to the $[\omega_L,\omega_R]$ 
alignment interval. In the right panels the dotted line corresponds to $\omega_{22}^{\rm GSF_{\rm break}}$, while the dashed line
indicates the adiabatic EOB LSO. The part of the GSF waveforms past the breakdown frequency is colored in light grey.}
\end{figure*}
%==============
%========================================
% Fig. 2 - all Q_omg and differences plotted together
%========================================
%\begin{figure}[t]
%\includegraphics[width=0.45\textwidth]{fig08.pdf} 
%\caption{\label{fig:all_Qomg} EOB-2GSF difference in $\Qo$. Notice that the evolution is cut at the 
%breakdown frequencies of 2GSF reported in Table~\ref{tab:Dphi}. Note that that the sign of the difference
%changes around $q=26$. For $q>26$ the GSF inspiral rate is faster than the EOB one.}
%\end{figure}
%========================================
%============
% New figure for q=15
%============
\begin{figure*}[t]
\center
\includegraphics[width=0.42\textwidth]{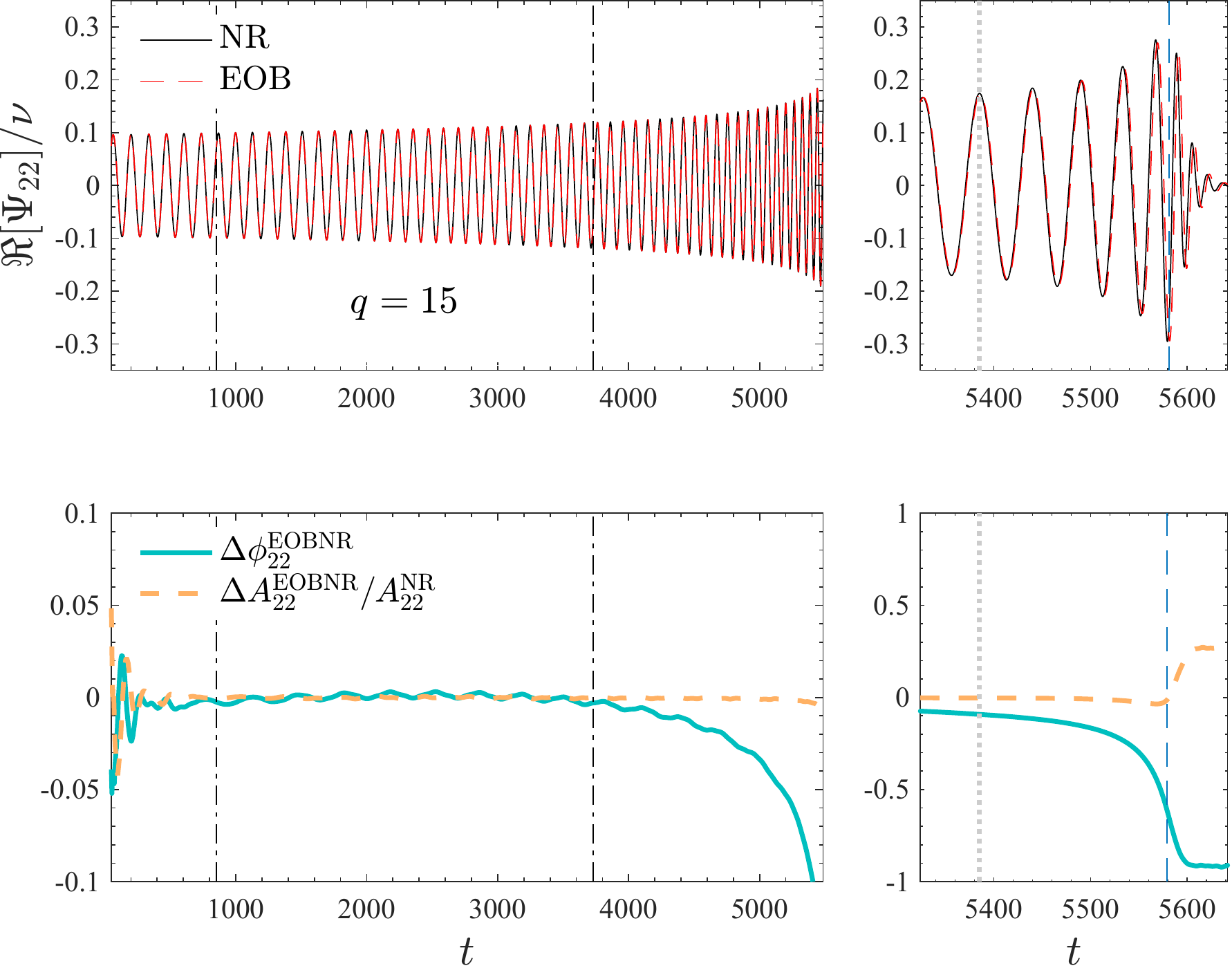} 
\qquad
\includegraphics[width=0.42\textwidth]{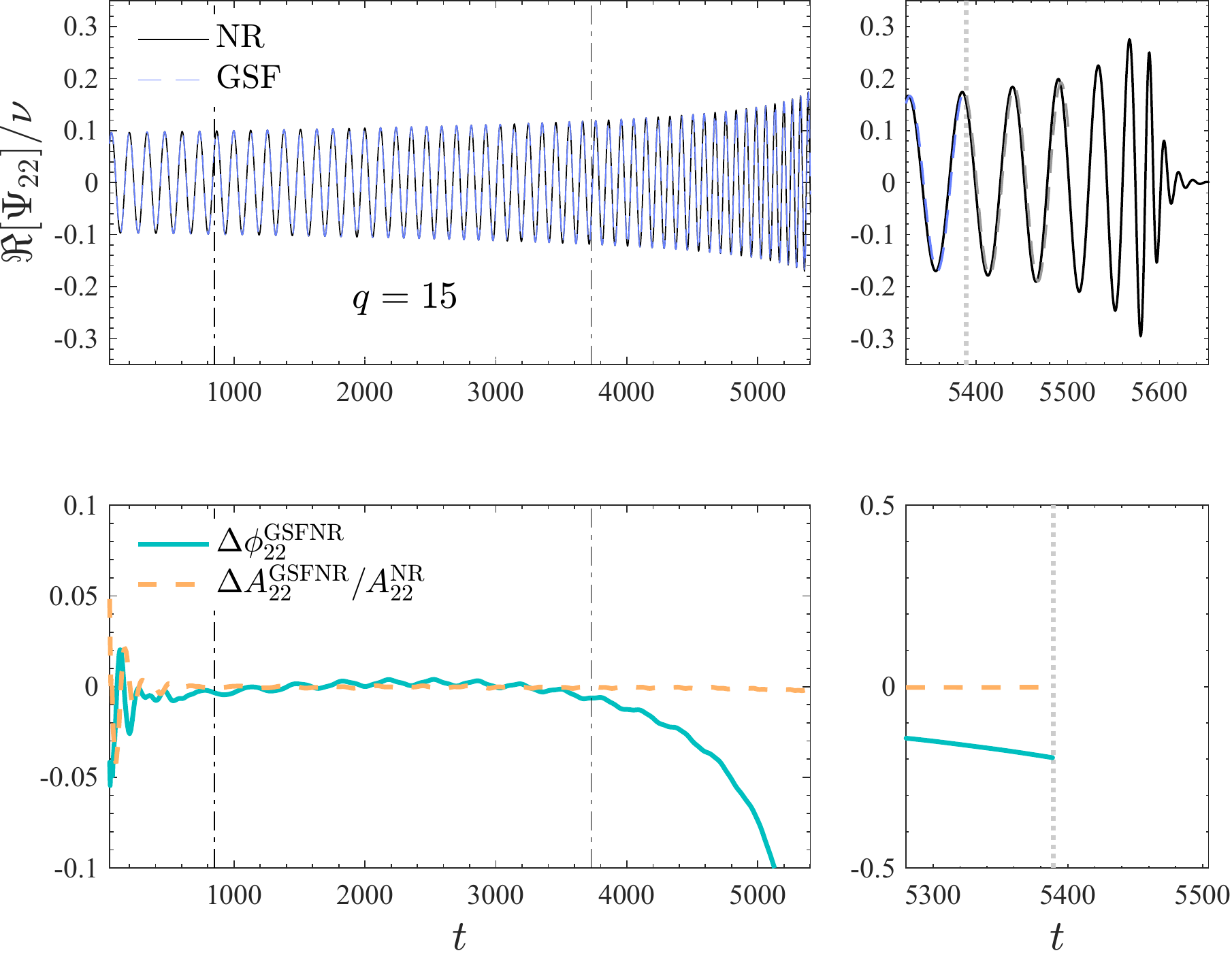} 
\caption{\label{fig:phasing_q15} 
EOB/NR and GSF/NR comparisons for $q=15$. Left: EOB/NR phasing using SXS:BBH:2247~\cite{Yoo:2022erv}.
The alignment frequency window is $[\omega_{\rm L}, \omega_{\rm R}] = [0.048, 0.063]$ (indicated by vertical dash-dotted lines), 
and the phase difference accumulated at the NR merger (dashed blue line) is $\Delta\phi^{\rm EOBNR}_{22}=-0.61$. 
The dotted vertical grey line indicates the time at which $\omega_{22}^{\rm NR} = \omega_{22}^{\rm GSF_{break}}$, 
and the EOB/NR phase difference at that point is -0.09.
Middle: GSF/NR phasing comparison using the {\it same} alignment window. One gets $\Delta\phi^{\rm GSFNR}_{22}\simeq -0.20$
at $\omega_{22}^{\rm GSF_{break}}$, and here the dotted vertical grey line indicates indeed the time at which 
$\omega_{22}^{\rm GSF} = \omega_{22}^{\rm GSF_{break}}$. Note that we show in grey the last part of the GSF waveform up to the critical frequency, 
but evaluate the phase difference at a time corresponding to the breakdown frequency (see Table~\ref{tab:Dphi}).}
%Right: EOB/GSF phasing with alignment window $[\omega_{\rm L}, \omega_{\rm R}] = [0.023, 0.028]$,
%that yields $\Delta\phi^{\rm EOBGSF}_{22}=0.75$ at $\omega_{22}^{\rm GSF_{break}}$.
%Here the dashed grey line indicates the EOB last stable orbit (LSO).}
\end{figure*}
%====
\subsection{EOB/GSF comparisons for intermediate-mass-ratio binaries, $q\geq 15$}
\label{sec:EOB-GSF-IMRI}
Let us turn now to {\it larger} mass ratios, in a regime that should be closer to the natural
domain of validity of 2GSF calculations and thus of the \PAT{} model. 
We focus here on four illustrative mass ratios, $q = (15,\, 32, \, 64, \,128)$.
These mass ratios are chosen for consistency with Ref.~\cite{Nagar:2022icd}, that presents
direct EOB/NR phasing comparisons in the IMR regime using the recent, breakthrough NR 
simulations of Refs.~\cite{Lousto:2020tnb} and~\cite{Yoo:2022erv}. We want to investigate here whether the \PAT{}
model can give us complementary information to the one obtained in~\cite{Nagar:2022icd}.
Reference~\cite{Nagar:2022icd} probed two things. On the one hand, using RIT~\cite{Lousto:2020tnb} data, 
it showed an excellent EOB/NR agreement, within the NR uncertainty, in the transition  
from late inspiral to plunge for $q=15$ and $q=32$, and similarly the consistency of late plunge 
and merger for $q=64$ and $q=128$. On the other hand, the use of a $q=15$ SXS long-inspiral
simulation~\cite{Yoo:2022erv} allowed to probe the \TEOBResumS{} waveform through full inspiral up to
merger, getting a $\simeq -0.6$~rad dephasing at merger (see Fig.~15 and~16 of~\cite{Nagar:2022icd}).
In addition, Ref.~\cite{Nagar:2022icd} also pointed out that the EOB/NR phasing disagreement
at merger can be reduced by $50\%$ by only incorporating an additional 6PN test-particle
correction in the $\ell=m=2$ multipole of the radiation reaction.
%Unfortunately, the inspiral portion of the NR waveforms used in Ref.~\cite{Nagar:2022icd} is not sufficiently 
%accurate or sufficiently long to use as a reliable benchmark. 1PAT1 waveforms thus serve as a test of the EOB inspiral waveform; 
%any deviations between the two waveforms can be benchmarked against the 1PAT1 error estimates in Secs.~\ref{sec:1PA accuracy}--\ref{sec:1PAT1 errors}. 
Thus, \TEOBResumS{} provides then a baseline test of the 1PAT1 waveforms, given that it is inherently 
accurate in the early inspiral (automatically recovering a high-order PN expansion there) and is NR-tested  
for $q=15$ and $q=32$.  A priori, since \TEOBResumS{} naturally incorporates a certain amount of test-particle information, 
we expect that the differences between 2GSF and EOB waveforms will reduce as $q$ is increased, until $q$ is 
sufficiently large that high-order-in-$\nu$ information becomes insignificant while small errors in low-order-in-$\nu$ 
terms become significant; for $q$ beyond that point, we  expect the dephasing between EOB and 2GSF waveforms 
to increase due to any failure of the current version of \TEOBResumS{} to precisely capture 0PA 
and 1PA effects (e.g., for the lack of the test-particle $\ell=m=2$ term pointed out above).
Our comparisons will bear out these expectations, consistently with the analysis of~\cite{Nagar:2022icd}.

We consider waveforms that start at rather low frequency and have many cycles. As discussed above, this will lead to 
larger cumulative errors in the GSF waveforms (as compared to the frequency interval used in our comparisons 
for comparable masses). But it provides the most stringent tests of our waveform models, and the errors in the GSF 
model can in any case be expected to decrease with increasing $q$.
Therefore, when aligning EOB to GSF, $\omega_L$ is always chosen very low. Then, $\omega_R$ (corresponding to 
the second vertical line in Fig.~\ref{fig:phasingsEOBGSF}) is increased progressively until the phase difference 
remains substantially flat on the scale of the plot. The so-obtained alignment intervals for each mass ratio are
displayed in Table~\ref{tab:Dphi}.
Figure~\ref{fig:phasingsEOBGSF} illustrates the high EOB/GSF consistency during the full inspiral, with phase differences accumulated at the time 
corresponding to $\omg_b$ of $(+0.38, -0.13,-0.52,-1.13)$~rad for $q=(15,32,64,128)$, respectively. 
These numbers are substantially confirmed by the $\Qo$ analysis, as shown in Table~\ref{tab:Dphi}.
In addition to the absolute magnitude of the phase differences reported in Table~\ref{tab:Dphi}, there is important
information in their {\it sign}: $\Delta\phi_{22}^{\rm EOBGSF}$ (computed either way) at $\omg_b$ 
is positive up to $q=15$, but it becomes negative for all other values of $q$. By simply inspecting
the values of $\Delta\phi_{22}^{\rm EOBGSF}$ at $\omg_b$ one deduces that 
$\Delta\phi_{22}^{\rm EOBGSF}\sim 0$ should occur at $q\simeq 26$.
Physically this means that up to $q\sim 26$ the gravitational interaction encoded within the EOB model is,
loosely speaking, {\it more attractive} (the phase acceleration is larger) than the one predicted by the GSF 
model. For $q>26$ it is the opposite. 
%This information is synthetically displayed in Fig.~\ref{fig:all_Qomg},
%that shows the differences between the EOB and GSF $\Qo$ extending from low frequencies to $\omg_b$.

The dephasings in Table~\ref{tab:Dphi} can be compared against the internal error estimates in the 1PAT1 model. 
If we assume that for $q\lesssim10$ a 1PA model's error is dominated by 2PA contributions, $\propto \nu$ [cf. Eq.~\eqref{phi expansion}], 
then we can estimate the error at larger $q$ as $\delta\phi^{\rm GSF}_{22}\approx \frac{\nu}{\nu_{q_0}}\delta\phi^{\rm GSF}_{22,q_0}$, 
where $q>q_0$. Using $q_0=7$ and $\delta\phi^{\rm GSF}_{22,q_0}=\Delta\phi_{22,q_0}^{\rm EOBGSF}$ 
(since the error in EOB is very small at this mass ratio), we obtain the 
error estimates $\delta\phi^{\rm GSF}_{22}=(0.68, 0.34, 0.17, 0.089)$~rad for $q=(15,32,64,128)$; 
using $q_0=10$, we obtain the broadly compatible estimates $\delta\phi^{\rm GSF}_{22}=(0.53, 0.26, 0.14, 0.069)$~rad. 
Crucially, these estimates assume the first-law binding energy is the correct one to use in the 1PA energy balance law. 
They also assume the same frequency interval is used for all mass ratios, while our  dephasing measurements 
$\Delta\phi_{22}^{\rm EOBGSF}$ use different breakdown frequencies. But we can show, using the near-LSO approximations 
from Sec.~\ref{sec:1PA accuracy}, that the accumulated error between two breakdown frequencies 
[approximately $ 2\int^{\omega^{\rm break}/2}_{\omega^{\rm break}_{q_0}/2}\phi'_2 d\hat\Omega$, from 
Eq.~\eqref{phi expansion}] is several orders of magnitude smaller than our estimated total cumulative 
error $\delta\phi^{\rm GSF}_{22}$. Based on our estimates of $\delta\phi^{\rm GSF}_{22}$,
 we can therefore say that the EOB-GSF dephasing $\Delta\phi_{22}^{\rm EOBGSF}$ may 
 be smaller than 1PAT1's error for $q=15$ and $q=32$, but  $\Delta\phi_{22}^{\rm EOBGSF}$ is 
 substantially larger than $\Delta\phi^{\rm GSF}_{22}$ for $q=64$ and $q=128$. This, combined 
 with our observations above, suggests that the turnover where 1PAT1 becomes more accurate 
 than \TEOBResumS{} likely lies somewhere in the range $26\lesssim q<64$.

We can glean more information by comparison with error-controlled NR simulations at mass ratios where they are available.
This can be done for $q=15$ and, to a certain extent, for $q=32$, building upon the results of Ref.~\cite{Nagar:2022icd}.
For $q=15$, Fig.~\ref{fig:phasing_q15} compares  the SXS waveform to the \TEOBResumS{} and \PAT{} ones,
using the same alignment window of Fig.~15 of Ref.~\cite{Nagar:2022icd}.  We find $\Delta\phi_{22}^{\rm EOBNR}\approx -0.09$~rad and $\Delta\phi_{22}^{\rm GSFNR}\approx -0.20$~rad 
at $\omg_b$ (dotted line in the right panels of the figure), that approximately occurs 2.5 orbits before merger. 
From this, we conclude that the \PAT{} model is a less faithful representation of the phasing up to $\omg_b$ 
than \TEOBResumS{}, in line with our expectation above. The error in both models is small, but we note that this dephasing is on a narrower frequency interval than the interval of our error estimate $\delta\phi^{\rm GSF}_{22}\approx 0.53-0.68$ obtained above.
This is analogous to the $q=(7,10)$ cases discussed above, where one has to be careful to compare dephasings over a consistent (frequency or time) interval.

The situation looks different for the $q=32$ case. Here, $\Delta\phi_{22}^{\rm EOBGSF}$ is globally smaller,
reaching only $\sim -0.13$~rad 5GW cycles ($\sim 2.5$ orbits) before merger. This value is consistent with our 
estimated error in \PAT{} as well as with the $\Delta\phi^{\rm EOBNR}_{22}$ phase difference for 
$q=32$ shown in Fig.~5 of Ref.~\cite{Nagar:2022icd}. 
We can therefore say that NR, GSF, and EOB are all consistent with one another at this mass ratio. 
Moreover, it appears that at this mass ratio \TEOBResumS{} {\it correctly bridges the gap} between the two 
very different approaches to the solution of Einstein's equations: GSF and NR. The 1PAT1 model can provide 
in principle very accurate inspirals (modulo the uncertainty in the binding energy), but only for sufficiently large mass ratios.
On the contrary, the RIT $q=32$ NR simulation of~\cite{Lousto:2020tnb}, compared with \TEOBResumS{} 
in Ref.~\cite{Nagar:2022icd}, delivers a robust and accurate description of the transition to merger and ringdown, 
but currently suffers from a rather large phase uncertainty ($\sim 0.6$~rad in total) during the whole simulated
inspiral of $\sim 12$~orbits. \TEOBResumS{} matches both of these models within their internal error estimates 
in their respective domains of validity, as well as providing the only complete inspiral-merger-ringdown model of the three at this mass ratio. 

Although this mutual consistency of the three approaches around $q\approx 30$ is reassuring, 
a more precise assessment of the numerical errors is needed. This is probably only possible with higher-accuracy, 
longer NR waveforms, as mentioned in Ref.~\cite{Nagar:2022icd}.
At present, for $q=64$ and $q=128$, we can say that $\Delta\phi_{22}^{\rm EOBGSF}$ is larger than 
$\delta\phi_{22}^{\rm GSF}$ and comparable to $\Delta\phi_{22}^{\rm EOBNR}$. For $q=64$, $\Delta\phi_{22}^{\rm EOBGSF}$ 
is loosely consistent with $\Delta\phi_{22}^{\rm EOBNR}$ reported in the bottom-left panel of Fig.~5 of Ref.~\cite{Nagar:2022icd}, 
although in this case (and in the $q=128$ case as well) it was not possible to deliver a robust estimate of the 
NR phase uncertainty because of the lack of a complete convergent series. 
Moreover, for $q=128$ the EOB/NR phase difference at  a point corresponding to $\omg_b$ ($\sim 8$ cycles before 
merger) is already too large (see again Fig.~5 of~\cite{Nagar:2022icd}) to allow us any additional 
quantitative assessment. Therefore, though we can estimate that \PAT{} is more accurate than \TEOBResumS{}  
for these mass ratios, and increasingly so for higher $q$, we cannot precisely quantify the accuracy of 
the \PAT{} waveforms beyond our rough internal error estimates. This is further complicated by the 
uncertainty in the \PAT{} model arising from the choice of binding energy. 

As a prelude to the next section, and building upon the finding of Sec.~VA of Ref.~\cite{Nagar:2022icd}, it is interesting 
to investigate how the EOB/GSF results above change when the $\ell=m=2$ 6PN test-mass coefficient is included in 
the radiation reaction. The corresponding EOB/GSF dephasings are listed in the last column of Table~\ref{tab:Dphi}.
The interesting finding is that the EOB/GSF dephasing {\it increases} for $q=15$, while it {\it decreases} for the other
mass ratios. This is thus a further indication of the correctness of our reasoning up to now, supporting the idea that
the EOB/GSF discrepancy for large mass ratios (say $\gtrsim 32$) is due to the analytical incompleteness 
of \TEOBResumS{}, while for smaller mass ratios (e.g., $q=15$) the EOB/GSF difference is due to errors in the 1PAT1 model.

Besides the sensitivity to the correction to the radiation reaction, 
\TEOBResumS{} incorporates only part of the known linear-in-$\nu$ analytical contributions and was designed 
primarily for comparable-mass binaries. This difference essentially lies in the EOB potentials, $(A,D,Q)$. 
The $A$ function includes analytical information {\it only} up to 4PN, while both $D$ and $Q$ contain information 
only up to 3PN. These functions are thus {\it different} from the {\it exact} GSF ones that incorporate the 
complete linear-in-$\nu$ information, and that were calculated in Ref.~\cite{Akcay:2015pjz}. The analysis 
of the next section will find evidence that this is likely among the causes of the EOB/GSF differences for large 
values of $q$, together with a needed upgrade of the dissipative sector of the model, as the last column of 
Table~\ref{tab:Dphi} already indicates.

%==============
% Q_omg fit
%==============
\section{On the origin of the GSF/EOB differences}
\label{sec:nu_dependence}
We have assessed, using two different methods, the existence of a nonnegligible phase 
difference between GSF and \TEOBResumS{} waveforms up to the GSF breakdown frequency.
Thanks to several EOB/NR/GSF comparisons, we can safely state that the \PAT{} description
of the inspiral is a less accurate representation of the true waveform than \TEOBResumS{} for $q\leq 15$.
By contrast, there seems to exist a region of mutual EOB/NR/GSF consistency in the range $25\lesssim q \lesssim 32$.
For larger values of $q$, the GSF model becomes increasingly more accurate than the EOB model.

Let us now attempt to investigate the origin of these differences by analyzing the structure of $\Qo$
as a function of $\nu$. We return to the asymptotic expansion~\eqref{Q expansion}, which we restate here for convenience:
\be
\label{eq:Qo_exp}
\Qo(\omega;\nu) = \frac{\Qo^{0}(\omega)}{\nu} + \Qo^1(\omega) + \Qo^2(\omega) \nu + O(\nu^2) .
\ee
Here the 0PA term, $\Qo^{0}$, is identical to $\Qo$ for a test-mass on a Schwarzschild background subject to 
leading-order dissipation (i.e., the order-$\nu$ dissipative self-acceleration or order-$\nu^2$ energy flux). 
The 1PA term, $\Qo^{1}$, incorporates the conservative contributions of the first-order self-acceleration as well 
as the first subleading dissipative contribution (i.e., the order-$\nu^2$ dissipative self-acceleration or order-$\nu^3$ 
energy flux, both of which are themselves affected by the full order-$\nu$ self-acceleration). Finally, the 2PA term, 
$\Qo^2$, contains the conservative contribution of the order-$\nu^2$ self-acceleration and third-order dissipative 
information (i.e., the order-$\nu^3$ dissipative self-acceleration or order-$\nu^4$ energy flux).

Given the resummed structure  of the EOB Hamiltonian, the actual $\Qo^{\rm EOB}$ has in fact 
an {\it infinite} number of $\nu$-dependent terms and Eq.~\eqref{eq:Qo_exp} is formally obtained by expanding in $\nu$.
As discussed previously, $\Qo^{\rm GSF}$ also has 
non-zero contributions from all higher-order $\Qo^n$ when expanded in powers of $\nu$, but it only exactly captures $\Qo^0$ and $\Qo^1$; this is straightforwardly seen from the expansion in Appendix~\ref{sec:exactQ0andQ1}.
Our aim here is to extract the three functions $\Qo^0$, $\Qo^1$ and $\Qo^2$ from \PAT{}
and \TEOBResumS{} and compare them. This will give us a more precise quantitative understanding 
of the differences between the two models.
To do so, we proceed as follows. We consider mass ratios\footnote{The $q = (26, 36)$ datasets have been exploited to have a more robust estimate of the fit coefficients. The $q = 36$ dataset is not considered 
elsewhere in this work since it does not yield additional significant information to the other comparisons.} $q=(7,10,15,26,32,36,64,128)$ and 
a range $[\omega_{\rm min},\omega_{\rm max}] = [0.023, 0.09]$ with spacing $\Delta \omega = 0.001$. 
Here, the maximum value is chosen so as to be sufficiently far from the 
possible breakdown of the underlying approximations in the \PAT{} model.
%=================================
% Fit at an illustrative frequency omg = 0.055
%=================================
\begin{figure}[t]
\center
\includegraphics[width=0.45\textwidth]{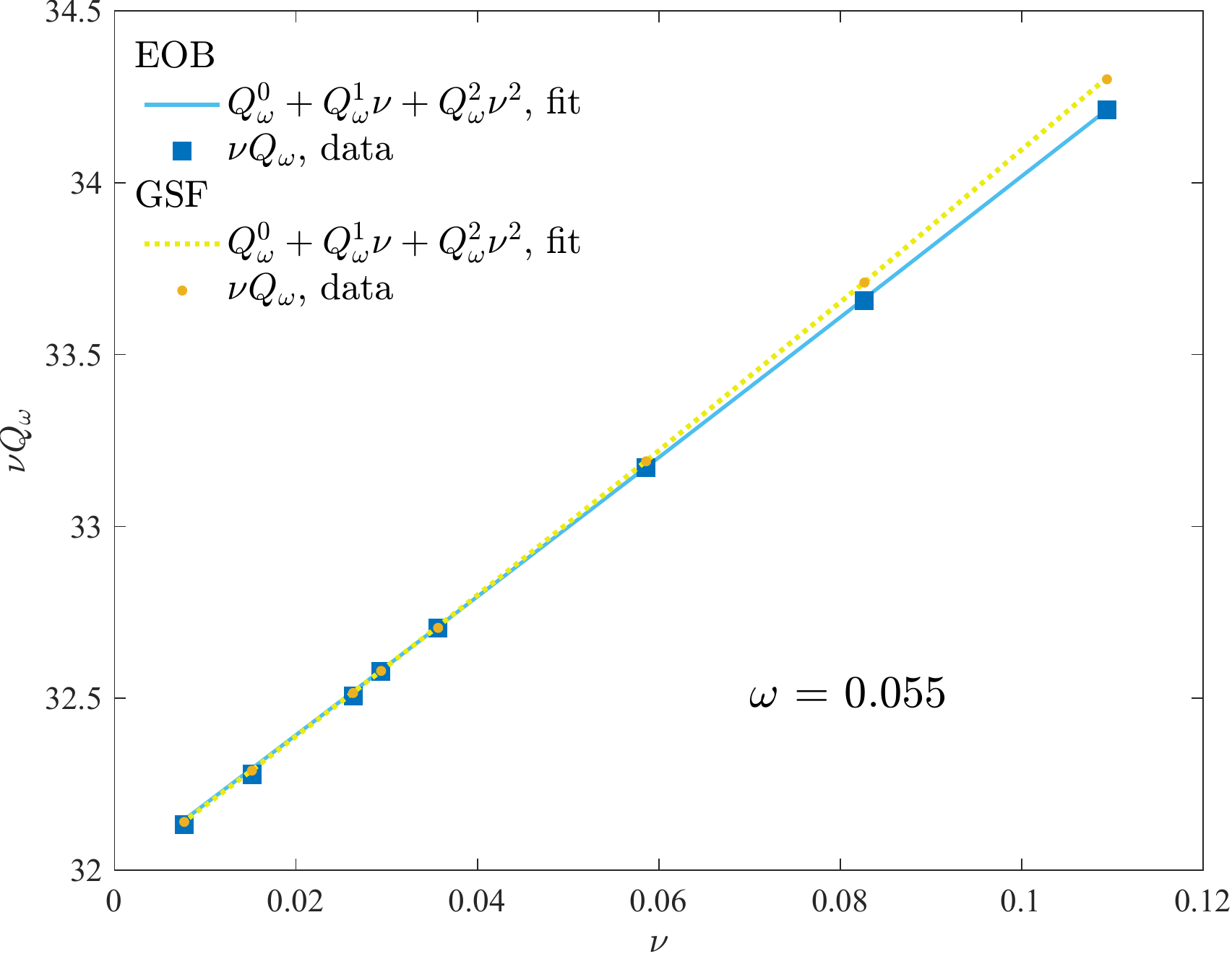} 
\caption{\label{fig:nuQomg} Numerical data and fits of $\nu \Qo$ for a set of mass ratios $q = (7,10,15,26,32,36,64, 128)$ at 
a fixed frequency, $\omega = 0.055$. One sees here how EOB and GSF (and the corresponding fits) are in agreement 
for $ q \gtrsim 26$ ($\nu \lesssim 0.036$).}
\end{figure}
%================================
For each value of $\omega$ we fit $Q_\omega(\omega;\nu)$ using Eq.~\eqref{eq:Qo_exp}. 
Figure~\ref{fig:nuQomg} shows the outcome of the fit versus $\nu$ for $\omega=0.055$. 
The same procedure is repeated for each value of $\omega$ within $[0.023, 0.09]$. 
This eventually gives the functions $\{\Qo^0(\omega),\Qo^1(\omega),\Qo^2(\omega)\}$, that are shown in Fig.~\ref{fig:Q0andQ1}. 

We also show in the same figure the ``exact'' $\Qo^0(\omega)$ and $\Qo^1(\omega)$, 
computed from 1GSF and 2GSF quantities using the formulas derived in Appendix~\ref{sec:exactQ0andQ1}. 
The fitted values of 1PAT1's $\Qo^0(\omega)$ and $\Qo^1(\omega)$ lie close to the exact values, 
broadly validating the fitting procedure, but they do begin to noticeably deviate at high frequencies. 
This might suggest that the fits are contaminated by the more complicated $\nu$ dependence of the transition to plunge, 
even significantly below $\omega^{\rm break}$. This is further testified by Fig.~\ref{fig:diffQ1GSF}, 
which shows how excluding mass ratios $q = \{7, 10\}$ from the fit, the result is closer to the exact one. 
However, the deviation is sufficiently small that it cannot alter our conclusions.

The left panel of Fig.~\ref{fig:Q0andQ1} indicates that there is very good EOB/GSF agreement in the $\Qo^0$ part.
This is not surprising given the highly accurate energy flux incorporated within \TEOBResumS{}, 
that builds upon~\cite{Damour:2008gu,Damour:2009kr}. The EOB flux includes all multipoles up to $\ell=8$. 
Each multipole is factorized and resummed following Ref.~\cite{Damour:2008gu} and currently includes up to (relative) 
test-mass 6PN information~\cite{Messina:2018ghh,Nagar:2019wds,Nagar:2020pcj}. The GSF $\Qo^0$ is fully 
determined by the first-order GSF flux through the horizon and infinity. In the \PAT{} model this was computed to 
machine precision by summing the fully relativistic modes up to $\ell=30$. Since the GSF calculation is 
effectively exact,\footnote{In practice, the GSF flux is only evaluated to a given number of digits. In this case 
we evaluated it to machine precision, but this can be pushed further by increasing the numerical accuracy
 to which the 1SF fluxes are computed.} we can be confident that the residual EOB/GSF difference is associated 
 with the fact that \TEOBResumS{} is {\it not} analytically complete, as already pointed out above; 
 as explained in Sec.~\ref{sec:EOB-GSF-IMRI} it is missing higher-order PN information and 
 higher-$\ell$ contributions. The smallness of the difference means it will only become significant 
 when it is comparable to $\Qo^1$ in absolute terms. Given that $\Delta \Qo^0 \sim 10^{-2}$ 
 and $\Qo^1 \sim 20$ this will only happen for large mass ratios $q \gtrsim 10^3$.

By contrast, the $\Qo^1$ and $\Qo^2$ terms point to more significant differences between \TEOBResumS{} and \PAT{}.
This is not unexpected given that there are approximations present in $\Qo^1$ in both models, and that \PAT{} is not 
directly controlling the error in $\Qo^2$ since it is neglecting potentially important 2GSF conservative and 3GSF 
dissipative contributions. We note that over much of the frequency range considered, the difference $\Delta Q^1_\omega$ 
is comparable to (or smaller than) the uncertainty in $Q^1_\omega$ stemming from the choice of binding energy, 
as shown in Fig.~\ref{fig:Q1-EFL-ESF}. It is therefore impossible to conclude which result lies closer to the 
true $Q^1_\omega$ for $\omega\lesssim 0.07$. For $\omega\gtrsim 0.07$, the picture is clearer, as the 
difference $\Delta Q^1_\omega$ becomes significantly larger than the uncertainty in the \PAT{} result. 
For $Q^2_\omega$, it is again not entirely clear which of the two models is more accurate, but in this case 
no credence should be given to the \PAT{} result: since all 2PA terms in $\dot\Omega$ are missing, \PAT{}'s $Q^2_\omega$ 
could be entirely incorrect. Similarly, since \TEOBResumS{} has been optimized for comparable-mass binaries, 
it may be possible that  \TEOBResumS{} contains significant errors in both $\Qo^1$ and $\Qo^2$ that effectively 
cancel one another for $q\lesssim 10$. On the other hand, \TEOBResumS{} should at least represent the 
correct $Q^2_\omega$ in the small-frequency limit, where it reduces to the PN value, leading us to 
infer that \TEOBResumS{}'s $Q^2_\omega$ is probably more reliable than \PAT{}'s.

Interestingly, Fig.~\ref{fig:diffQomg_GSF} shows that both models have only small contributions 
from $Q_\omega^n$ beyond $n=2$. In other words, in the range of masses and frequencies considered in this 
section, $Q_\omega^{\rm EOB}$ and $Q_\omega^{\rm GSF}$ are well represented by only 
the first three terms in the expansion~\eqref{eq:Qo_exp}. Therefore our analysis of those three 
terms should provide a fairly complete picture of the two models.

%======================
% Plotting the Qomg functions
%======================
\begin{figure*}[t]
\center
\includegraphics[width=0.32\textwidth]{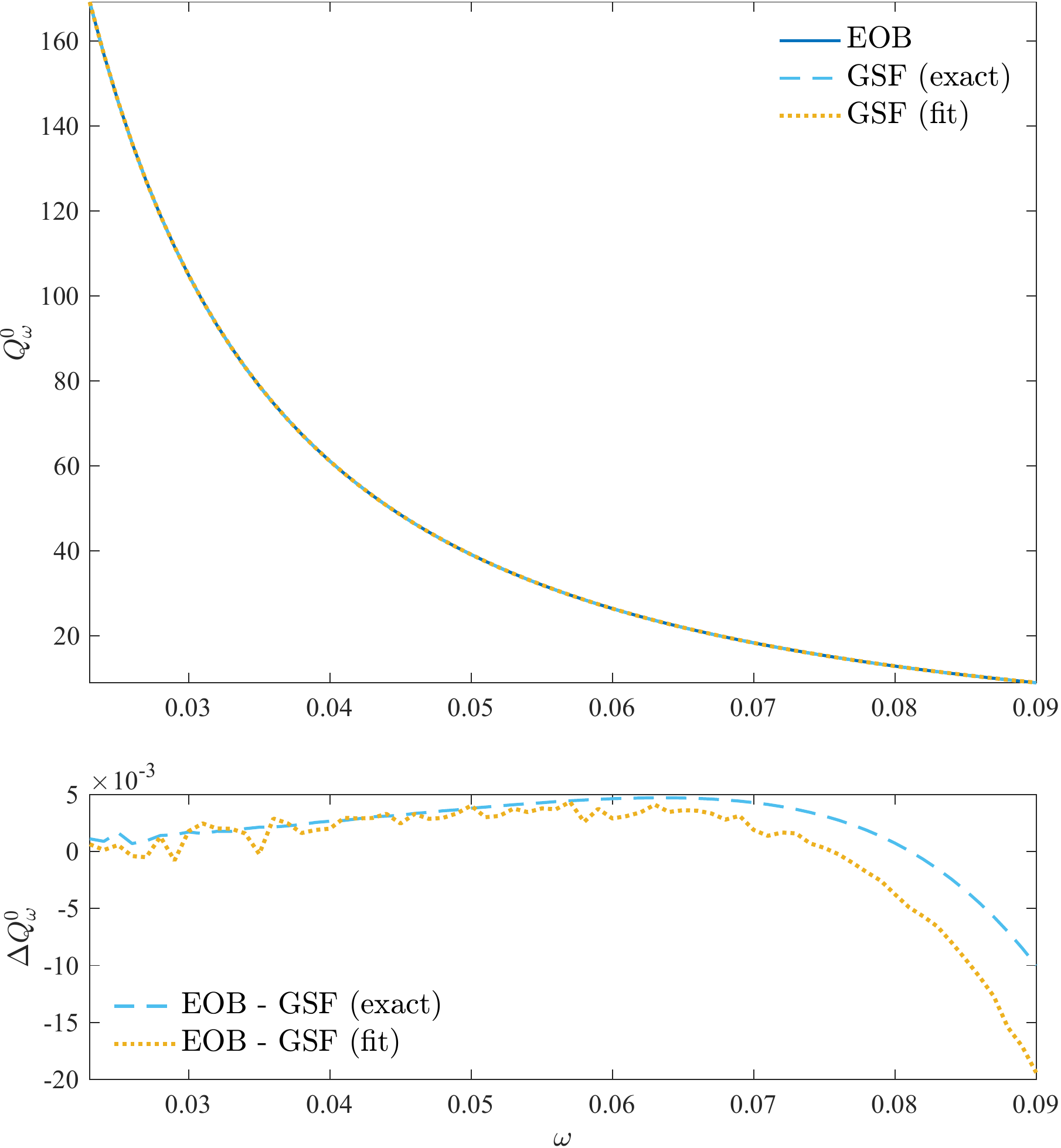} 
\includegraphics[width=0.32\textwidth]{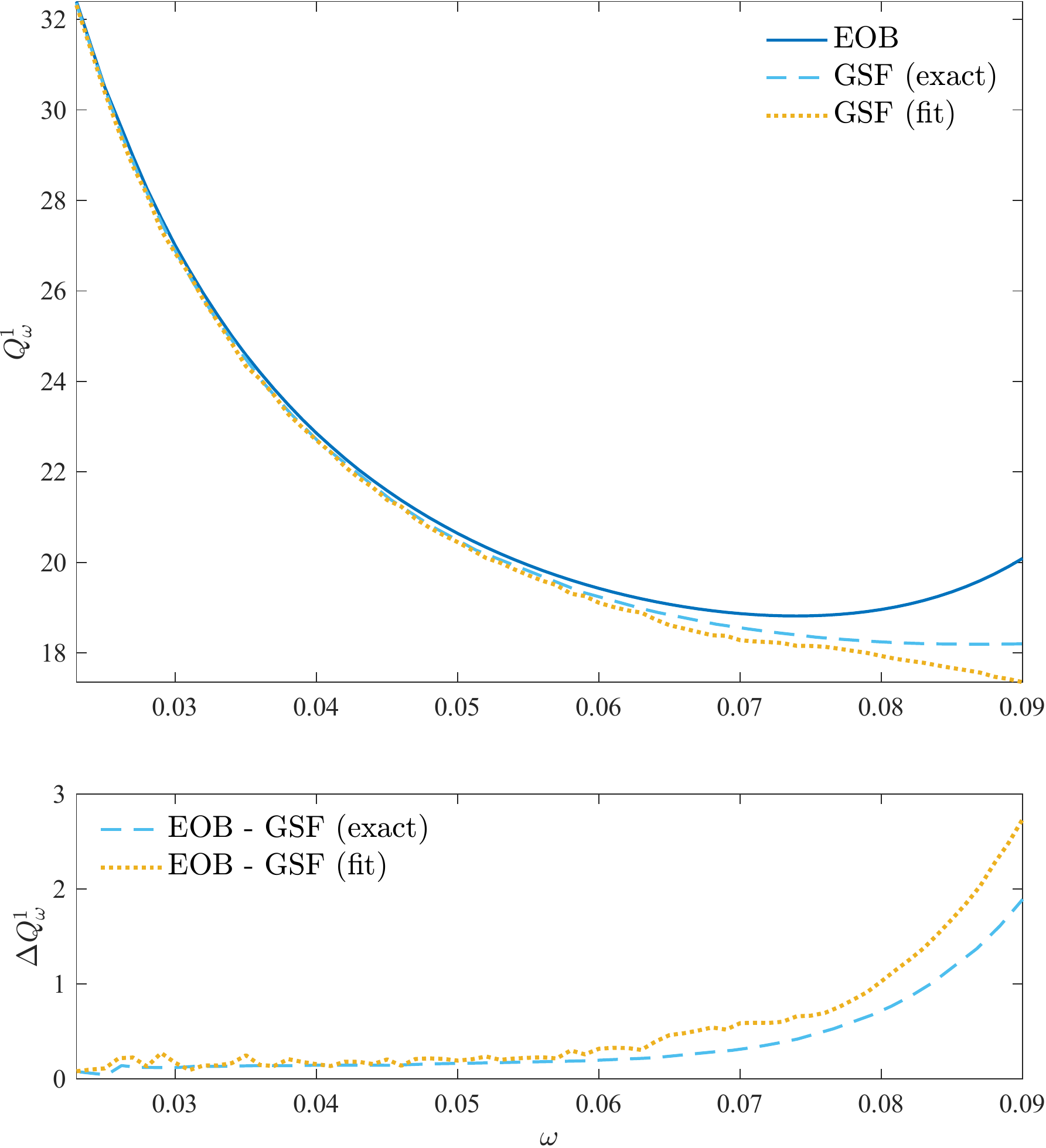} 
\includegraphics[width=0.32\textwidth]{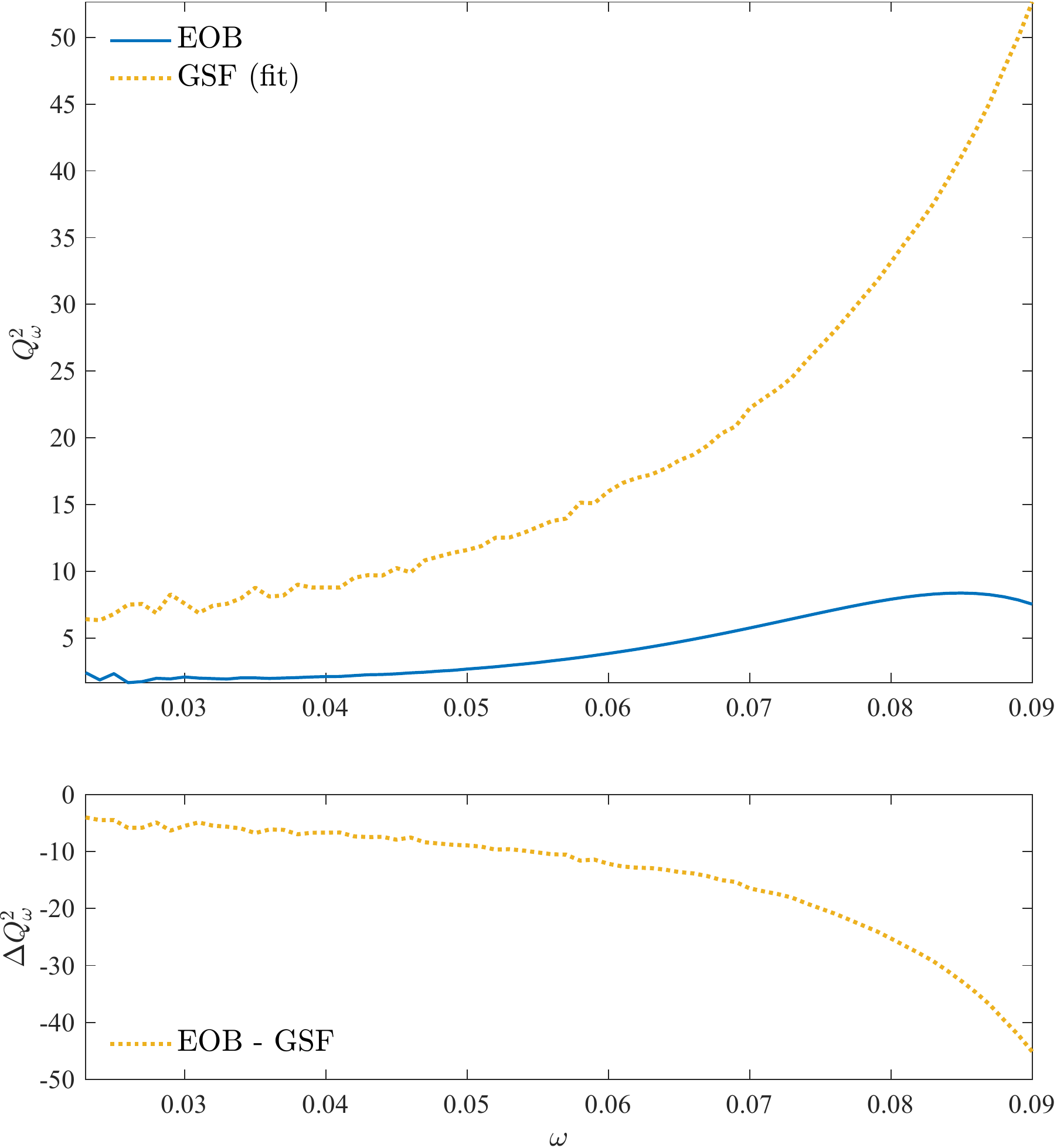} 
\caption{\label{fig:Q0andQ1} The coefficients $\Qo^0$ (left), $\Qo^1$ (center) and $\Qo^2$ (right) from the expansion 
\eqref{eq:Qo_exp}, fitted for a set of values of $\omega$ and $\nu$. 
The lower panels display the EOB-GSF difference. The two models show good agreement for $Q_0$,
but differ in $\Qo^1$ and $\Qo^2$. This slight disagreement in $\Qo^0$ is related to the fact that the EOB potentials implemented in \TEOBResumS{} do not incorporate the full linear-in-$\nu$ knowledge.}
\end{figure*}
%

%==============================================
% Difference between various Q1GSF (changing mass ratios)
%==============================================
\begin{figure}[t]
\center
\includegraphics[width=0.45\textwidth]{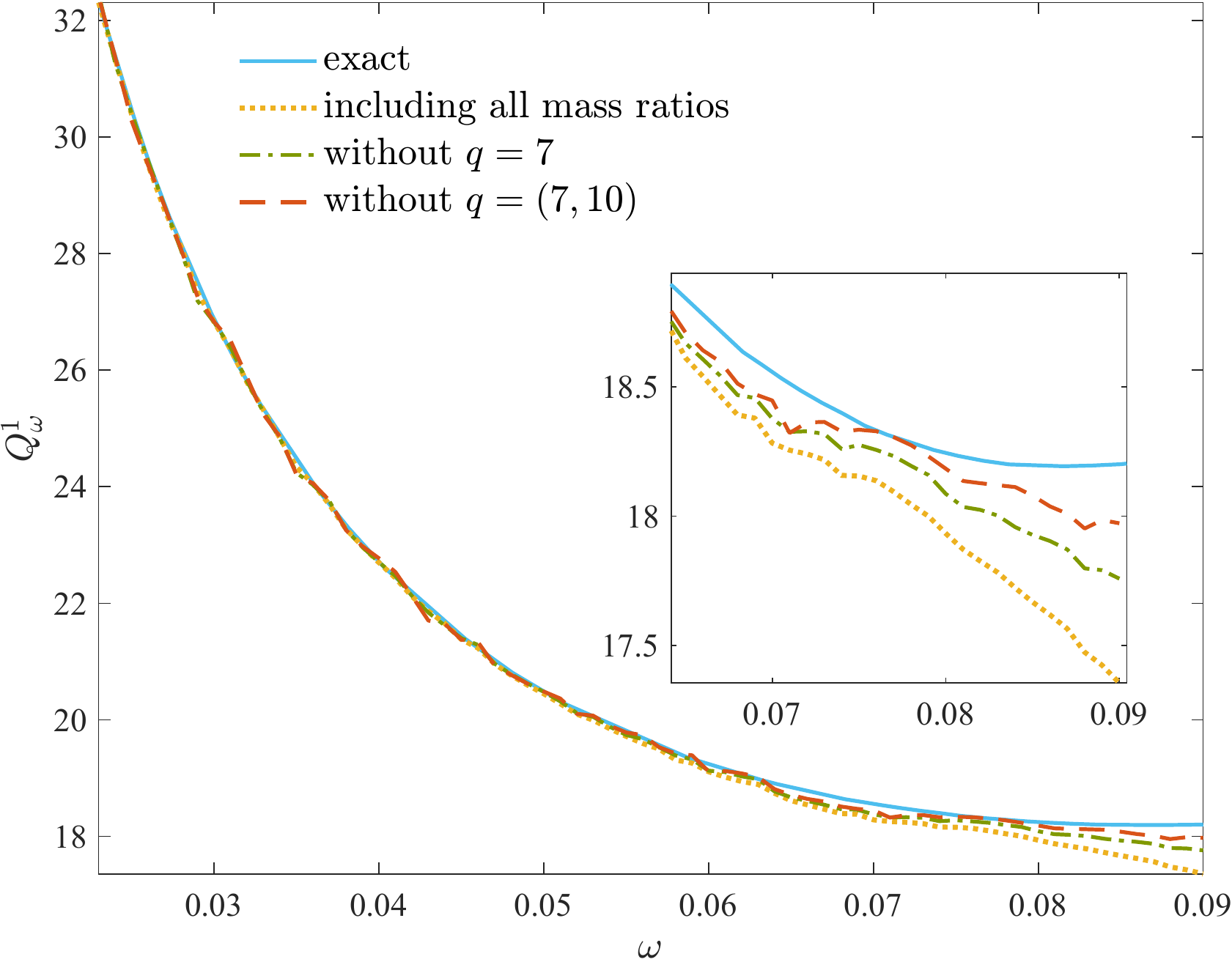}
\caption{\label{fig:diffQ1GSF} Changing the mass ratios included in the GSF fit: the inclusion of $q=(7,10)$ makes the fit less reliable
and drives it away from the exact result at higher frequencies.}
\end{figure}

%=========================================
% Difference between the GSF fit & the complete Qomg
%=========================================
\begin{figure}[t]
\center
\includegraphics[width=0.45\textwidth]{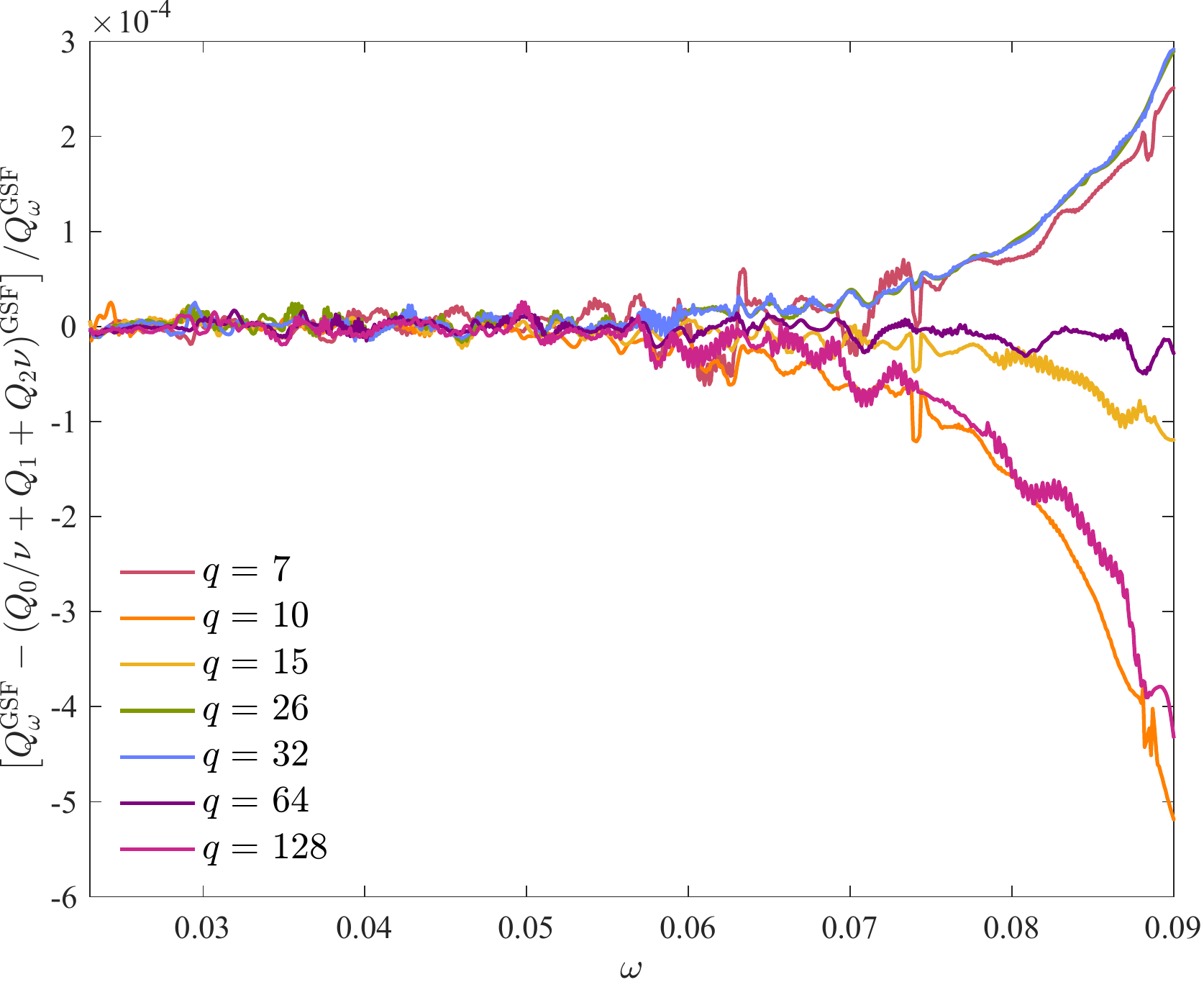} \\
\vspace{0.5mm}
\includegraphics[width=0.45\textwidth]{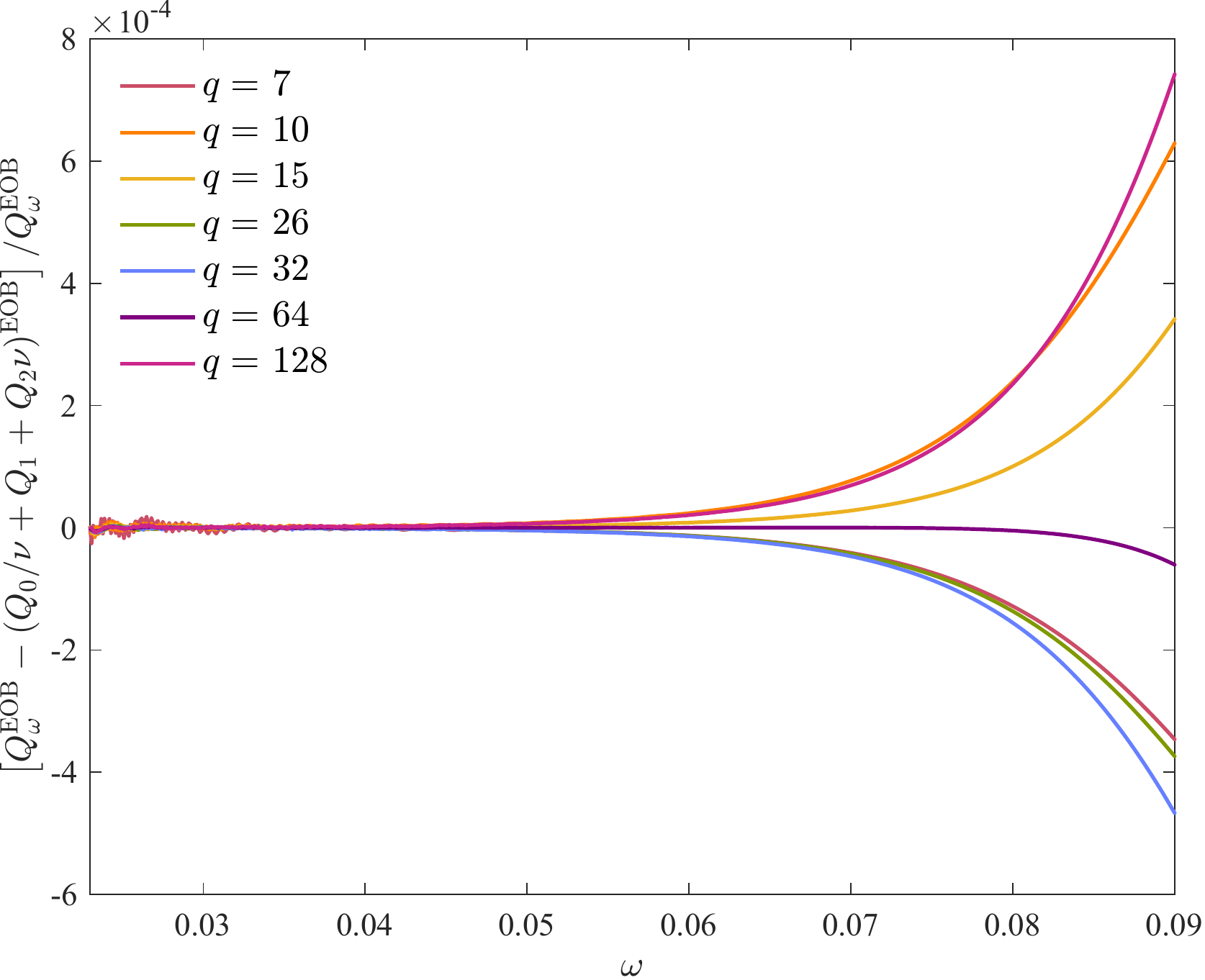} 
\caption{\label{fig:diffQomg_GSF}  The relative difference between the $\Qo$ function and the corresponding fits
evaluated from the \PAT{} waveforms (top) and the \TEOBResumS{} ones (bottom). 
Interestingly, the magnitude of the difference is small, suggesting that both models
have a small contribution to $\Qo$ beyond $\Qo^2$. }
\end{figure}

To assess how much each of these three terms in the expansion of $\Qo$ impact on the phasing, 
we can estimate three contributions  to the phase difference on the frequency interval $(\omega_1,\omega_2)$:
\begin{align}
\label{eq:dphis}
\Delta \phi_0 &\equiv \frac{1}{\nu}  \int_{\omega_1}^{\omega_2}  \left( \Qo^{0, \rm EOB} -  \Qo^{0, \rm GSF} \right) d\log\omega,\\
\Delta \phi_1 &\equiv \int_{\omega_1}^{\omega_2}  \left( \Qo^{1, \rm EOB} -  \Qo^{1, \rm GSF}\right) d\log\omega, \\
\Delta \phi_2 &\equiv \nu \int_{\omega_1}^{\omega_2}  \left( \Qo^{2, \rm EOB} -  \Qo^{2, \rm GSF}\right) d\log\omega,
\end{align}
so that the total phase difference between $(\omega_1,\omega_2)$ is
\be
\Delta\phi^{\rm EOBGSF}_{(\omega_1,\omega_2)}=\Delta\phi_0 + \Delta\phi_1 + \Delta\phi_2.
\ee
The result of this calculation over the frequency interval $(\omega_1,\omega_2)=(0.023,0.09)$ is displayed 
in Table~\ref{tab:dphi_Qis}. For comparison, we note that the uncertainty in the choice of binding energy contributes an uncertainty $\Delta\phi_1^{\rm GSF}\approx 0.45$~rad in 1PAT1's 1PA phase $\phi_1$ on this frequency interval, which is not dramatically smaller than the EOB-GSF difference $\Delta \phi_1$. We also stress again that these phase differences \textit{cannot} be compared to those
obtained integrating $\Qo$ on a fixed time interval, namely they should not be contrasted to the ones in Table~\ref{tab:Dphi}.
They can instead be compared to the ones in Table~\ref{tab:Dphi_Q} for $q = (7,10)$, with which they are consistent, given the larger frequency interval used here. 

\begin{figure}[t]
\center
\includegraphics[width=0.45\textwidth]{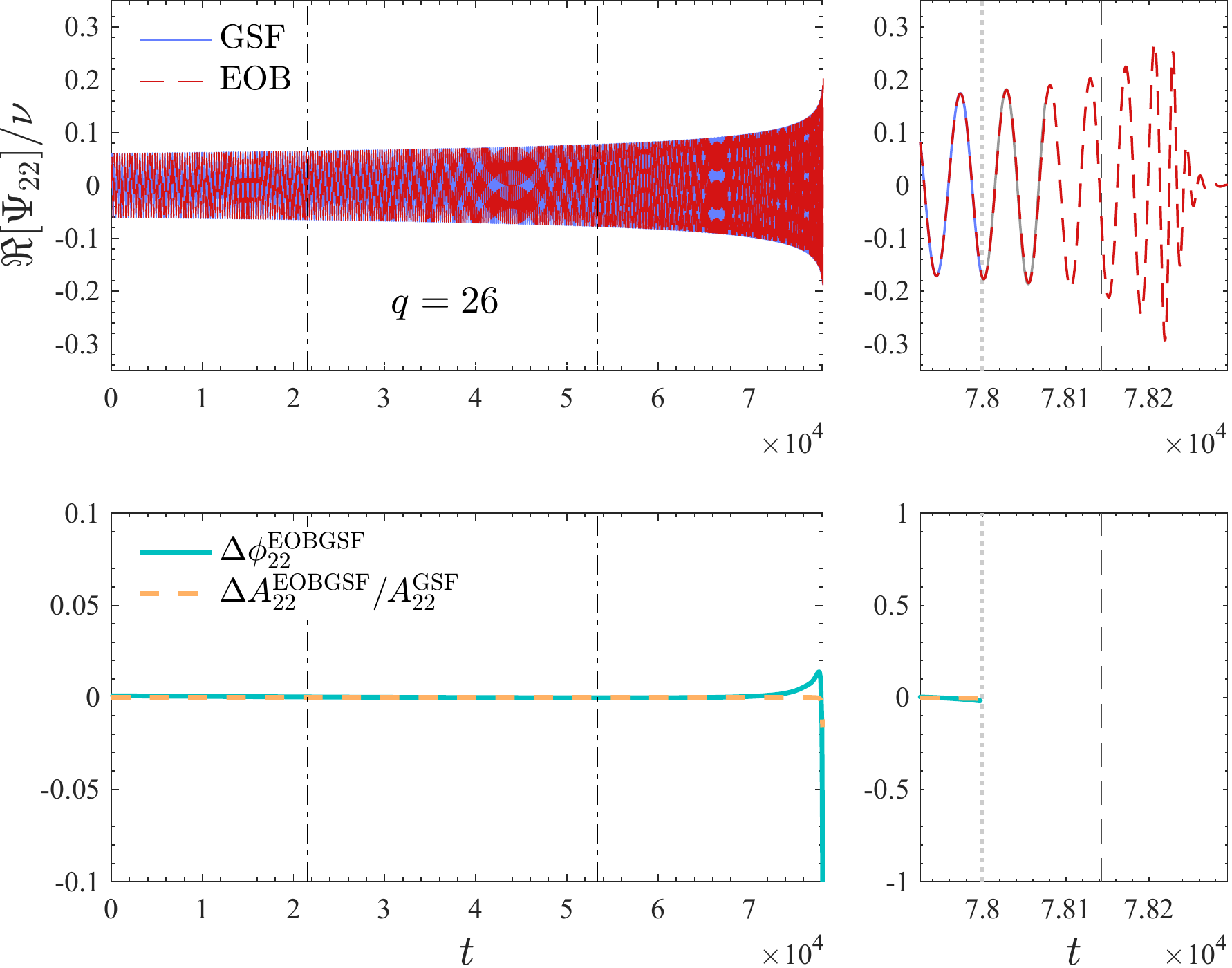} 
\caption{\label{fig:phasing26} EOB/GSF time-domain phasing for $q = 26$. The alignment interval
is indicated by the dash-dotted lines in the left panels. The accumulated dephasing up to the GSF breakdown frequency is $\Delta \phi^{\rm EOBGSF}_{22, t} = -0.0194$,
confirming the high agreement between EOB and GSF for this mass ratio.}
\end{figure}

However, Table~\ref{tab:dphi_Qis} yields a deeper understanding of why EOB and GSF apparently agree best 
around $q \sim 26$, as was seen within the time-domain analysis\footnote{We also verify this conclusion in Fig.~\ref{fig:phasing26}, which shows the time-domain phasing for $q = 26$. The accumulated phase difference between \TEOBResumS{} and \PAT{} waveforms at the GSF breakdown 
frequency ($\omega_{\rm break} = 0.11050$) is $\Delta \phi^{\rm EOBGSF}_{22, t} = -0.0194$. 
The alignment interval we use here is $[0.025, 0.033]$, and the integration of $\Qo$ yields $\Delta \phi^{\rm EOBGSF}_{22, \Qo} = -0.0204$. When adding the 6PN term in the EOB flux, the accumulated phase difference becomes $\Delta \phi^{\rm EOBGSF}_{22, t} =0.1321$.}. In fact, this empirical 
deduction is a simple consequence of the fact that the contributions 
$\Delta \phi_1$ and $\Delta \phi_2$ largely cancel for this mass ratio; for smaller mass ratios the dephasing is dominated
by $\Delta \phi_2$ while for larger mass ratios it is dominated by $\Delta \phi_1$. From the perspective of EOB, this corresponds to errors in EOB's 1PA term fortuitously cancelling higher-PA terms. From the perspective of GSF, it corresponds to the 1PA model's 2PA error terms becoming sufficiently small that they are comparable to the errors in EOB and NR (in line with the discussion in the previous section). The cancellation point will change if the first-law binding energy turns out to be the incorrect choice for the 1PA evolution, but this overall picture should remain the same.

Quite generally, then, we learn from Table~\ref{tab:dphi_Qis} that the impact of $\Qo^2$ 
decreases when increasing $q$, which is of course expected since it is multiplied by $\nu$. We also learn that 
 the errors in $\Qo^0$ and $\Qo^1$ contribute more than $\Qo^2$ when $q\gtrsim 26$.
Therefore, the takeaway messages are (i) that EOB can be improved for $q \gtrsim 26$ by including more information in
$\Qo^0$ and $\Qo^1$; (ii) that the error in the GSF model is probably dominated by its incorrect 2PA term $\Qo^2$, even for $q\lesssim10$ where 3PA and higher terms might have been significant; (iii) that the two formalisms approximately meet each other at $q \sim 26$ as a result of fortuitous cancellation of the dephasings coming from $\Qo^1$ and $\Qo^2$; and (iv) to ensure the 1PA model's accuracy in the small-mass-ratio regime and obtain more reliable internal estimates of its error, we must determine the correct binding energy.

Point (i) is specifically useful on the EOB side, since it allows us to detect the weaknesses of
the current model if one wants to push it to the IMR regime. In particular, the improvement at 1PA can be 
achieved by the implementation of GSF-informed potentials. As the mass ratio increases past $q \gtrsim 10^3$, however, the impact of the 0PA term on the dephasing
will prevail over all others. This means that for EMRI systems the most relevant and urgent update of the EOB model concerns the 0PA flux,
implying the need of incorporating more test-mass information into the radiation reaction. Both the implementation of
GSF-informed potentials and of a different flux in an EOB model specifically targeted for EMRIs will be presented and compared to
2GSF in an upcoming work~\cite{Albertini:2022dmc}.

Because of the uncertainty in 1PAT1's value of $Q^1_\omega$, it is hard to say the extent to which \TEOBResumS{}'s value of $Q^1_\omega$ must be improved. However, we note that this uncertainty does not affect our ability to incorporate 1PA information into the EOB model. The flux ${\cal F}^{(2)}_\infty$, for example, does not make use of the binding energy (or utilize the other two approximations described in Sec.~\ref{sec:1PAT1 errors}); it should therefore be exact up to numerical error. Similarly, the EOB potentials can be informed by independent, conservative 1GSF information without appealing directly to the 1PAT1 model.

Point (ii) gives analogous useful insight on the GSF side. The effective $\Qo^2$ in 1PAT1 is probably a significant overestimate of the true value, and this overestimate might dominate the model's error. %If \TEOBResumS{}'s value of $Q^1_\omega$ is reasonably close to correct, then the values of $\Delta \phi_2$ in Table~\ref{tab:dphi_Qis} provide a rough estimate of 1PAT1's phase error (in addition to its possible error in the choice of binding energy). This error estimate would be somewhat larger than our estimates in the previous section, and they would indicate that 1PAT1 errors remain nonnegligible until $q\gtrsim 100$. However, 
This could suggest that alternative formulations of the 1PA evolution equations with smaller contributions to $\Qo^2$ may significantly improve the phase accuracy at lower mass ratios. However, Figs.~\ref{fig:Qomg_all} and \ref{fig:DQomg} clearly show that the true value of $\Qo^2$ is not negligible, meaning a model that simply sets it to zero may incur similar levels of error as 1PAT1. We leave a more detailed study of this for future work.

Ultimately, we return again to the need for longer, higher-accuracy, lower-eccentricity, smaller-mass-ratio NR simulations. With such simulations, one could hope to obtain independent estimates of the true values of $Q_\omega^1$ and $Q_\omega^2$, helping to lift the uncertainties discussed in this section.

We note, finally, that the considerations above hold assuming that a small mass ratio expansion yields a faithful representation
of the waveform. Interestingly, Fig.~\ref{fig:diffQomg_GSF} suggests that this may be the case. 
The figure shows 
that for both GSF and EOB models $\Qo$ is largely encapsulated in the three coefficients $\Qo^0$, $\Qo^1$, and $\Qo^2$, with only a 
small residual accounted for by higher-order terms. 

%===================================================
% Accumulated EOB/GSF phase differences between 0.023 and 0.09
%===================================================
\begin{table}[t]
\begin{center}
\begin{ruledtabular}
\begin{tabular}{c c c c c}
$q$ &  $\Delta \phi_0$ & $\Delta \phi_1$ & $\Delta \phi_2$ & $\Delta \phi_{(\omega_1, \omega_2)}^{\rm EOBGSF}$\\
\hline
\hline
7 & 0.011 & 0.538 & $-1.690$ & $-1.141$ \\
10 & 0.015 & 0.538 & $-1.277$ & $-0.724$ \\
15 & 0.021 & 0.538 & $-0.905$ & $-0.347$ \\
\hline
26 & 0.034 & 0.538 & $-0.551$ & $0.021$ \\
32 & 0.042 & 0.538 & $-0.454$ & $0.125$ \\
64 & 0.081 & 0.538 & $-0.234$ & $0.384$ \\
128 & 0.159 & 0.538 & $-0.119$ & $0.578$ \\
\end{tabular}
\end{ruledtabular}
\end{center} 
\caption{From left to right, the columns report: the mass ratio $q$, the phase differences due to the first three term in
the expansion of $\Qo$, and the sum of these latter. The $\Delta \phi$'s are obtained using the definition~\eqref{eq:dphis},
integrating over the frequency interval $(\omega_1, \omega_2) = (0.023, 0.09)$.}
\label{tab:dphi_Qis}
\end{table}
%============
% Figure for q=26
%============

\section{Conclusions}
\label{sec:conclusions}
We have provided a comprehensive comparison between $\ell=m=2$ gravitational waveforms obtained
with a 2GSF-based approach~\cite{Wardell:2021fyy} and the state-of-the-art EOB model \TEOBResumS{}~\cite{Nagar:2020pcj,Riemenschneider:2021ppj,Albertini:2021tbt}.
Among the two available EOB models (the other being {\tt SEOBNRv4HM}~\cite{Cotesta:2018fcv}),  
\TEOBResumS{} shows the highest level of NR faithfulness and has been checked to be consistent 
with the plunge and ringdown phase of state-of-the-art NR simulations~\cite{Lousto:2020tnb}  for large mass ratios up to $q=128$~\cite{Nagar:2022icd}.
On the 2GSF side, we work with the time-domain \PAT{} model introduced in~\cite{Wardell:2021fyy}.
This model is limited to the inspiral phase, since it has not yet incorporated a model for the transition from inspiral
to plunge. The \PAT{} waveforms are reliable up to some frequency before the Schwarzschild LSO GW frequency, 
$\omega_{\rm LSO}^{\rm Schw}=0.136$, where the two-timescale approximation on which the model is
built ceases to be valid. Our analysis is thus limited to the  inspiral waveform only, up to dimensionless 
GW frequency $\omega\lesssim 0.1$. Note that this frequency is always smaller than the LSO frequency $\omega_{\rm LSO}^{\rm EOB}(\nu)$ 
predicted by the EOB model for any mass ratio, and one has $\omega_{\rm LSO}^{\rm EOB}(\nu)>\omega_{\rm LSO}^{\rm Schw}$~\cite{Buonanno:2000ef}.
We also benchmarked our findings with NR waveform data, similarly but more thoroughly than was done in Ref.~\cite{Wardell:2021fyy}, and we provided a detailed analysis of the 1PAT1 model's sources of error and domain of validity.

Our conclusions are as follows:
\begin{enumerate}[label=(\roman*)]
\item We have found that effects of the transition to plunge are significant over a larger frequency interval than one might expect, restricting 1PAT1's domain of validity to orbital frequencies much smaller than the ``breakdown frequency'' $\approx\Omega^{\rm Schw}_{\rm LSO}-0.026\nu^{1/4}$. Similarly, we have stressed that GSF models should not be pushed too far into the weak-field regime, as they will accumulate arbitrarily large error when the initial frequency approaches zero (though the frequency interval can be broadened for smaller $\nu$). We have also highlighted the use of the first-law binding energy as a source of significant uncertainty in 1PAT1's phasing, $\sim 0.5$~rad for all mass ratios.
\item We have revisited the 2GSF/NR comparison of Ref.~\cite{Wardell:2021fyy} in more detail using the gauge-invariant description
of the gravitational phase provided by the function $\Qo=\omega^2/\dot{\omega}$ extracted from the Weyl scalar $\psi_4^{22}$. 
The use of this quantity is {\it crucial} to have access to a reliable description of the NR $\Qo$, as noted long ago 
in Ref.~\cite{Damour:2012ky}.
We focus on mass ratios $q=7$ and $q=10$, and
our novel analysis allows us to conclude that for these mass ratios the \PAT{} waveform introduces accumulated dephasings $\lesssim 1$~rad up to frequency $\sim 0.1$. As expected, larger phase differences are found for smaller values of $q$, as described in detail 
in Appendix~\ref{sec:eobnrgsf_q}.
\item Focusing again on mass ratios $q=7$ and $q=10$, we have similarly extensively compared time-domain and  
frequency-domain phasing analysis using \PAT{}, NR and \TEOBResumS{}, in order to eliminate possible systematics that may arise when choosing the alignment window. 
\item We have explored the level of agreement between \PAT{} and \TEOBResumS{} for mass ratios $q=(15,32,64,128)$.
We performed several types of EOB/GSF phasing comparison both in the time domain and using $\Qo$, notably carefully
cross-checking the results obtained with the two approaches.
Thanks to complementary information gained from a recent EOB/NR comparison~\cite{Nagar:2022icd}, and also considering a long-inspiral $q = 15$
SXS dataset, we concluded that \PAT{}
is less accurate than \TEOBResumS{} up to frequency $\sim 0.1$ also for $q=15$, in analogy with 
the $q\leq 10$ mass ratios mentioned above, though in this case the dephasing between the two models is much smaller, 0.38~rad over a long inspiral (i.e., large frequency interval).
\item By contrast, we found a region of excellent EOB/GSF phase agreement around $q\sim 26$, although the 2GSF/EOB 
differences are found to increase  again for larger mass ratios up to $q = 128$. Simple error estimates suggest that the 1PA model's error should be significantly below the disagreement between the two models for $q=64$ and $q=128$, implying that the 1PAT1 waveform should be more accurate than the EOB model for these mass ratios, and increasingly so for higher mass ratios. However, this is complicated by (i) the uncertainty in 1PAT1 due to its choice of binding energy, and (ii) our limited knowledge of the magnitude of the true 2PA coefficient in the phase. Since this is a region where no long-inspiral, error-controlled NR simulations are available, it is therefore difficult to state precisely the
limitations of both \TEOBResumS{} and \PAT{}. 
\item To attempt a partial clarification of these issues, we provided a novel analysis of the contributions to the phasing, analyzing both 
the 2GSF and the EOB $\Qo$'s as expansions in $\nu$. This allowed us to single out quantitatively the main differences between the 
two approaches in the small-$\nu$ regime. We found that the two models do differ ($\sim 0.5$~rad) already at the level of $\Qo^1$, but this is again complicated by the uncertainty due to choice of binding energy; the difference in $\Qo^1$ between the two models' only becomes larger than the uncertainty at high frequencies.
For $q\sim 26$ there is a compensation between the difference in $\Qo^1$ and a contribution $\sim -0.5$~rad from $\Qo^2$ that largely cancels it to give an overall good 2GSF/EOB agreement.
\item For larger values of $q\gtrsim 10^3$, small differences in $\Qo^0$ that are negligible for comparable mass ratios become 
more and more relevant. These can be attributed to incomplete analytical information in \TEOBResumS{} and point to an important area for future improvement.
\item 2PA terms in $\Qo$ are significant at least for mass ratios $q\lesssim30$. While the 1PAT1 model includes an effective $\Qo^2$, its value appears to be a large overestimate. If the model's choice of binding energy is shown to be correct, than this overestimate of  $\Qo^2$ is likely the dominant source of error for all mass ratios up to a point at sufficiently large $q$ when small numerical errors in the 0PA or 1PA terms dominate. %, though it contributes only a $\lesssim 0.1$~rad.
\end{enumerate}
%========================================
% Table of GSF frequencies and phasing accumulated
%========================================
\begin{table*}[t]
\begin{center}
\begin{ruledtabular}
\begin{tabular}{c  c c c c c c c c c c}
$q$ &  $\omega_1$ & $\omega_2$ & $n_1$ & $n_2$ & $n_3$ & $n_4$ & $n_5$ & $d_1$ & $d_2$ & $d_3$ \\
\hline
\hline
7 & 0.04522 & 0.10477 & $-219.98419$ & 2265.56464 &  $-890.00345$ & $-22823.02252$ & $35185.66602$ & $86.79492$ & 289.17836 & $-6440.91349$ \\
10 & 0.04522 & 0.10477 & 0.10756 &  $-160.50258$ &  165.53483 & 1726.39116 & $-3008.63838$ & $-21.20574$ & 92.70890 & 247.75797 \\
\end{tabular}
\end{ruledtabular}
\end{center} 
\caption{Coefficients enterting the fitting function for the Newton-rescaled $\hat{Q}_\omega$ function obtained from the curvature waveform $\psi_4^{22}$
for the SXS:BBH:0298 ($q=7$) and SXS:BBH:0303 ($q=10$) NR datasets. We use waveforms extrapolated to infinity with $N=3$ extrapolation order.
In the second and third columns we reports the values of the boundaries of the frequency interval $(\omega_1,\omega_2)$ on which the fits are performed.
See Figs.~\ref{fig:Qomg_clean_7}-\ref{fig:Qomg_clean_10} for the visual behavior of the fit.}
\label{tab:Qomg_coeff}
\end{table*}
%========================================
%==================
% Cleaning Qomg: q=7
%==================
\begin{figure*}[t]
\center
\includegraphics[width=0.35\textwidth]{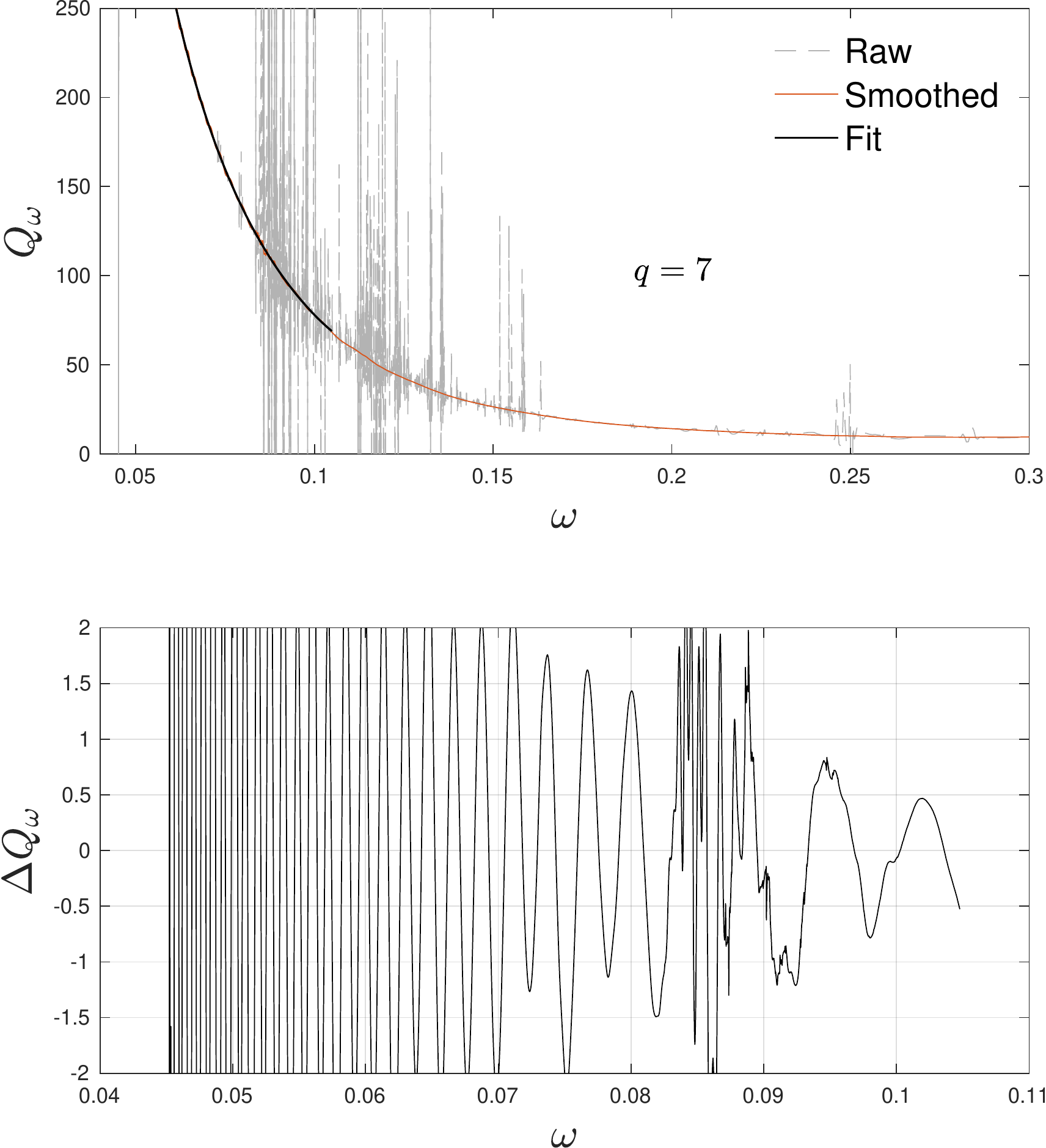} 
\hspace{5 mm}
\includegraphics[width=0.35\textwidth]{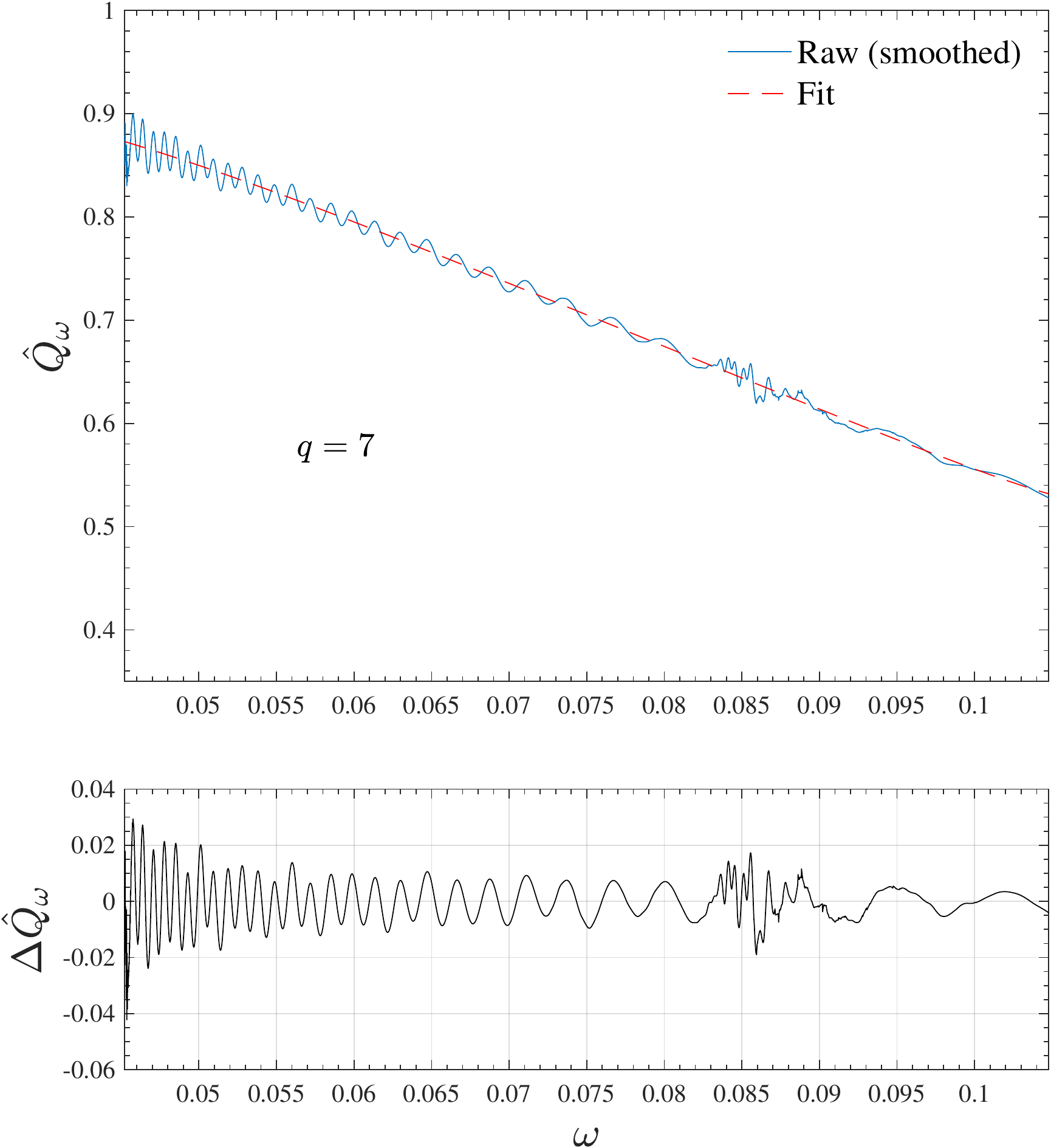} 
\caption{\label{fig:Qomg_clean_7} Removal of high-frequency and low-frequency oscillations from the $Q_\omega$ function for 
SXS:BBH:0298 ($q=7$). The residual, low-frequency,
oscllations in $\Delta \hat{Q}_\omega$ average to zero. The coefficients of the fit to $\hat{Q}_\omega$ are listed in Table~\ref{tab:Qomg_coeff}.}
\end{figure*}
%==================
% Cleaning Qomg: q=10
%==================
\begin{figure*}[t]
\center
\includegraphics[width=0.35\textwidth]{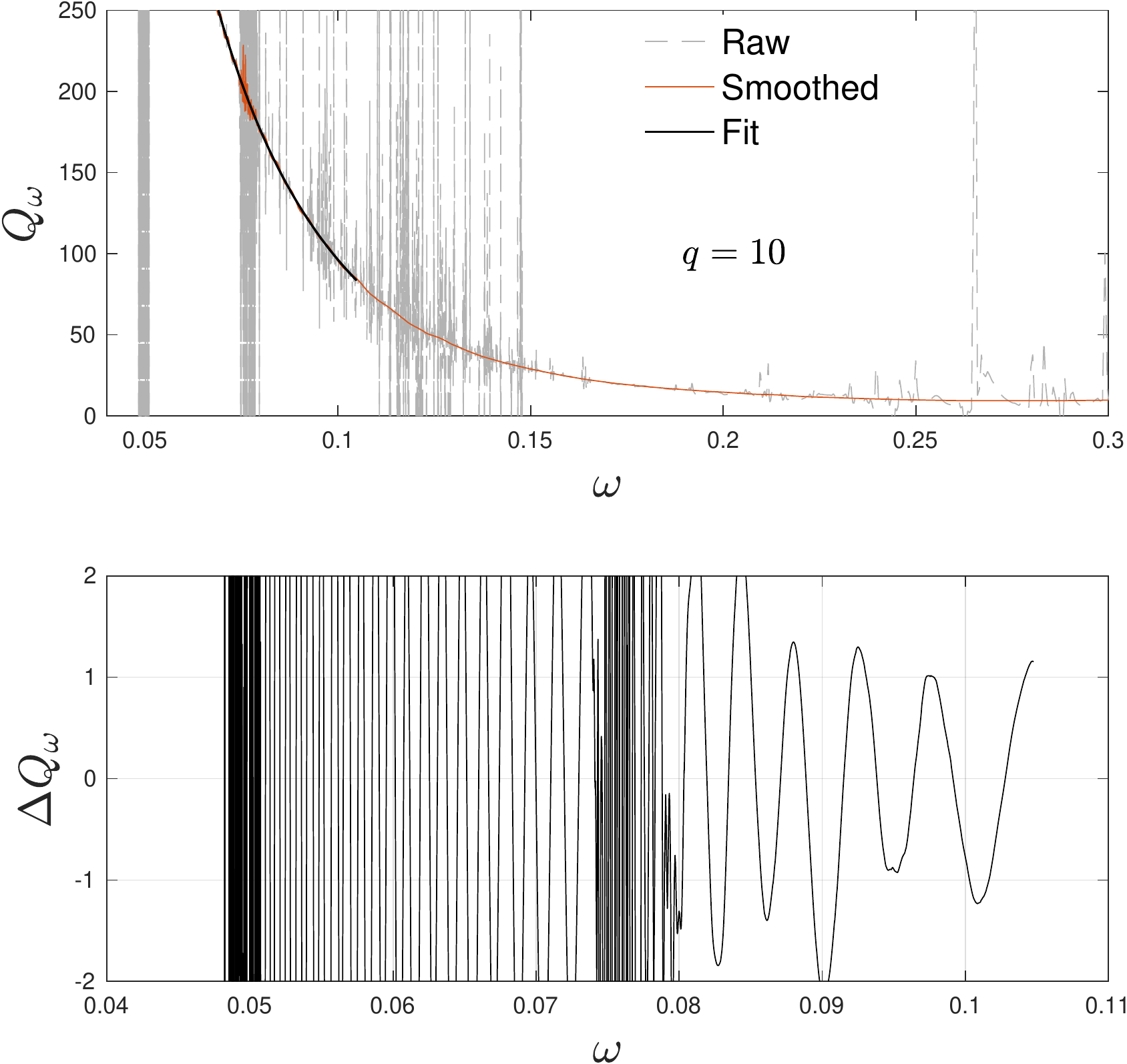} 
\hspace{5 mm}
\includegraphics[width=0.35\textwidth]{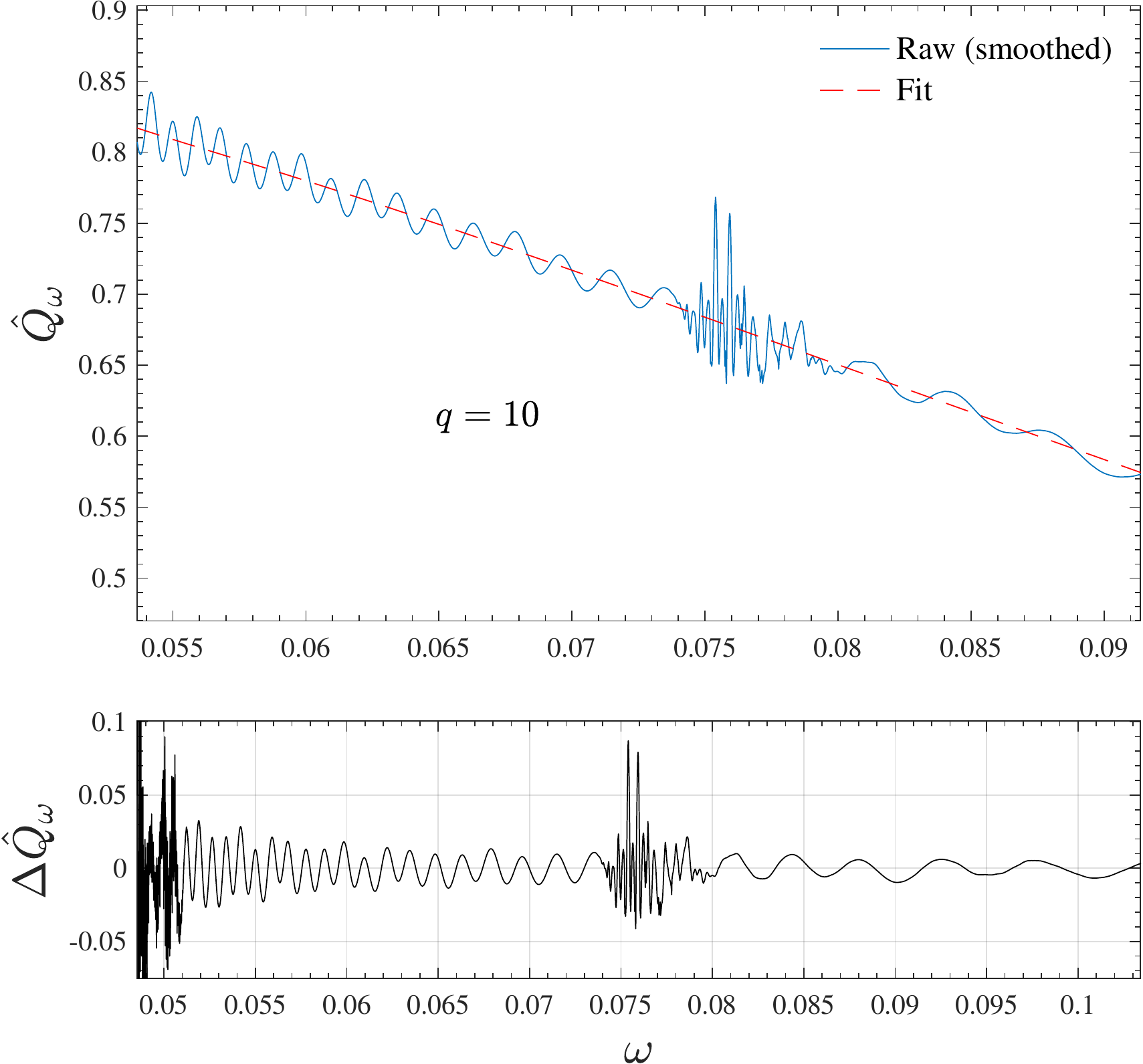} 
\caption{\label{fig:Qomg_clean_10}Removal of high-frequency and low-frequency oscillations from the $Q_\omega$ function for 
SXS:BBH:0303 ($q=10$). The residual, low-frequency,
oscllations in $\Delta \hat{Q}_\omega$ average to zero. The coefficients of the fit to $\hat{Q}_\omega$ are listed in Table~\ref{tab:Qomg_coeff}.}
\end{figure*}

%===========

In broad terms, our analysis gives cause for optimism that EOB and 2GSF models can ultimately provide reliable waveforms in the entire $q\gtrsim 10$ regime, both for the LVK collaboration and for use in third-generation ground-based detectors such as Einstein Telescope~\cite{Hild:2009ns} or Cosmic Explorer~\cite{Evans:2021gyd}. The two models we considered currently agree within $\sim 0.5$~rad over a large frequency interval for mass ratios in the range $15\lesssim q\lesssim 64$, and there are clear paths to improvement both within and beyond that range.

On the EOB waveform modeling side, the next challenge will be to improve \TEOBResumS{} to build a new, GSF-faithful EOB 
model that is closer in phasing to the \PAT{} model for large mass ratios. This will substantially happen by including 1GSF information 
in the conservative nonspinning dynamics, building upon the results of Ref.~\cite{Akcay:2015pjz}. Note, however, that our 
$\Qo$ analysis indicates that improvements in $\Qo^0$ are {\it also} needed, i.e. concerning the 0PA flux, and these improvements are going to be 
progressively more important as the mass ratio goes into the extreme-mass-ratio regime. The development of such a GSF-informed
model and the evaluation of its performance against \PAT{} will be presented in an upcoming work~\cite{Albertini:2022dmc}.

Our work also suggests several needed improvements on the GSF waveform modelling side. The most critical is the inclusion of the final plunge, merger, and ringdown. Similarly, to model waveforms of any length, the model must incorporate small-frequency, PN information (though this is a lower priority, as EOB already provides a robust framework for combining weak- and strong-field information). It is also clear that the 1PAT1 model has an unnecessarily large (and incorrect) $\Qo^2$, and that this is likely the model's dominant source of phase error. Alternatives to this model that include a more faithful $\Qo^2$ will be considered in future work. To make the model fully reliable in the inspiral phase,  we must also calculate the internally consistent binding energy, revisiting the calculation in Ref.~\cite{Pound:2019lzj}, or calculate the 1PA term in $\dot\Omega$ using the local second-order self-force; ultimately, to be entirely confident in these calculations, we should obtain consistent values for $\dot\Omega$ using both methods. However, we note that the improved accuracy of the 1PAT1 model at larger $q$ (e.g., when compared to a $q=15$ SXS waveform) suggests that the first-law binding energy probably lies very close to the true value.

%Our waveform analysis suggests that the accuracy of \TEOBResumS{} in $\Qo^0$ and $\Qo^1$ should be improved 
%to get a better match with \PAT{} and consequently provide a more accurate representation of the long-inspiral waveform for those
%mass ratios where, a priori, 2GSF is expected to give a reliable description of the dynamics and the waveform. 

Finally, we stress that long-inspiral, highly-accurate NR simulations with $q\geq 10$ are needed to achieve a precise evaluation of the accuracy of GSF and EOB models in this regime. Only a sparse number of  
simulations of typical SXS accuracy are required. All that is needed is sufficient data to clearly see the behaviour $Q_\omega=Q^0_\omega(\omega)/\nu+Q^1_\omega(\omega)+\nu Q^2_\omega(\omega)+O(\nu^2)$ and determine an order-of-magnitude estimate of $Q^2_\omega$, which should enable sufficiently precise estimates of the error in a 1PA approximation. This kind of procedure is well established and already possible using simulations with $q\leq10$~\cite{vandeMeent:2020xgc,Albalat:2022vcy}, but the conclusions would be far more robust with higher-$q$ data; as our analysis has shown, small-$\nu$ fits of $q\lesssim 10$ data can be problematic at high frequencies. Simulations of longer inspirals for $q\sim 10$ would also provide an important additional check of GSF's low-frequency behaviour. For now, EOB waveforms provide the only independent benchmark on the early inspiral phase of 2GSF waveforms; and conversely, 2GSF waveforms provide the only benchmark on the large-$q$, strong-field inspiral phase of EOB waveforms.

\acknowledgements
We are grateful to Rossella Gamba for critical observations and comments
on the manuscript.
We thank J. Yoo, V. Varma, M. Giesler, M. Scheel, C. Haster, H. Pfeiffer, L. Kidder, and
M. Boyle for sharing with us the $q=15$ waveform of Ref.~\cite{Yoo:2022erv} before having it available through the SXS catalog.
A.A. has been supported by the fellowship Lumina Quaeruntur No.
LQ100032102 of the Czech Academy of Sciences.
A.P. acknowledges the support of a Royal Society University Research Fellowship. N.W. acknowledges support from a Royal Society - Science Foundation Ireland University Research Fellowship via grants UF160093 and RGF\textbackslash R1\textbackslash180022.
This work makes use of the Black Hole Perturbation Toolkit \cite{BHPToolkit} and Simulation Tools \cite{SimulationToolsWeb}.

\appendix

\section{Computation of $Q_\omega$ from NR data}
\label{sec:Qomg_clean}
Let us report here some technical details about the removal of both the high and low-frequency oscillations
that are present in the NR curvature (i.e. from $\psi_4^{22}$) $\Qo$ functions for datasets SXS:BBH:0298
($q=7$) and SXS:BBH:0303 ($q=10$) discussed in the main text. We take here the highest resolution available 
and use $N=3$ extrapolated data, including the correction for the spurious motion of the center of mass.
We then apply the procedure of Ref.~\cite{Damour:2012ky}, briefly reviewed in Sec.~\ref{sec:gsf_nr}, to obtain a reliable $\Qo$ that does not present oscillations (either of low-frequency or of high-frequency)
and that is qualitatively and quantitatively consistent with the EOB one. The procedure is applied on the
Newton-normalized function, $\hat{Q}_\omega$, as defined in Eq.~\eqref{eq:hatQ}. Useful quantitative information
is reported in Figs.~\ref{fig:Qomg_clean_7} and ~\ref{fig:Qomg_clean_10}. For each mass ratio, the left
panel of the figure shows: (i) the raw $Q_\omega$; (ii) the smoothed $Q_\omega$ obtained
applying a low-pass filter to remove high-frequency numerical noise that appears in
the computation of $\omega$ and $\dot{\omega}$; (iii) the final results after the fit
to a rational function.
The right panels of Figs.~\ref{fig:Qomg_clean_7} and~\ref{fig:Qomg_clean_10} 
report the Newton-normalized $\hat{Q}_\omega$ function, with the low-frequency 
oscillations, and the fitting function that averages them.
In the bottom row we also show  the difference between the original $\hat{Q}_\omega$ and
the fitted one: the fact that the differences oscillate about zero is a good indication of the reliability
of the procedure. The coefficients entering the fitting function of Eq.~\eqref{eq:Qom_ratio} in
the main text are listed in Table~\ref{tab:Qomg_coeff}.

As for numerical accuracy, it can be estimated by comparing the highest and second highest
resolutions available. This was done in Ref.~\cite{Damour:2012ky} (see Fig.~7 therein), 
where it is shown how the difference between the $\Qo$ derived from the two resolutions
is of order $\sim 0.1$.

\section{Exact formulas for $Q_\omega^0$ and $Q^1_\omega$}
\label{sec:exactQ0andQ1}

In this Appendix we derive exact formulas for the 0PA and 1PA coefficients $Q_\omega^0$ and $Q^1_\omega$ in the expansion~\eqref{eq:Qo_exp}. We do so by relating the waveform frequency $\omega$ to the orbital frequency $\Omega$ and then appealing to Eq.~\eqref{eq:Omegadot} for $\dot\Omega$.

The waveform frequency of the $\ell=2,m=2$ strain waveform is defined as given below Eq.~\eqref{eq:RWZnorm}. Equivalently, we write
\begin{equation}
h_{22} = |h_{22}| e^{-i\phi_{22}}
\end{equation}
and $\hat\omega\equiv M\omega=M\dot\phi_{22}$. This allows us to easily relate $\omega$ to the 1PA waveform~\eqref{1PAT1 hlm - nu}. Using $\phi_{22}=-\arctan\frac{{\rm Im} h_{22}}{{\rm Re} h_{22}}$, substituting Eq.~\eqref{1PAT1 hlm - nu}, and using Eq.~\eqref{eq:Omegadot} for $\dot\Omega$, we find
\begin{equation}\label{Omega to omega}
\hat\omega = 2\hat\Omega +\nu\hat\omega_1(\hat\Omega) +O(\nu^2),
\end{equation}
with
\begin{equation}\label{omega1}
\hat\omega_1 = \frac{F_0}{|R^{(1)}_{22}|^2}\left({\rm Im}R^{(1)}_{22} {\rm Re}\partial_{\hat\Omega}R^{(1)}_{22}-{\rm Re}R^{(1)}_{22} {\rm Im}\partial_{\hat\Omega}R^{(1)}_{22}\right).
\end{equation}
Conveniently, the 1PA correction here only involves 0PA amplitudes. We also find in practice that $\hat\omega_1$ is numerically very small.

Consistent with our 1PAT1 model, we approximate $M$ as constant (though the extension to non-constant $M$ is straightforward). A derivative of Eq.~\eqref{Omega to omega} then reads
\begin{equation}
\dot\omega = 2\dot\Omega +\nu\dot\Omega\partial_{\hat \Omega}\hat\omega_1(\hat\Omega)+O(\nu^3).
\end{equation} 
Substituting $\hat\omega_1$ from Eq.~\eqref{omega1} and again appealing to Eq.~\eqref{eq:Omegadot} for $\dot\Omega$, we obtain $\dot\omega$ as an expansion in powers of $\nu$ at fixed $\Omega$. We then substitute the inverse of Eq.~\eqref{Omega to omega}, $\hat\Omega = \frac{1}{2}\left[\hat\omega-\nu\hat\omega_1\big(\tfrac{\hat\omega}{2}\big)+O(\nu^2)\right]$, to obtain
\begin{multline}\label{omegadot}
\dot\omega = \frac{\nu}{M^2}\bigg\{2 F_0\big(\tfrac{\hat\omega}{2}\big)+\nu\Big[2 F_1\big(\tfrac{\hat\omega }{2}\big) -\hat\omega_1\big(\tfrac{\hat\omega }{2}\big)F_0'\big(\tfrac{\hat\omega}{2}\big)\\
 +F_0\big(\tfrac{\hat\omega}{2}\big)\omega_1'\big(\tfrac{\hat\omega }{2}\big)\Big]+O(\nu^2)\bigg\}.
\end{multline}
Primes here denote differentiation with respect to the function's argument.

We finally obtain $Q_\omega=\omega^2/\dot\omega$. Using Eq.~\eqref{omegadot} and expanding, we find $Q_\omega=\frac{1}{\nu}Q_\omega^0(\hat\omega) + Q_\omega^1(\hat\omega)+O(\nu)$, where
\begin{align}
Q_\omega^0 &= \frac{\hat\omega^2}{2F_0\big(\tfrac{\hat\omega }{2}\big)},\label{Q0 exact}\\
Q_\omega^1 &= -\frac{\hat\omega^2}{4\big[F_0\big(\tfrac{\hat\omega }{2}\big)\big]^2} \Big[2 F_1\big(\tfrac{\hat\omega }{2}\big) - \omega_1\big(\tfrac{\hat\omega}{2}\big) F_0'\big(\tfrac{\hat\omega }{2}\big)\nonumber\\
&\qquad\qquad\qquad\quad+F_0\big(\tfrac{\hat\omega}{2}\big) \omega_1'\big(\tfrac{\hat\omega }{2}\big)\Big].\label{Q1 exact}
\end{align}

Carrying out the same expansion for the curvature waveform~\eqref{psi4 phase} yields identical results for $Q_\omega^0$ and $Q_\omega^1$; the two versions of $Q_\omega$ only begin to differ at 2PA order.

\section{EOB/NR/GSF phasing comparison: comparable mass case}
\label{sec:eobnrgsf_q}
\begin{figure*}[t]
\center
\includegraphics[width=0.31\textwidth]{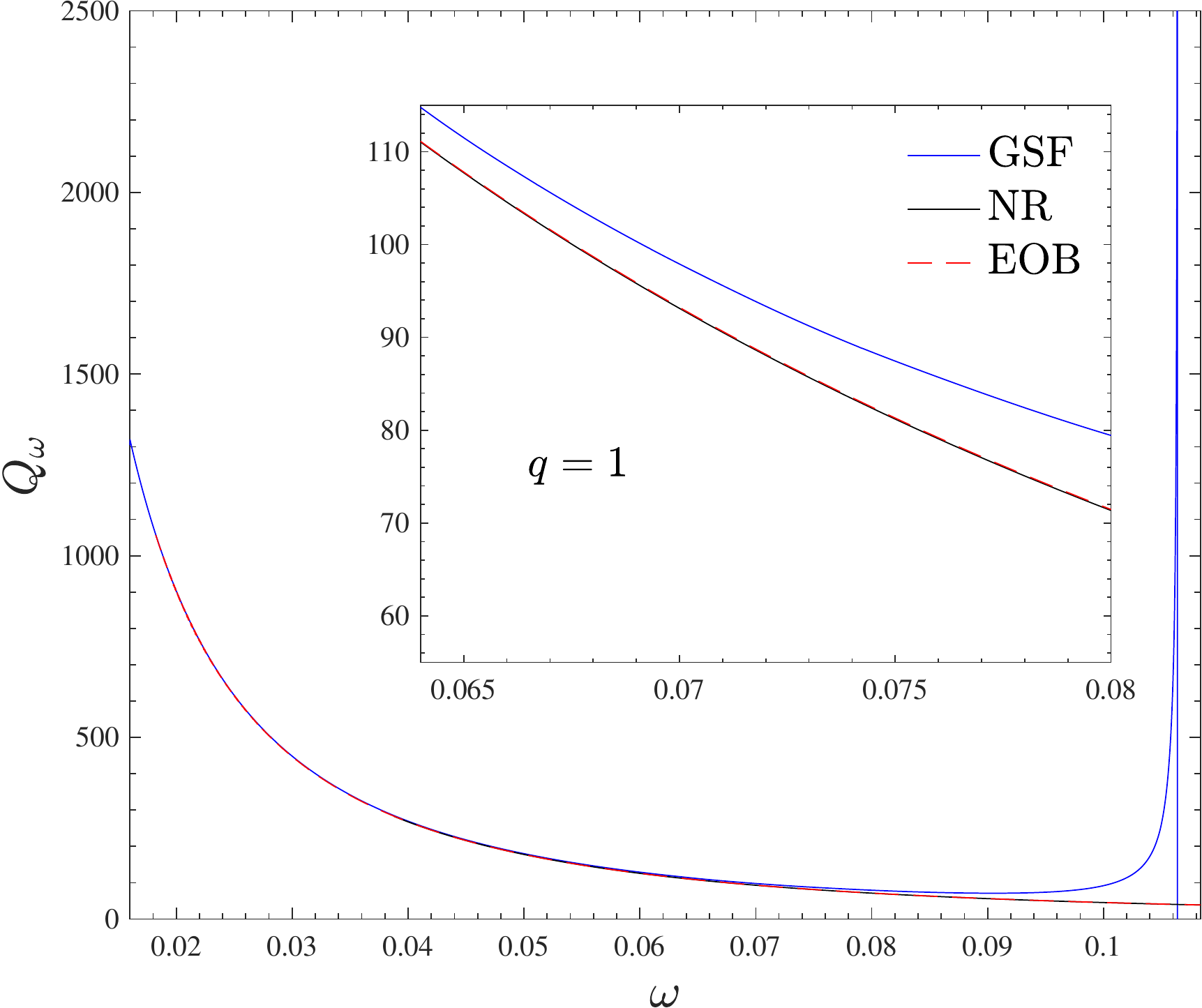} 
\hspace{2mm}
\includegraphics[width=0.31\textwidth]{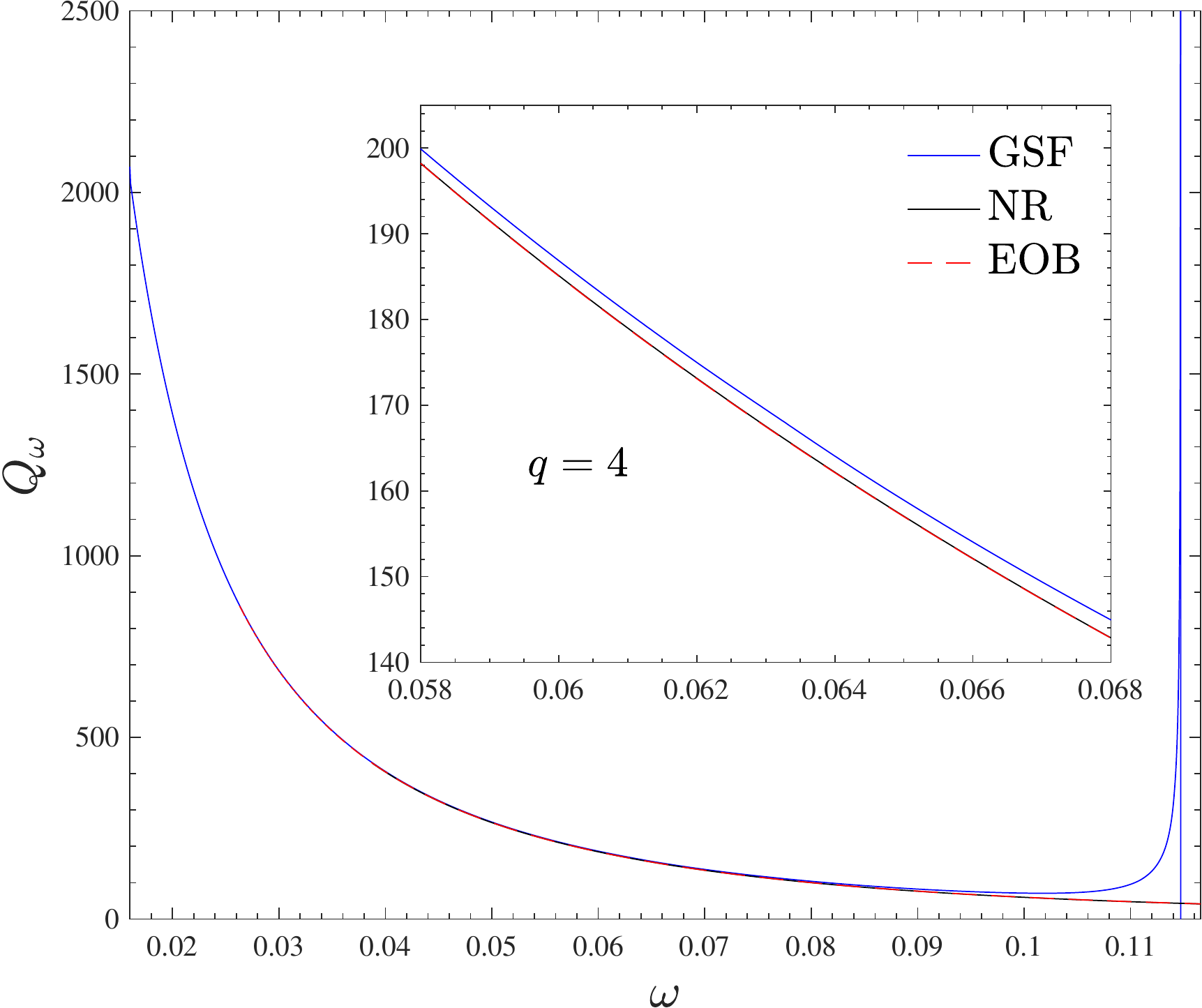}  
\hspace{2mm}
\includegraphics[width=0.31\textwidth]{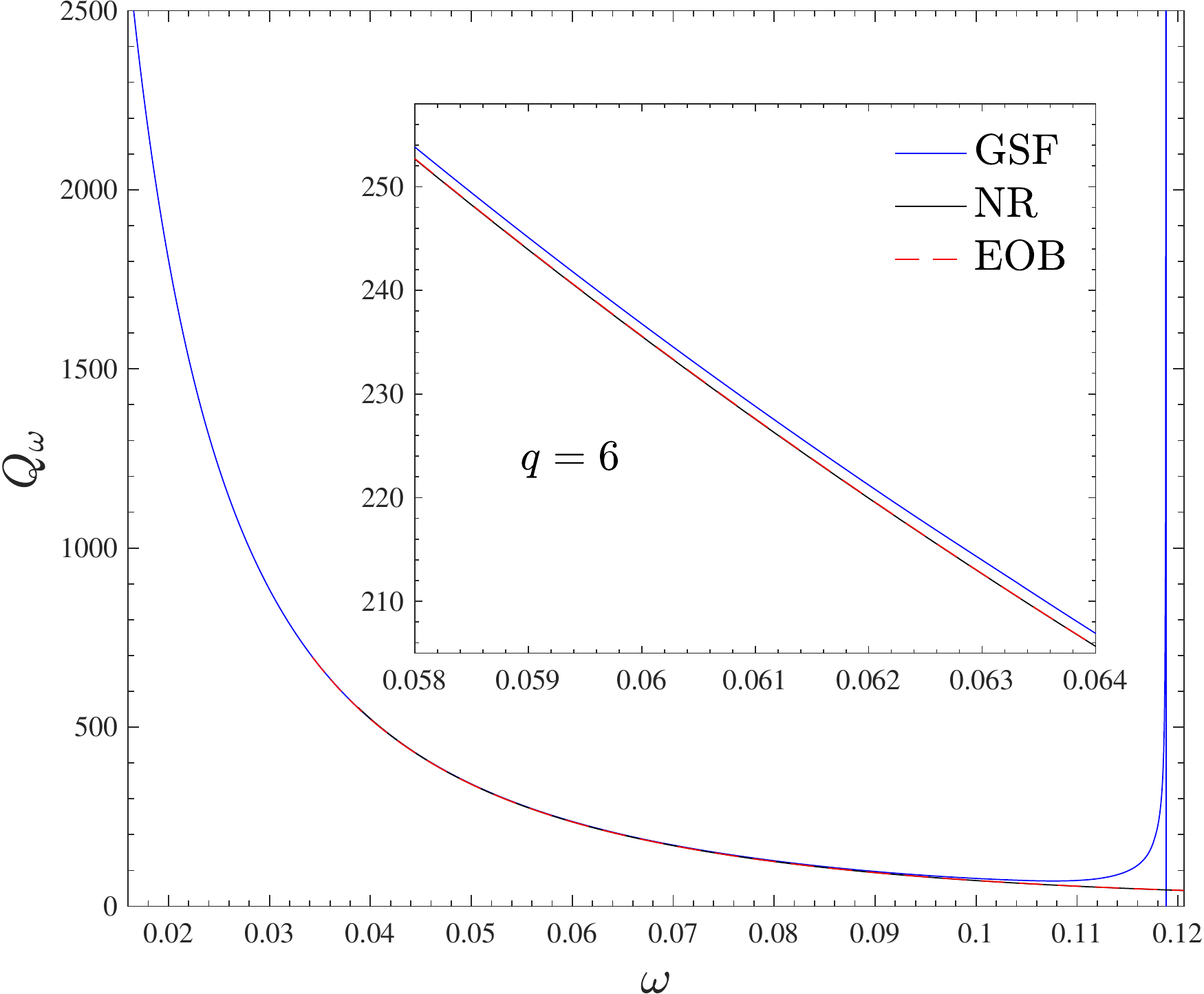} \\
\vspace{4mm} 
\includegraphics[width=0.31\textwidth]{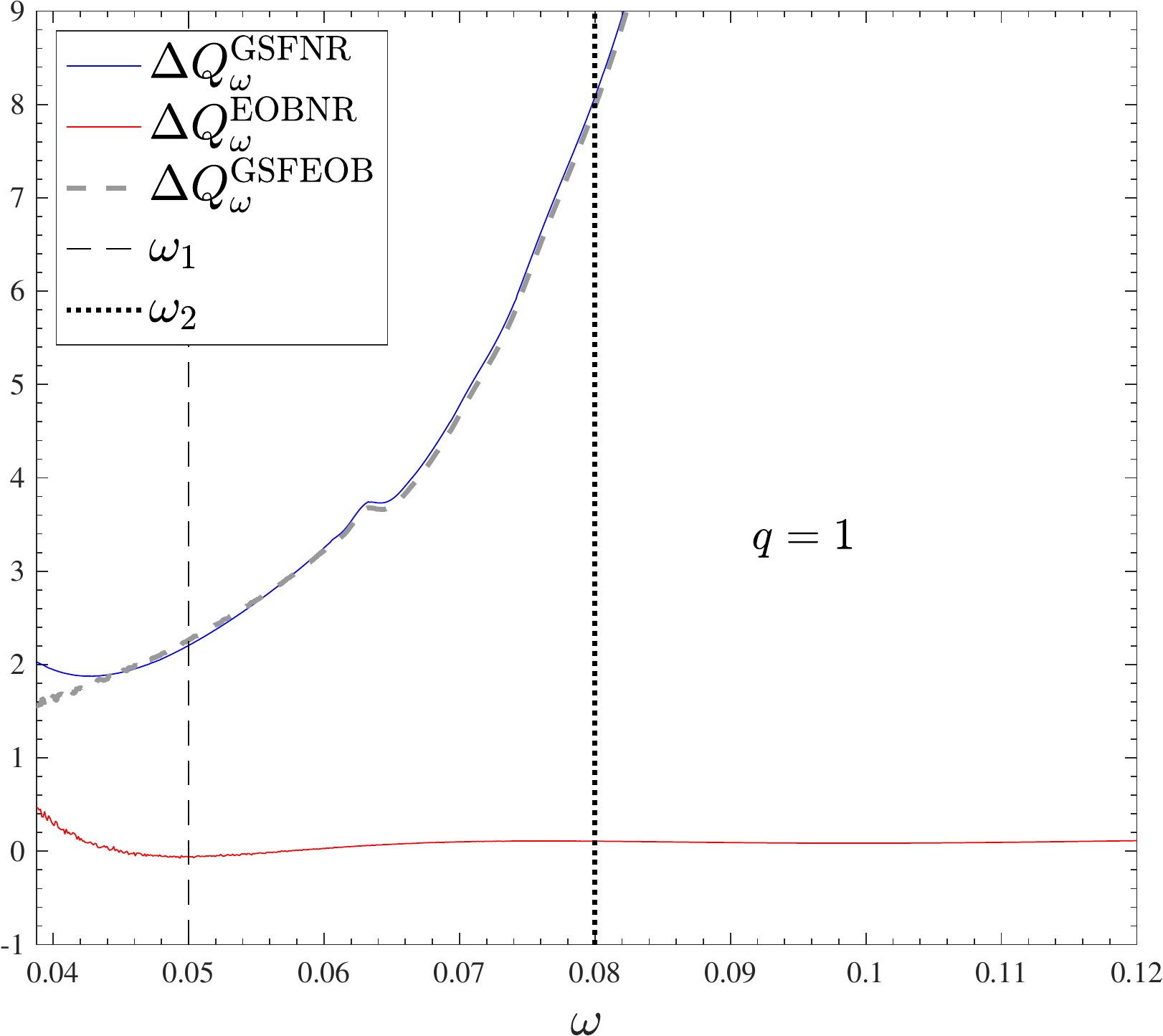}
\includegraphics[width=0.31\textwidth]{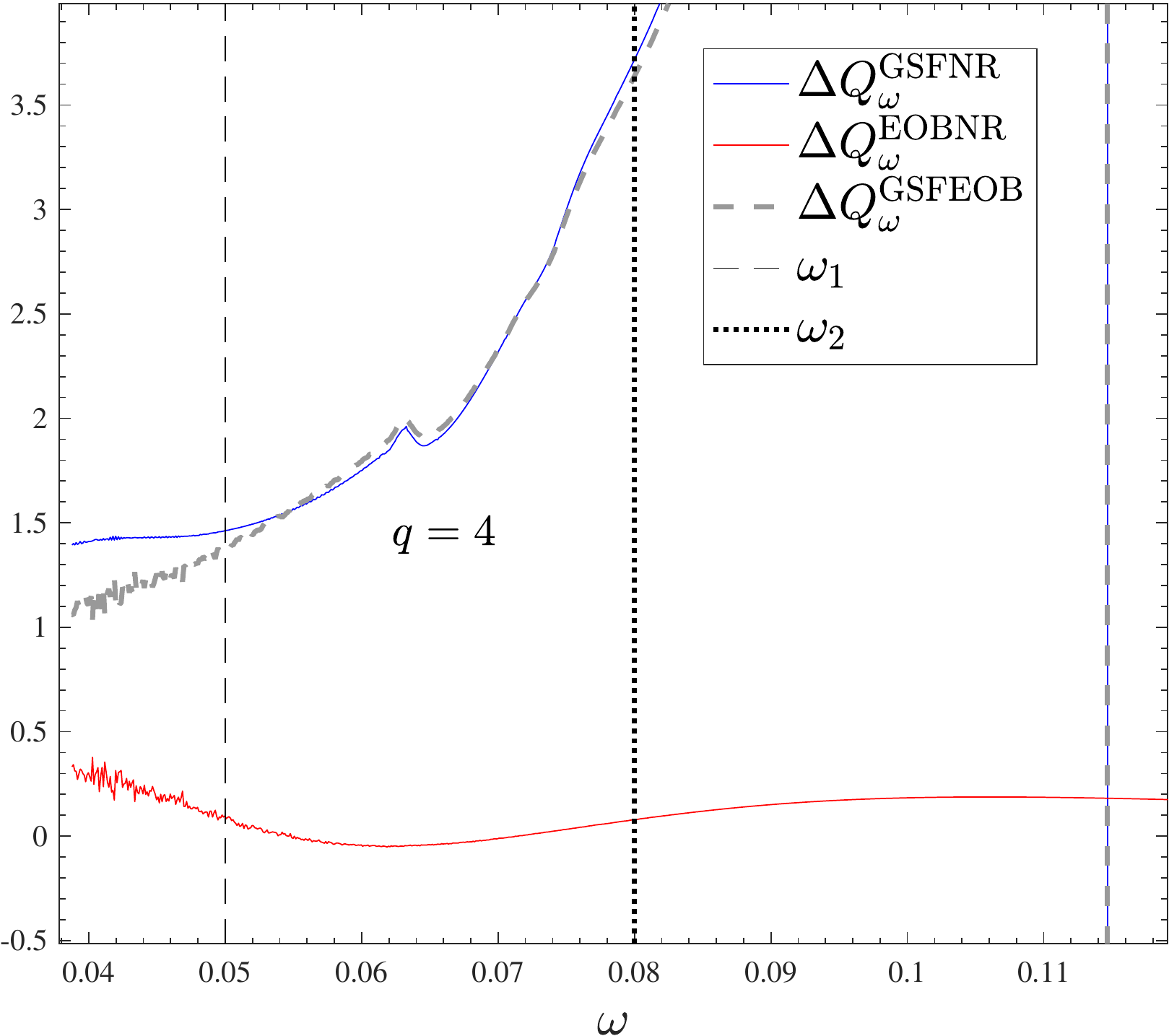} 
\hspace{1mm}
\includegraphics[width=0.31\textwidth]{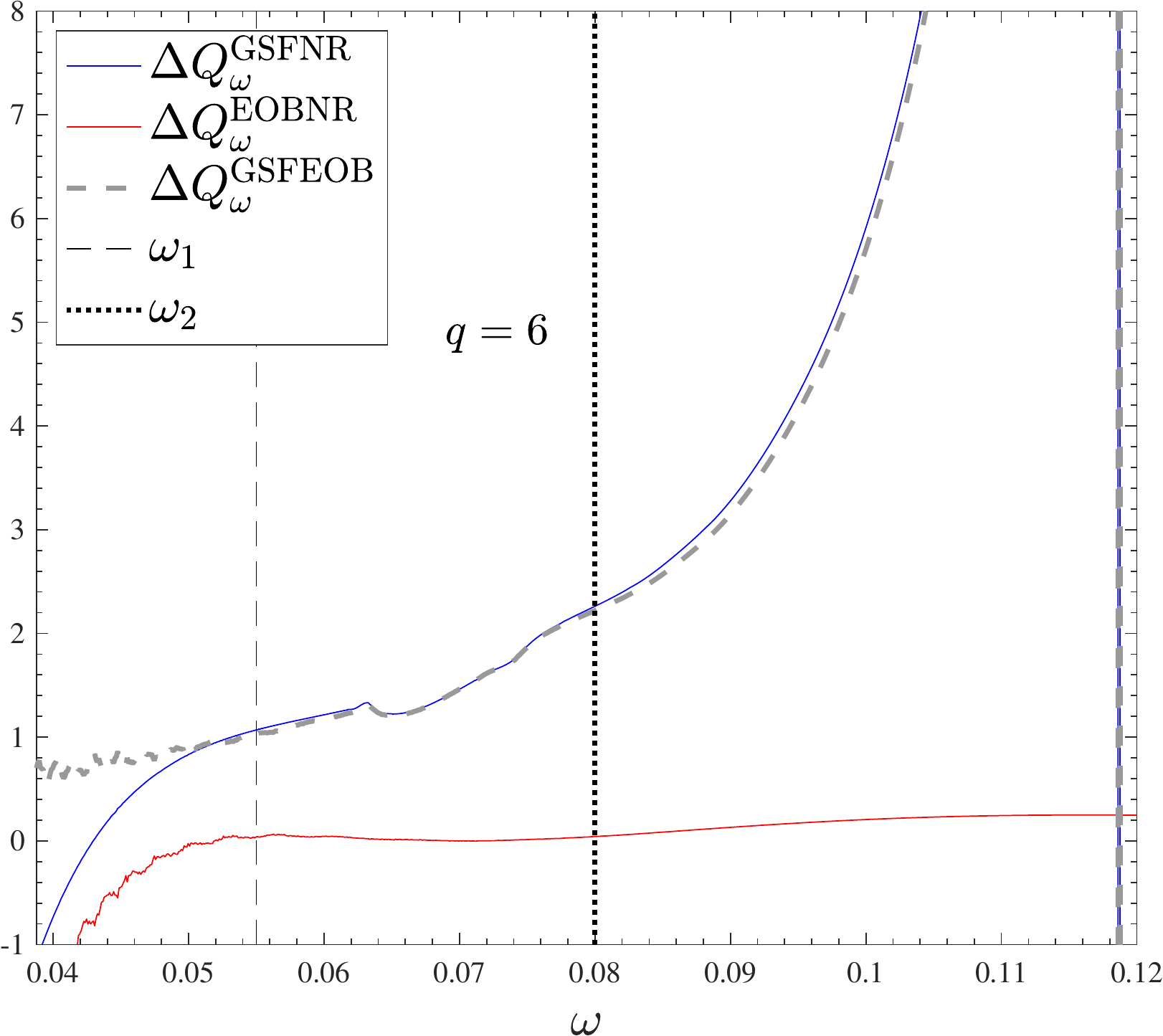}  
\caption{\label{fig:DQomg_q}EOB/NR/GSF phasing comparisons for mass ratios $q=(1,4,6)$. Consistently with the analysis for
$q=7$ and $q=10$ presented in Fig.~\ref{fig:Qomg_all}, the \PAT{} curve is always above either the NR or EOB ones. The dephasings accumulated over the interval $(\omega_1,\omega_2)$ are listed in Table\ref{tab:Dphi_comparable}.}
\end{figure*}

In this Appendix we present the EOB/GSF/NR phasing comparison for comparable-mass binaries
up to $q=1$. The results presented here complement the analysis for $q=7$ and $q=10$ in the main
text. We work with the $Q_\omega$ obtained from the $\psi_4^{22}$ phase. The \PAT{} curves are
obtained by straight differentiation and then application of a low-pass filter to remove unphysical 
high-frequency noise\footnote{This noise is mostly due to the derivation of the interpolated GSF phase.} and get smooth curves. 
The NR curves are directly obtained from the 
coefficients of Table~III of Ref.~\cite{Damour:2012ky} used in our Eq.~\eqref{eq:Qom_ratio} above.
The top panels of Fig.~\ref{fig:DQomg_q} compare the EOB, GSF and NR $\Qo$. The bottom
panels show the various differences. Consistently with the $q=7$ and $q=10$ cases, while NR and EOB are
in very good agreement, the GSF curves are always above them with nonnegligible differences 
at all frequencies. Note that in the $q=6$ case the NR curve is not reliable below
$\omega=0.05$ due to edge effects in the fitting procedure. This is not a problem here, but one has to remember that the focus of Ref.~\cite{Damour:2012ky}
was to obtain a reliable $\Qo$ for high-frequencies and not for low frequencies (where the NR noise
is typically larger) and thus the cleaning interval was optimized for this.

%===================
% Fig: Qomg comparisons
%===================
%======================
%===================
% Fig: phasing for q = 1
%===================
\begin{figure}[t]
\center
\includegraphics[width=0.45\textwidth]{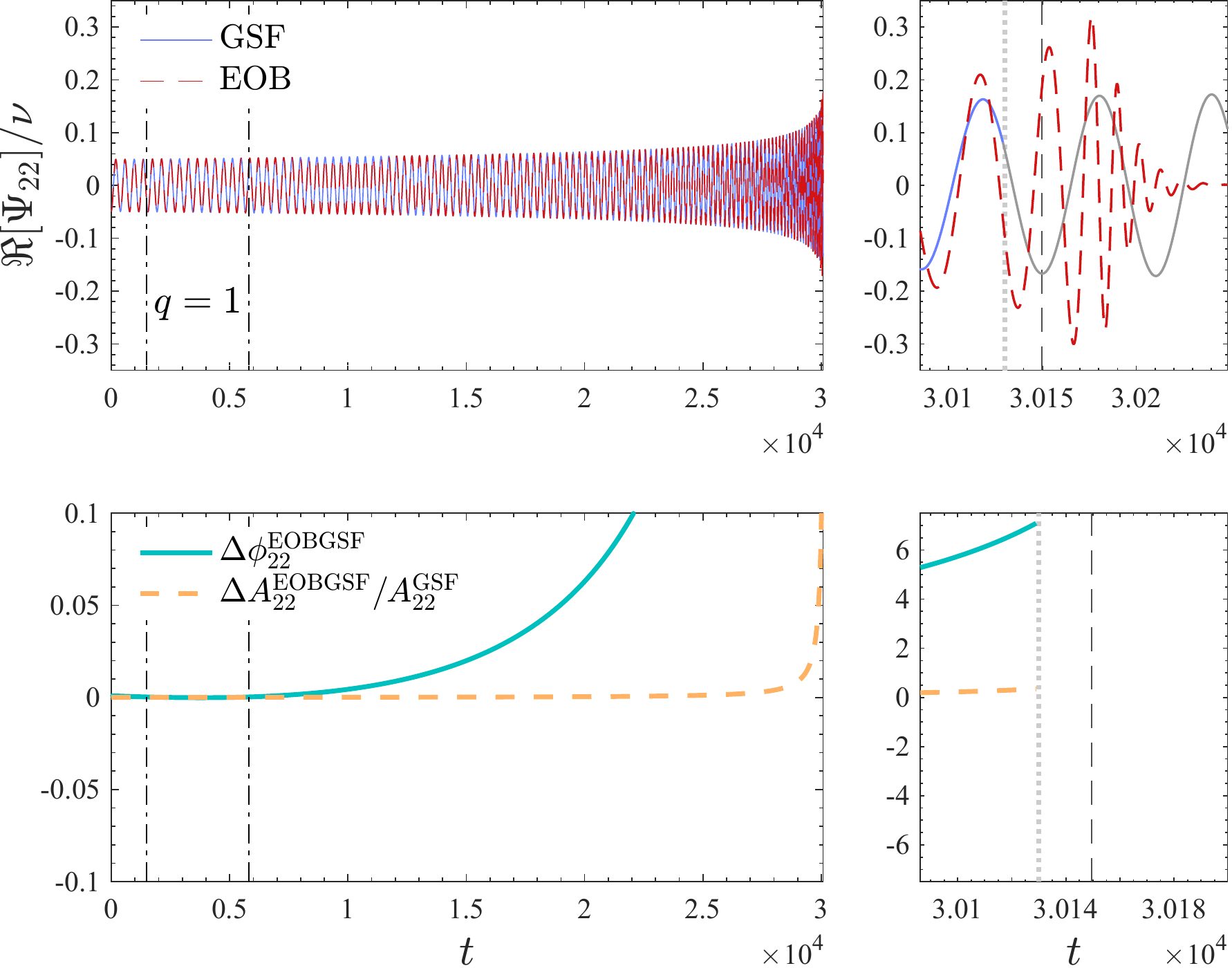}  
\caption{\label{fig:phasing_q1} EOB/GSF time-domain phasing for $q = 1$. The alignment interval is 
$[\omega_L, \omega_R] = [0.0163, 0.0173]$, the breakdown frequency is  $\omega_{22}^{\rm GSF_{\rm break}} = 0.0993$,
and the accumulated phase difference up to this frequency is $\Delta\phi_{22}^{\rm EOBGSF} = 7.1498$ rad.} 
\end{figure}
%======================
%======================================
% Figs: Comparison comparable mass case
%======================================
\begin{figure*}[t]
\center
\includegraphics[width=0.45\textwidth]{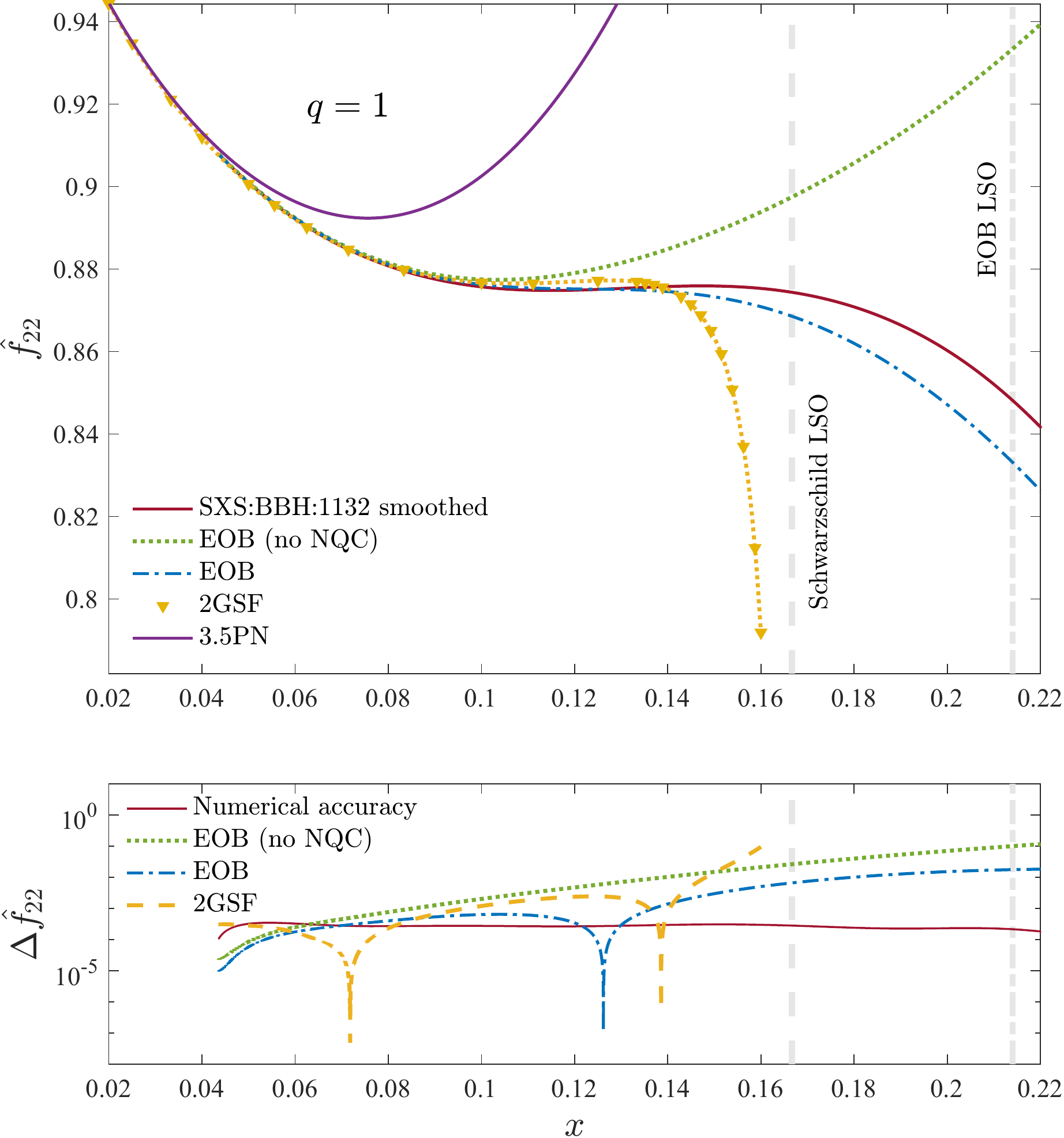} 
\qquad
\includegraphics[width=0.45\textwidth]{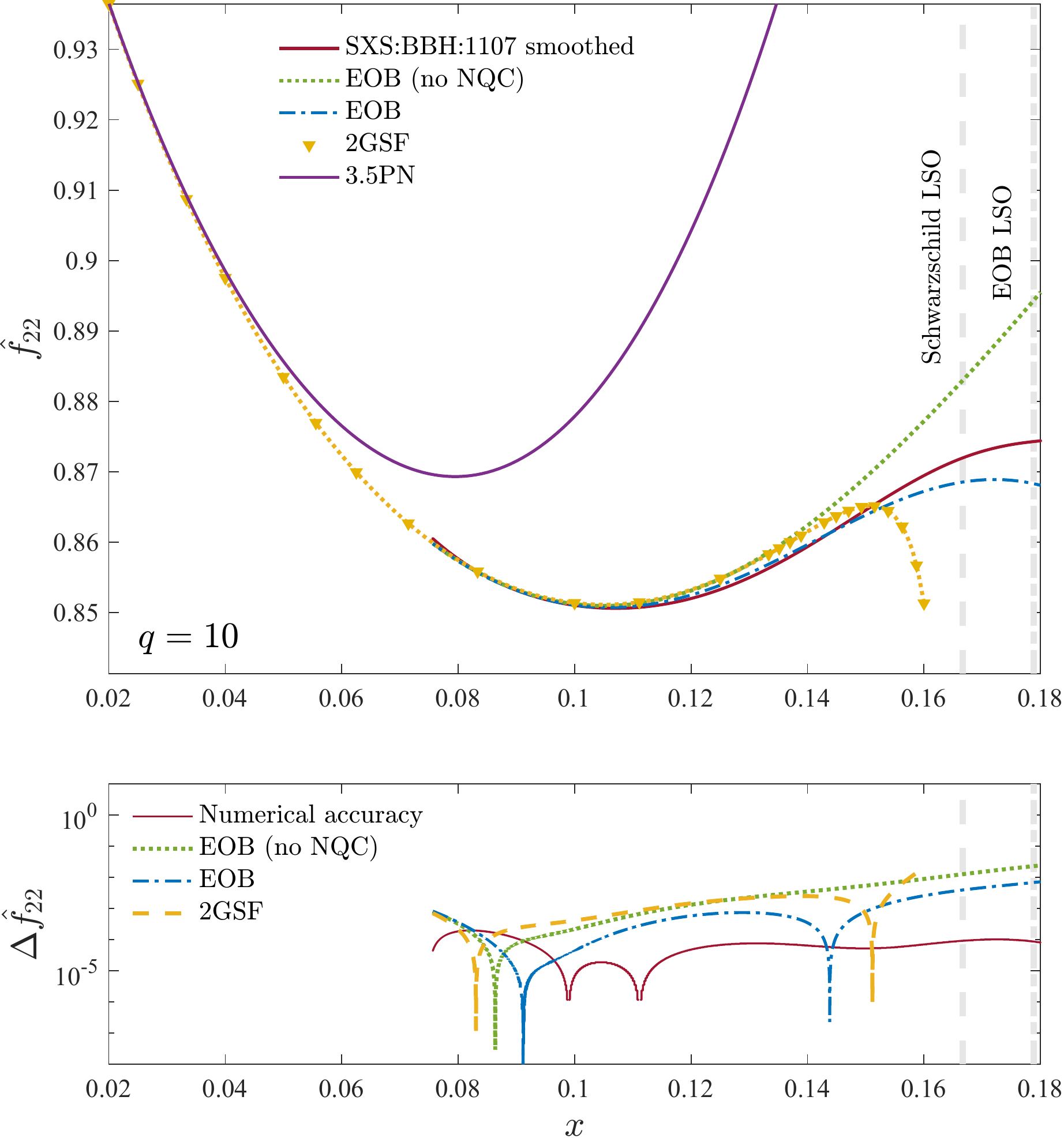} 
\caption{\label{fig:nonspin_fluxes}Newton normalized energy fluxes: EOB/NR/GSF and 3.5PN $\ell=m=2$ comparison for 
the two configurations of Table~\ref{SXSdata}. Note that the NR curves were post-processed to remove low-frequency 
spurious oscillations following Ref.~\cite{Albertini:2021tbt}.
The sharp downward trend of the GSF fluxes near the Schwarzschild LSO is due to the breakdown of the two-timescale approximation in this region.
The bottom panel shows the fractional difference between each
analytic description and the NR flux. The NR uncertainty is obtained by taking the difference between the highest and second 
highest resolutions available. Note the EOB/NR consistency also during the late-inspiral and plunge phase. }
\end{figure*}
%=======================================
%========
% Fig: IMR fluxes
%========
\begin{figure*}[t]
\center
\includegraphics[width=0.326\textwidth]{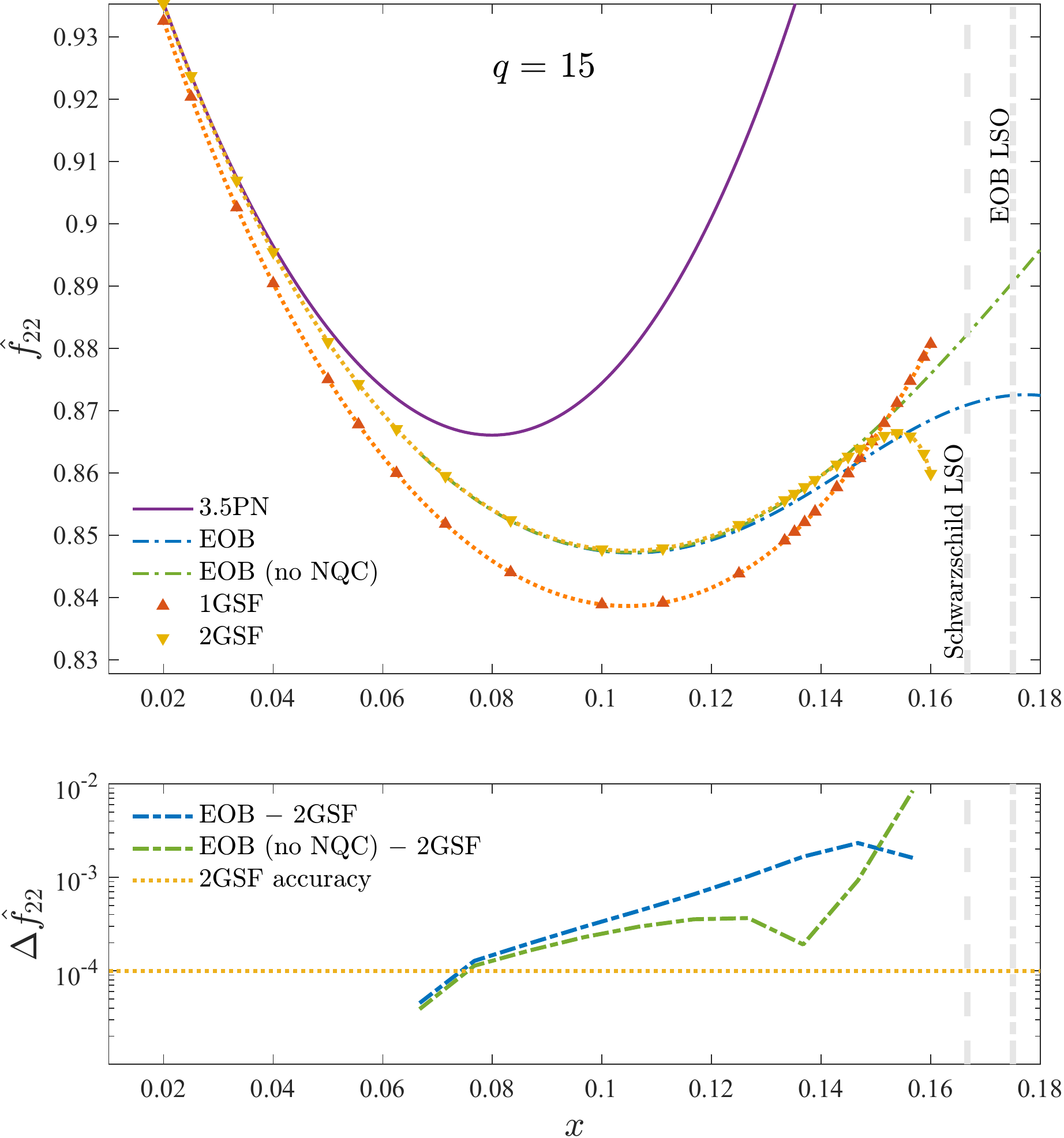} 
\includegraphics[width=0.326\textwidth]{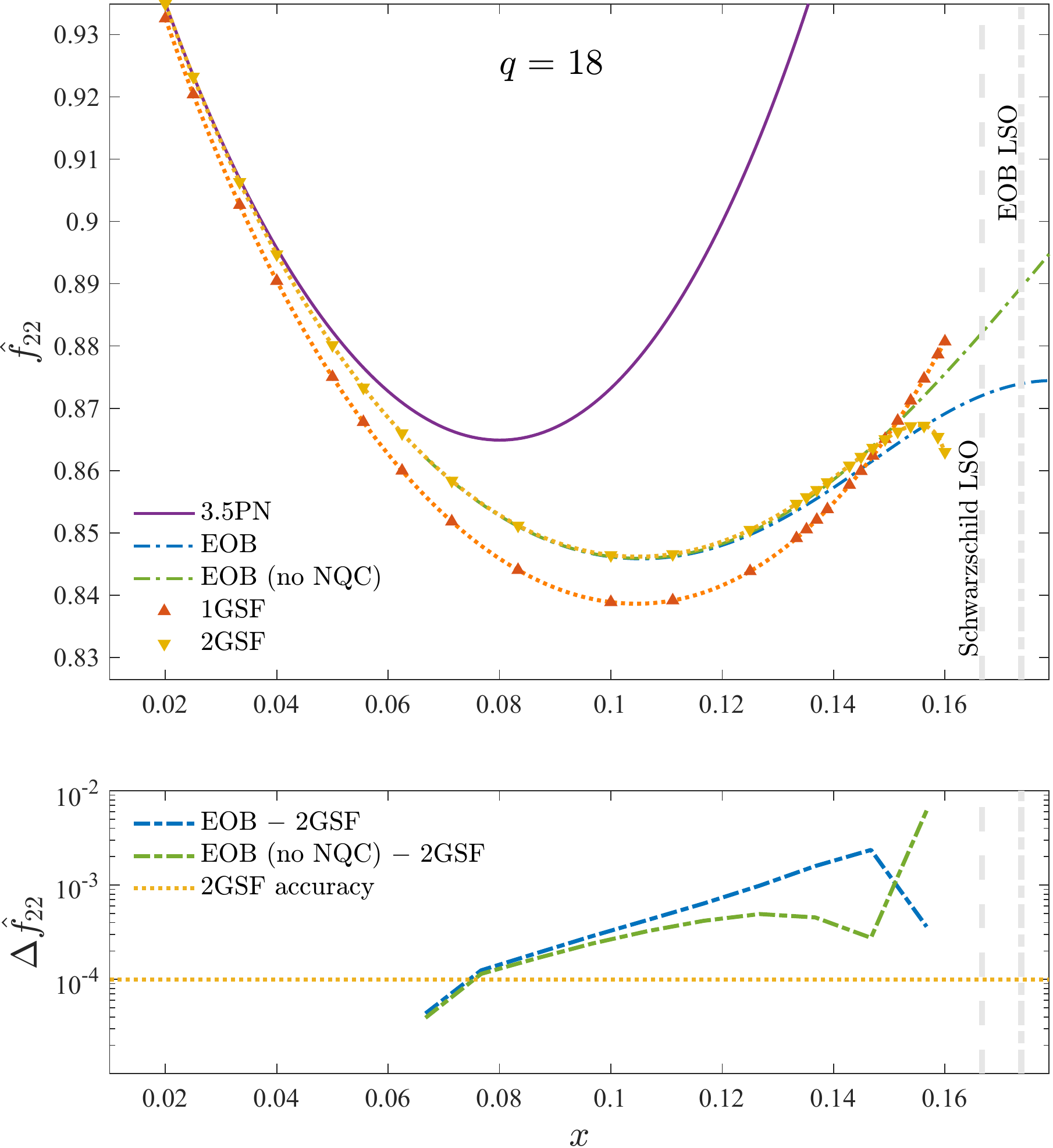} 
\includegraphics[width=0.326\textwidth]{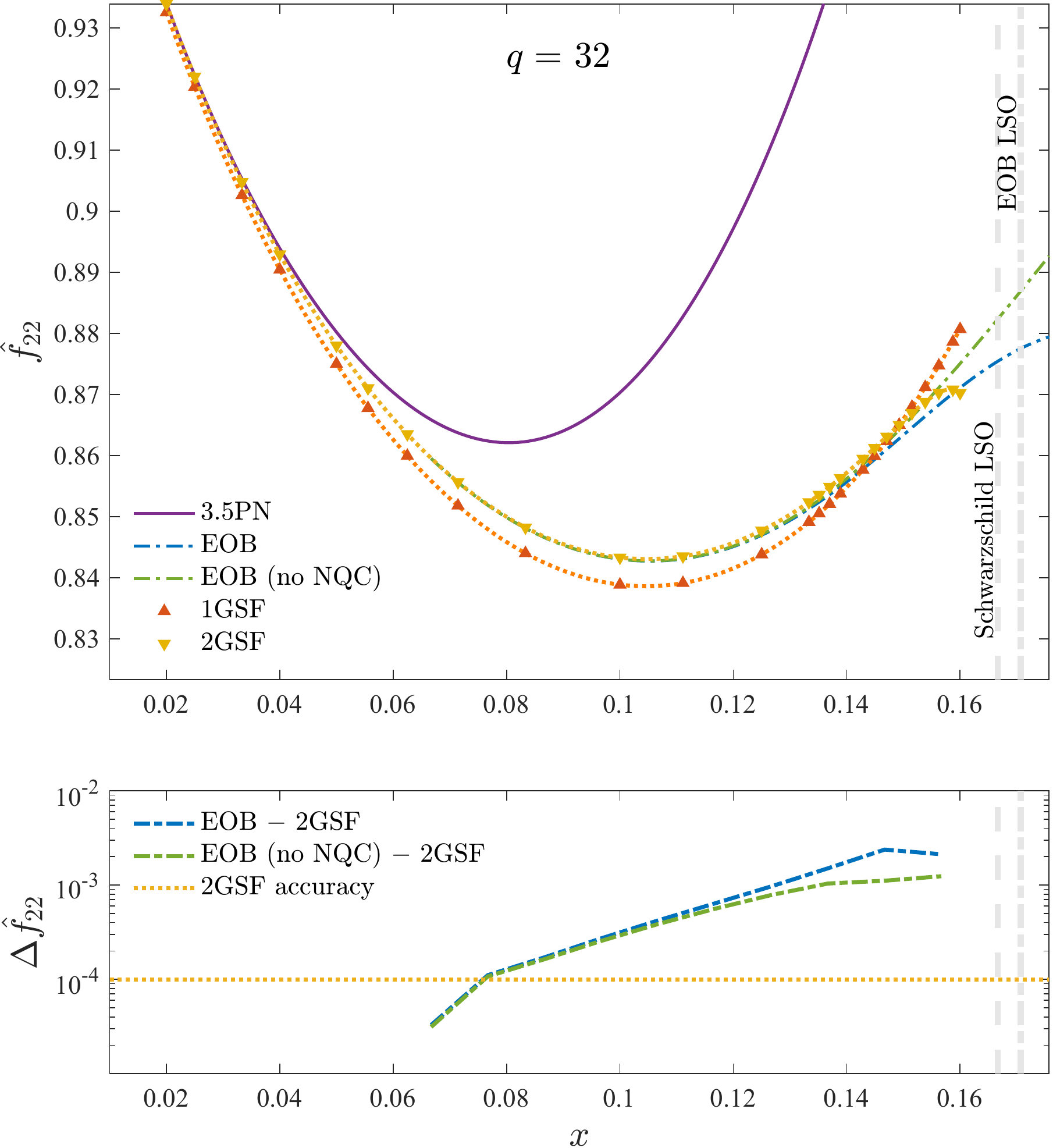}  \\
\includegraphics[width=0.326\textwidth]{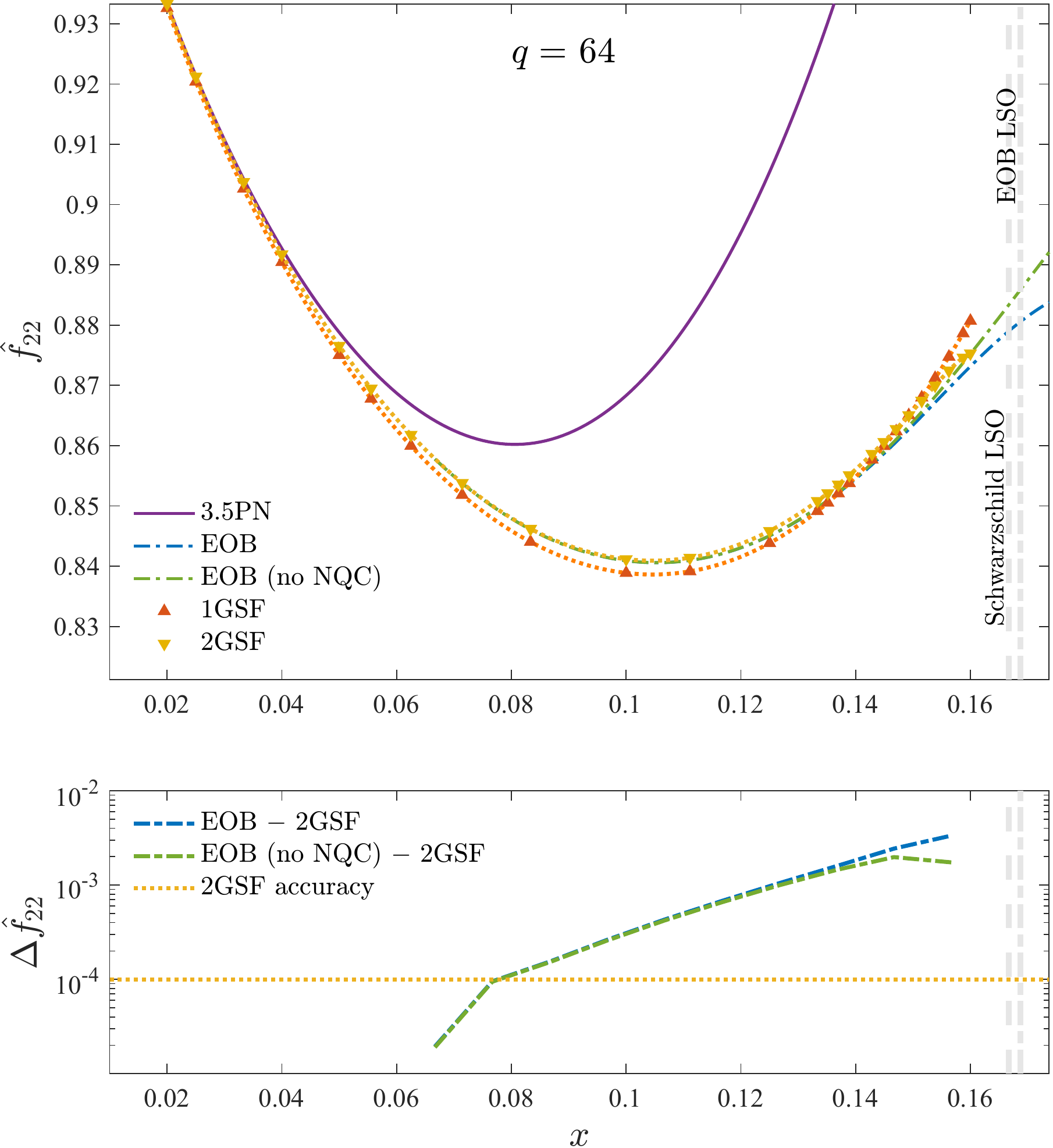}  
\includegraphics[width=0.326\textwidth]{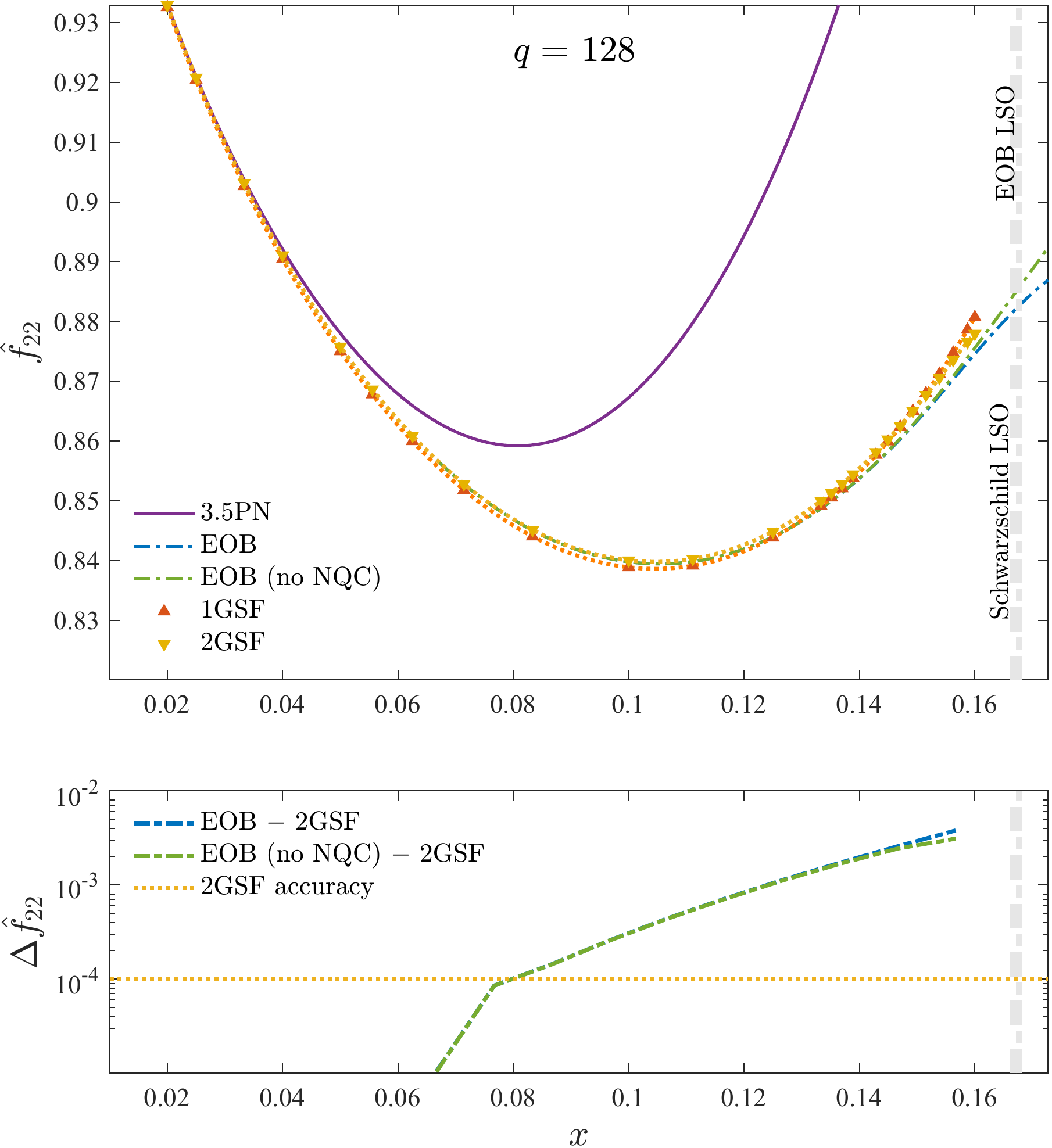}  
\includegraphics[width=0.326\textwidth]{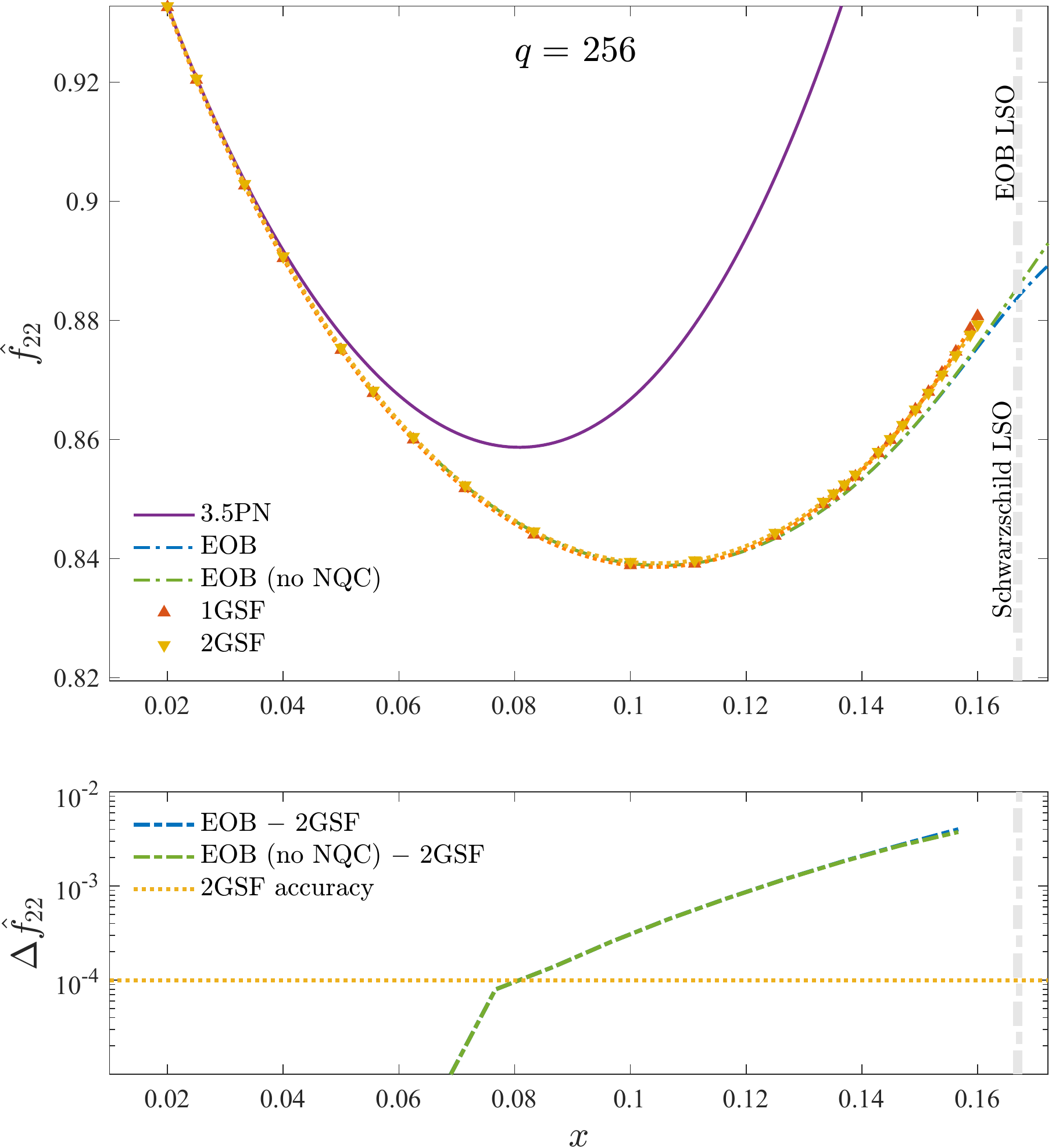}  
\caption{\label{fig:q_high}Newton-normalized energy fluxes for larger mass ratios, not covered by SXS simulations (apart from the $q=15$ one). 
We are including (i) the EOB fluxes with and without the NQC factor, (ii) both the 1GSF and the 2GSF results, and (iii) the 3.5PN result.}
\end{figure*}
%======================

%========================================
% Table of GSF frequencies and phasing accumulated
%========================================
\begin{table}[t]
\begin{center}
\begin{ruledtabular}
\begin{tabular}{c c c c  c}
$q$ & $(\omega_1,\omega_2)$ & $\Delta\phi^{\rm GSFNR}$ & $\Delta\phi^{\rm GSFEOB}$ & $\Delta\phi^{\rm EOBNR}$ \\
\hline
\hline
1 &  $(0.05,0.08)$ & 1.926 & 1.904 & 0.021\\
4 & $(0.05,0.08)$ & 0.988  & 0.989  &$ -0.00062$ \\
6 & $(0.055,0.08)$ & 0.547  & 0.538 & 0.0087  
\end{tabular}
\end{ruledtabular}
\end{center} 
\caption{\label{tab:Dphi_comparable}Phase differences (in radians) accumulated integrating the curves 
of Fig.~\ref{fig:DQomg_q} between $(\omega_1,\omega_2)$.}
\end{table}
%=========================================
The visual disagreement between GSF and NR visible in Fig.~\ref{fig:DQomg_q} leads to the phase
differences listed in Table~\ref{tab:Dphi_comparable}, which were obtained integrating by the $\Delta\Qo$ differences over the interval
$(\omega_1,\omega_2)$. This confirms that, over the frequency range considered in the figure, the \PAT{}
model has a phase error of order $\sim 1$~rad.

Given the consistency between EOB and NR for nonspinning comparable-mass 
binaries~\cite{Damour:2009kr,Damour:2012ky,Nagar:2017jdw,Nagar:2018zoe,Nagar:2019wds,Nagar:2020pcj},
we assume that \TEOBResumS{} also yields a faithful representation of the motion for low frequencies
and use it to benchmark the GSF inspiral at lower frequencies than are available in NR simulations.
The time-domain phasing for $q=1$ is shown in
Fig.~	\ref{fig:phasing_q1}, with an accumulated phase difference $\Delta\phi_{22}^{\rm EOBGSF} = 7.1498$ rad 
up to the breakdown frequency  $\omega_{22}^{\rm GSF_{\rm break}} = 0.0993$. The integration
of $\Qo$ in the frequency domain yields a final dephasing $\Delta\phi_{22,\Qo}^{\rm EOBGSF} = 7.1489$ rad,
consistent with the time-domain result. This reinforces the danger of applying the 1PAT1 model over frequency intervals extending far into the weak field, as discussed in the body of the paper: though the dephasing is slower in the weak field, it becomes unbounded as the initial frequency tends to zero.

Like the results for $q=7$ and $q=10$, the results in this section can be compared to (and are compatible with) Fig.~4 of Ref.~\cite{Wardell:2021fyy}.

\section{Analysis of energy fluxes}
\label{sec:flux}
Let us finally collect some results concerning comparisons between EOB, GSF and NR fluxes.
The main purpose of this analysis
is to compare the description of the $\ell=m=2$ mode of the flux yielded by \TEOBResumS{}, \PAT{} and NR.
This analysis should be seen as a complement to the NR/PN/GSF comparison recently
presented in Ref.~\cite{Warburton:2021kwk}, and as a consistency 
check of the waveform analysis discussed in the main text. One advantage of this analysis is that the calculation of 2GSF fluxes in  Ref.~\cite{Warburton:2021kwk} did {\it not} use any of the ``additional'' approximations described in Sec.~\ref{sec:1PA approximations}; the fluxes should therefore be exact at 1PA order (up to numerical error).

\subsection{Definitions}
%================
% Table of fluxes used
%================
\begin{table}[t]
\begin{center}
\begin{ruledtabular}
\begin{tabular}{c c c c c c c r}
SXS ID & $q$  & $\text{Lev}_h$ & $\text{Lev}_l$ & $n$ & $\langle \Delta \hat{f}_{22}^{\rm R-C} \rangle$ \\
\hline
\hline
SXS:BBH:1132 & 1  & 4 & 3 & 9 & $7.08 \cdot 10^{-5}$  \\
SXS:BBH:1107 & 10  & 4 & 3 & 7 & $5.06 \cdot 10^{-5}$  \\
\end{tabular}
\end{ruledtabular}
\caption{\label{SXSdata}SXS simulations used to check the EOB/GSF/NR flux consistency. The table reports: the SXS ID, the mass ratio,
the highest  (Lev$_h$) and second highest (Lev$_l$) resolutions available, the order of the polynomial used to fit the cleaned flux 
(see Ref.~\cite{Albertini:2021tbt} for details), and the average of the difference between the raw and the cleaned flux. 
We used $N =4$ extrapolation order to infinite extraction radius for each dataset.}
\end{center} 
\end{table}
%===============
The energy flux at infinity is given as a sum over all $m > 0$ multipoles as
\begin{equation}
\dot{E}^{\infty} = \frac{1}{8\pi} \sum_{\ell,m}  |\dot{h}_{\lm}|^2 , 
\end{equation}
so that the $\ell=m=2$ contribution reads
\begin{equation}
\label{fluxes_def}
\dot{E}^{\infty}_{22} = \frac{1}{8\pi} |\dot{h}_{22}|^2 \ .
\end{equation}
The GSF implementation of this formula is expanded in powers of $\nu$ and truncated at order $\nu^3$ as described in Sec.~\ref{sec:1PA approximations}.

To meaningfully compare fluxes obtained with different approaches, we consider them as functions of the frequency 
parameter
\begin{equation}
x = \left(\dfrac{\omega_{22}}{2}\right)^{2/3} \ ,
\end{equation}
that allows us to construct the (formal) Newtonian circular-orbit flux
\begin{equation}
\dot{E}_{22}^{\rm Newt} = \frac{32}{5} \nu^2 x^5 \ .
\end{equation}
It is then natural to compare the Newton-normalized quadrupolar flux
\begin{align}
\hat{f}_{22}\equiv \dfrac{\dot{E}_{22}}{\dot{E}^{\rm Newt}_{22}} \ .
\end{align}

We will compute, and compare to NR and GSF results, two types of EOB energy fluxes:
(i) the straight one making use of the \textit{full} waveform including inspiral merger and ringdown 
from Eq.~\eqref{eq:h_eob} and (ii) a non-NQC-corrected flux, obtained by only using 
$h_{22}^{\rm Newt} \hat{h}_{22}$. The Newtonian term used to normalize the second flux is a
function of $x=(\omega_{22}/2)^{2/3}$ with $\omega_{22}$ evaluated from 
$h_{22}^{\rm Newt} \hat{h}_{22}$ instead of the full waveform.

\subsection{NR/GSF/EOB/PN comparisons}
To start with, let us focus on the two NR simulations of Table~\ref{SXSdata}.
The raw numerical fluxes have spurious oscillations during the inspiral, which we remove by
a method described in Ref.~\cite{Albertini:2021tbt}. As a last step, this approach represents the flux as
a  polynomial in $x$, whose order is also displayed in Table~\ref{SXSdata}, together with the average of 
the difference between the raw and the cleaned flux. This allows for a direct check of the accuracy of
the procedure. Note that, to avoid other NR-related systematics during the inspiral, we here 
use $N = 4$ extrapolation order for all NR datasets.

Figure~\ref{fig:nonspin_fluxes} compares the fluxes for the configurations of Table~\ref{SXSdata},
also including the 3.5PN flux as a benchmark for very low frequencies.
On the basis of our discussion in Sec.~\ref{sec:1PA accuracy},
we do not expect the 2GSF fluxes to be accurate at high frequencies where transition-to-plunge effects become important.
Indeed, this is what we find, with the GSF curve diverging from the NR one at high frequencies.

There is also an appreciable difference between GSF and NR curves at lower frequencies, away from the transition-to-plunge region. In particular, $\omega\sim 0.1$, which we typically identify as an acceptable upper limit on the
reliability of the 2GSF evolution, corresponds to $x=0.1357$. For both $q=1$ and $q=10$ the flux 
difference at this frequency is visible on the plot, and is larger than the difference between EOB (with NQC) and NR.
This is consistent with our previous conclusion that EOB is closer to NR than GSF for comparable mass systems.

%==============
% Fluxes: IMR
%==============
Finally, in Fig.~\ref{fig:q_high} we display the EOB/GSF energy flux comparison for higher mass ratios,
$q =(15, 18, 32, 64, 128, 256)$.
Again we include the two types of EOB curves described above and the 3.5PN result, but also the 1GSF result. 
The latter clearly gives inconsistent results for the lower mass ratios, while its curve is drawn nearer to the 2GSF 
one as $q$ is increased. Likewise, with increasing $q$ the EOB curves tend toward the GSF ones over much of the parameter space. However, the EOB and GSF curves clearly separate near the LSO. This does not appear to be a symptom of the two-timescale expansion's breakdown at the transition to plunge; the separation between the EOB and GSF curves begins below the breakdown frequency~\eqref{Omega break} (which corresponds to $x\approx 0.123$ for $q=256$). Moreover, even at the breakdown frequency we only expect the GSF error terms ($\sim \nu^4$) to be comparable to the 2GSF ($\sim \nu^3$) term in the flux, while the EOB-GSF difference is very significantly larger than the contribution of the 2GSF term at $q=256$. We therefore conclude that the EOB-GSF difference is most likely due to missing 0PA information in the EOB model (and to a much lesser extent, missing 1PA information).

\bibliography{refs20221102.bib,local.bib}

\end{document}